\documentclass[twocolumn]{aastex62}

\newcommand{\BAYMAX}{{\tt BAYMAX}}
\shortauthors{Foord et al.}
\usepackage{relsize}

\usepackage{relsize}
\makeatletter
\providecommand{\ion}[2]{#1$\;$\textsmaller{\@Roman{#2}}}
\makeatother

\usepackage{amsmath,amssymb} %% \ltsim \la, that kind of thing.
\usepackage{graphicx,mathptmx} %% best image inclusion and best font.
\usepackage{natbib} %% bibliography styling.

\newcommand{\beq}{
\begin{equation}
}
\newcommand{\eeq}{
\end{equation}
}

\newcommand{\beqa}{
\begin{eqnarray}
}
\newcommand{\eeqa}{
\end{eqnarray}
}

\begin{document}

\title{A Second Look at 12 Candidate Dual AGNs using \BAYMAX{}}

\author[0000-0002-1616-1701]{Adi Foord}
\affil{Department of Astronomy and Astrophysics, University of Michigan, Ann Arbor, MI 48109}

\author[0000-0002-1146-0198]{Kayhan G\"{u}ltekin}
\affil{Department of Astronomy and Astrophysics, University of Michigan, Ann Arbor, MI 48109}

\author[0000-0002-2397-206X]{Rebecca Nevin}
\affil{Department of Astrophysical and Planetary Sciences, University of Colorado, Boulder, CO, 80309}

\author[0000-0001-8627-4907]{Julia M. Comerford}
\affil{Department of Astrophysical and Planetary Sciences, University of Colorado, Boulder, CO, 80309}

\author[0000-0002-2397-206X]{Edmund Hodges-Kluck}
\affil{Department of Astronomy, University of Maryland, College Park, MD 20740}
\affil{Code 662, NASA Goddard Space Flight Center, Greenbelt, MD 20771}

\author[0000-0001-8627-4907]{R. Scott Barrows}
\affil{Department of Astrophysical and Planetary Sciences, University of Colorado, Boulder, CO, 80309}

\author{Andrew D. Goulding}
\affil{Department of Astrophysics, Princeton University, Princeton, NJ, 08540}

\author{Jenny E. Greene}
\affil{Department of Astrophysics, Princeton University, Princeton, NJ, 08540}



\begin{abstract}
We present an analysis of 12 optically selected dual AGN candidates at $z < 0.34$.  Each candidate was originally identified via double-peaked [\ion{O}{3}] $\lambda$5007 emission lines, and have received follow-up \emph{Chandra} and \emph{HST} observations.  Because the X-ray data are low-count ($<100$ counts) with small separations ($<1\arcsec$), a robust analysis is necessary for classifying each source. Pairing long-slit [\ion{O}{3}] observations with existing \emph{Chandra} observations, we re-analyze the X-ray observations with \BAYMAX{} to determine whether the X-ray emission from each system is more likely a single or dual point source. We find that 4 of the 12 sources are likely dual X-ray point source systems.  We examine each point source's spectra via a Monte Carlo method that probabilistically identifies the likely origin of each photon. When doing so, we find that (i) the secondary X-ray point sources in 2 of the systems have $L_{\mathrm{X}}<10^{40}$ erg s$^{-1}$, such that we cannot rule out a non-AGN origin, (ii) one source has a secondary with $L_{\mathrm{X}}>10^{40}$ erg s$^{-1}$ but a spectrum that is too soft to definitively preclude being X-ray emitting diffuse gas that was photoionized by the primary AGN, and (iii) one system (SDSS J1126+2944) is a dual AGN. Additionally, using complementary \emph{HST} observations, we analyze a sub-sample of systems that are visually identified as merging. Our results suggest that dual AGNs may preferentially reside in mergers with small separations, consistent with both simulations and observations.
\end{abstract}

%%%%%%%%%%%%%%%%%%%%%%%%%%%%%%%%%%%%%%%%%%
%%%%%%%%%%%%%% INTRODUCTION %%%%%%%%%%%%%%
%%%%%%%%%%%%%%%%%%%%%%%%%%%%%%%%%%%%%%%%%%
\section{Introduction}
\label{sec:intro}
Dual Active Galactic Nuclei (AGNs) are systems comprised of two actively accreting supermassive black holes (SMBHs) whose host galaxies are in the process of merging.  Given that all massive galaxies are likely to have a central supermassive black hole \citep{Kormendy&Richstone1995},  dual SMBHs are thought to be a natural consequence of hierarchical galaxy formation (e.g., \citealt{WhiteandReese1978}).  Dual SMBH systems represent the earliest stages of the merger, where the SMBHs are at kiloparcsec separations and not yet gravitationally bound (see, e.g., \citealt{Begelman1980}).  As the system loses energy through dynamical friction, the separation between the two SMBHs decreases with time as both sink towards the center of the gravitational potential-well.
\par Whether or not galaxy--galaxy interactions trigger accretion onto the central SMBHs remains a topic of debate. Similar-mass (with mass ratios $>$1:4), gas-rich galaxy mergers have been shown to provide a favorable environment for the assembly of AGN pairs (e.g., \citealt{Volonteri2003, Hopkins2005, DiMatteo2005}) and this hypothesis has been supported by studies of nearby galaxies (e.g., \citealt{Koss2010, Ellison2017, Satyapal2014, Goulding2017}).  However, other studies that target higher-redshift ($z>1$) galaxies over a wide range of AGN luminosity conclude there is no special relation between SMBH activity and host galaxy interactions (e.g., \citealt{Cisternas2011, Kocevski2012, Schawinski2012, Villforth2014}).  It is likely that these contradictory results are due to the variability of the AGNs activity during the lifetime of the merger \citep{Goulding2017}, as the AGNs may not be `on' during the entire merger event.  In this framework, the probability of AGN observability should increase as a function of decreasing separation, which has been supported by both simulations and observations (e.g. \citealt{Koss2012, Blecha2013, Ellison2013, Goulding2017, Capelo2017, Barrows2017}).
\par Understanding which environmental factors are most important for dual SMBH activity allows for a better understanding about black hole growth and its relation (or lack thereof) to galaxy--galaxy interactions.  Additionally, as progenitors to SMBH mergers, the rate of dual AGNs has implications for the rate of expected gravitational wave events that will be detected by pulsar timing arrays (e.g., \citealt{Mingarelli2019}) and future space-based interferometry.
\par There exist many multi-wavelength techniques to detect dual AGN candidates, each with their own caveats.  The most popular technique is to use optical spectroscopy to search for double-peaked narrow line emission regions (which can sometimes be spatially resolved; see, e.g., \citealt{Zhou2004, Gerke2007, Comerford2009, Liu2010b, Fu2012, Comerford2012, Comerford2013, Barrows2013}).  Dual AGN systems can display two sets of narrow line emission regions, such as [\ion{O}{3}] $\lambda$5007, during the period of the merger when their narrow line regions (NLRs) are well separated in velocity.  Here, the separation and width of each peak will depend on parameters such as the distance between the two AGNs. However, double-peaked emission features are known to originate from other processes, such as bipolar outflows and rotating disks  (\citealt{Greene&Ho2005a, Rosario2010, MullerSanchez2011, Smith2012,Nevin2016}).
\par Thus, confirmation of dual AGN systems requires spatially resolving each individual AGN; beyond $z>0.05$ high-resolution imaging is necessary, which can be accomplished with both radio or X-ray observations. Radio observations can resolve radio-emitting cores on the smallest spatial scales (see \citealt{Rodriguez2006, Rosario2010, Tingay&Wayth2011, Fu2011, Fu2015, Deane2014, Gabanyi2014, Wrobel2014a, Wrobel2014b, MullerSanchez2015, Kharb2017}), however this technique is only efficient for radio-loud AGNs ($\approx 15\%$ of the AGN population; \citealt{Hooper1995}), and AGNs can only be differentiated from jet components at radio frequencies if they are compact and have flat or inverted spectral indices (see, e.g., \citealt{Burke-Spolaor2011,Hovatta2014}).  Indeed, this is further complicated by the fact that regions of intense starbursts can mimic both compactness and brightness temperatures of AGNs; thus complementary IR data may be necessary to properly classify the source (see, e.g., \citealt{Varenius2014})
\par A more efficient method is to use X-ray observations taken with the \emph{Chandra X-ray Observatory} (\emph{Chandra}).  X-rays are one of the most direct methods of finding black holes, as AGNs are one of the few sources that emit at X-ray luminosities above $10^{40}$ erg s$^{-1}$ \citep{Lehmer2010, Lehmer2019}.  However, the detection of the most closely separated pairs is limited by the instrument's Point Spread Function (PSF).  Even with \emph{Chandra}'s superior angular resolution (where the half-power diameter of \emph{Chandra}'s Advanced CCD Imaging Spectrometer, ACIS, is $\sim$0\farcs{8} at $\sim$1 keV), systems with physical separations less than 1 kpc become difficult to resolve beyond $z\ge0.05$.  
\par Currently, many analyses on \emph{Chandra} observations of dual AGN candidates implement the Energy-Dependent Subpixel Event Respositioning (EDSER) algorithm \citep{Li2004}.  EDSER reduces photon impact position uncertainties to subpixel accuracy, and in combination with \emph{Chandra}'s dithering can resolve sub-pixel structure down to the limit of the \textit{Chandra} High Resolution Mirror Assembly.  However, without a robust and statistical approach to analyze the \emph{Chandra} observations, the interpretation of dual AGNs with separations $<1\arcsec$ can lead to false negatives and false positives, even after undergoing EDSER reprocessing. This issue is amplified in the low-count regime ($<$ 100 counts), where even dual AGNs with large separations are difficult to identify.
\par As a result, very few dual AGNs have been confirmed to date, with the majority of systems at separations $>$ 1 kpc. (see \citealt{Deane2014}).  Thus, we have developed a {\tt PYTHON} tool \BAYMAX{} (\textbf{B}ayesian \textbf{A}nal\textbf{Y}sis of \textbf{A}GNs in \textbf{X}-rays) that allows for a rigorous analysis of whether a source in a given \emph{Chandra} observation is more likely composed of one or two X-ray point sources (see \citealt{Foord2019}).  \BAYMAX{} is capable of detecting dual X-ray point source systems for systems with low flux ratios between the primary and secondary, as well for systems with angular separations smaller than \emph{Chandra}'s half-power diameter. 
\par In this paper we present an analysis of 12 optically selected dual AGN candidates that have existing archival \emph{Chandra} data. The \emph{Chandra} observations of these 12 targets were originally analyzed in \cite{Comerford2015}, using a simpler PSF model and source identifier technique. Using this approach, one of the twelve systems was classified as a likely dual AGN \citep{Comerford2015}.  We now re-analyze the \emph{Chandra} observations using \BAYMAX{}, with the goal of identifying other dual AGN systems using a robust statistical analysis.  Combining the X-ray observations with archival \emph{Hubble Space Telescope} (\emph{HST}) and \emph{Wide-field Infrared Survey Explorer} (\emph{WISE}) observations, we aim to learn more about the preferential environments of each dual AGN candidate.
\par The remainder of the paper is organized into 5 sections. In section 2 we introduce the sample and the existing multi-wavelength coverage. In section 3 we review Bayesian inference, Bayes factor and how \BAYMAX{} calculates the likelihoods. In section 4 we present our results from running \BAYMAX{} on the \emph{Chandra} observations, review the best-fit parameters for each model, and quantify the strength of each result. In section 5 we discuss the nature of each dual AGN candidate by evaluating the spectral fits and discussing possible sources of contamination.  In section 6 we discuss the sensitivity and limitations of \BAYMAX{} across parameter space, and compare environmental properties between the dual AGN candidates and the single AGN candidates. Lastly, we summarize our findings in section 7. Throughout the paper we assume a $\Lambda$CDM universe, where $H_{0}=69.6$, $\Omega_{M}=0.286$, and $\Omega_{\Lambda}=0.714$

%%%%%%%%%%%%%%%%%%%%%%%%%%%%%%%%%%%%%
%%%%%%%%%%%%%% SAMPLE %%%%%%%%%%%%%%
%%%%%%%%%%%%%%%%%%%%%%%%%%%%%%%%%%%%%
\section{Sample}
\label{sec:sample}
The sample of galaxies studied in this paper was created from a larger parent sample of 340 AGNs, which all have double-peaked narrow emission lines identified via the Sloan Digital Sky Survey (SDSS; \citealt{Wang2009, Liu2010a, Smith2010}).  Using follow-up long-slit spectroscopy with the Lick 3 m telescope; the Apache Point Observatory 3.5 m
telescope; the Palomar 5 m telescope; the MMT 6.5 m telescope; and the Magellan II 6.5 m telescope, galaxies were chosen if their [\ion{O}{3}] $\lambda$5007 emission components were separated by $\Delta x_{\mathrm{[O~III]}}$ $>$0\farcs{75} \citep{Greene2011, Shen2011, Comerford2012}, making them more easily resolved by \emph{Chandra}.  The sample was further filtered by enforcing a $2$--$10$ keV flux limit of $F_{2-10} > 8\times10^{-15}$ erg cm$^{-2}$ s$^{−1}$, where $F_{2-10}$ was estimated using the [\ion{O}{3}] $\lambda$5007 fluxes \citep{Heckman2005}. For more details regarding the [\ion{O}{3}] $\lambda$5007 data analysis and sample cuts, we refer the reader to \cite{Comerford2015} and references therein.
%
%%%%%%%%%%%%%%%%%%%%%%%%%%%%%%%%%%%%%%%%%%%%%%
%%%%%% TABLE GALAXY INFO %%%%%%%%%%%%%%%%%%%%%
%%%%%%%%%%%%%%%%%%%%%%%%%%%%%%%%%%%%%%%%%%%%%%
\begin{table*}[t]
\begin{center}
\caption{Galaxy Sample Properties}
\label{tab:info}
\small
\begin{tabular*}{0.76\textwidth
}{cccccc}
	\hline
	\hline
	\multicolumn{1}{c}{Galaxy Name} & \multicolumn{1}{c}{Redshift} &
	\multicolumn{1}{c}{$D_{A}$ (Mpc)} & \multicolumn{1}{c}{\emph{Chandra} Obs. ID} & \multicolumn{1}{c}{\emph{Chandra} Exp. Time (s)} & \multicolumn{1}{c}{\emph{HST}}\\
	\multicolumn{1}{c}{(1)} & \multicolumn{1}{c}{(2)} & \multicolumn{1}{c}{(3)} & \multicolumn{1}{c}{(4)} & \multicolumn{1}{c}{(5)} & \multicolumn{1}{c}{(6)} \\
	\hline
	SDSS J014209.01$-$005050.0 & 0.133 & 490.8 & 13959 & 19804 & Yes\\
	SDSS J075223.35+273643.1 & 0.069 & 273.8 & 12826 & 29650 & No \\
	SDSS J084135.09+010156.2 & 0.111 & 419.9 & 13950 & 19801 & Yes \\
	--- & --- & --- & 18199 & 21940 & --- \\
	SDSS J085416.76+502632.0 & 0.096 & 369.4 & 13956 & 20078 & Yes \\
	SDSS J095207.62+255257.2 & 0.339 & 1007.0 & 13952 & 19807 & Yes \\
    SDSS J100654.20+464717.2 & 0.123 & 459.0 & 13957 & 19783 & Yes \\
    SDSS J112659.54+294442.8 & 0.102 & 389.8 & 13955 & 19798 & Yes \\
    SDSS J123915.40+531414.6 & 0.201 & 688.6 & 13953 & 19804 & Yes \\
    SDSS J132231.86+263159.1 & 0.144 & 524.9 & 13958 & 19807 & Yes \\
    SDSS J135646.11+102609.1 & 0.123 & 459.0 & 13951 & 19804 & Yes \\
    --- & --- & --- & 17047 & 34840 & --- \\
    --- & --- & --- & 18826 & 42870 & --- \\
    SDSS J144804.17+182537.9 & 0.038 & 156.4 & 13954 & 19807 & Yes \\
    SDSS J160436.21+500958.1 & 0.146 & 531.1 & 12827 & 29582 & No \\

	\hline 
\end{tabular*}
\end{center}
Note. -- Columns: (1) SDSS galaxy designation; (2) redshift; (3) angular diameter distance; (4)  \emph{Chandra} Observation ID; (5) exposure time of \emph{Chandra} observation; (6) \emph{HST}/WFC3 data available. 
\end{table*}
%%%%%%%%%%%%%%%%%%%%%%%%%%%%%%%%%%%%%%%%%%%%%%
%%%%%%%%%%%%%%%%%%%%%%%%%%%%%%%%%%%%%%%%%%%%%%

\par Ultimately, the final sub-sample is composed of 13 galaxies, each of which received \emph{Chandra} observations over two separate programs (GO1-12142X, PI: Gerke; GO2-13130X, PI: Comerford).  The analysis of one of these galaxies, SDSS J171544.05+600835.7, was presented in \cite{Comerford2011}, where they confirm that the system is likely a dual AGN. The \emph{Chandra} observations of the 12 remaining systems were analyzed in \cite{Comerford2015}, where 1 of the 12 systems (SDSS J112659.54+294442.8, hereafter SDSS J1126+2944) was classified as a dual AGN. Here, we re-visit the 12 galaxies presented in \cite{Comerford2015}; using \BAYMAX{} we aim to (i) identify new dual X-ray point sources and (ii) re-evaluate the true nature of SDSS J1126+2944.  The galaxies are located at redshifts $0.04 < z < 0.34$, and two of them are classified as Type 1 AGNs by their SDSS spectra (SDSS J095207.62+255257.2 and SDSS J123915.40+531414.6, hereafter SDSS J0952+2552 and SDSS J1239+5314) while the others are classified as Type 2 AGNs.  In addition to \emph{Chandra} data, 11 of the galaxies were also observed with multiband HST/WFC3 imaging to examine the host galaxies (see \citealt{Comerford2015}). For more information about each source, please see  Table~\ref{tab:info}.
\subsection{X-ray Data Analysis}
For each galaxy, the \emph{Chandra} exposure times were chosen such that both AGNs in a given dual AGN candidate should have at least 15 counts. They were observed with over the course of two programs, GO1-12142X (PI: Gerke) and GO2-13130X (PI: Comerford).  We looked for additional archival \emph{Chandra} observations for these targets, and found them for SDSS J084135.09+010156.2 (hereafter SDSS J0841+0101; PI: Satyapal) and SDSS J135646.11+102609.1 (hereafter SDSS J1356+1026; PI: Greene).  Incorporating these newer observations (see Section~\ref{sec:methods}) increases the total number of counts to analyze and gives \BAYMAX{} greater sensitivity across parameter space. 
\par Each galaxy observation was on-axis and placed on the back-illuminated S3 chip of the ACIS detector.  We follow a similar data reduction as described in previous \emph{Chandra} analyses on AGNs (e.g., \citealt{Foord2017, Foord2017b, Foord2019}), using \emph{Chandra} Interactive Analysis of Observations ({\tt CIAO}) v4.8 \citep{Fruscione2006}. Further, all observations are reprocessed with EDSER.
\par We first correct for astrometry, cross-matching the \emph{Chandra}-detected point-like sources with the SDSS Data Release 9 (SDSS DR9) catalog. The \emph{Chandra} sources used for cross-matching are detected by running {\tt wavdetect} on the reprocessed level-2 event file. We require each observation to have a minimum of 3 matches with the SDSS DR9, and each matched pair to be less than 2\arcsec~from one another.  Each galaxy meets the criterion for astrometrical corrections, and the resultant astrometric shifts are shift less than 0\farcs5.  Background flaring is deemed negligible for each observation, as there are no time intervals where the background rate is 3$\sigma$ above the mean level. We then rerun {\tt wavdetect} to generate a list of X-ray point sources.  For each observation, {\tt wavdetect} identifies an X-ray point source coincident with the SDSS-listed optical center. 
%

%%%%%%%%%%%%%%%%%%%%%%%%%%%%%%%%%%%%%
%%%%%%%%%%%%%% METHODS %%%%%%%%%%%%%%
%%%%%%%%%%%%%%%%%%%%%%%%%%%%%%%%%%%%%
\section{Methods}
\label{sec:methods}
\BAYMAX{} uses a Bayesian approach to analyze a given \emph{Chandra} observation and estimate the likelihood that is it better described by one or multiple point sources.
In the following section we review \BAYMAX{}'s capabilities with regards to our 12 specific systems. In general, however, \BAYMAX{} is flexible to include other models and/or prior distributions. For a more detailed review on the statistical techniques behind \BAYMAX{}'s calculations, please see \cite{Foord2019}. 
%%%%%% TABLE SYMBOLS %%%%%%
\begin{table*}[t]
\begin{center}
\caption{Symbols}
\label{tab:symbols}
\small
\begin{tabular*}{0.8\textwidth
}{l@{\extracolsep{\fill}}l}
	\hline
	\hline
	\multicolumn{1}{c}{Symbol} & \multicolumn{1}{c}{Definition} \\
	\multicolumn{1}{c}{(1)} & \multicolumn{1}{c}{(2)} \\
	\hline
	($x_{i}, y_{i}$) & Sky coordinate of photon $i$ \\
	$E_{i}$ & Energy of photon $i$,  in keV \\
	$n$ & Total counts of given source \\
	$\mu$	& Central position of given point source in sky coordinates (2D; $\mu = [\mu_{x}, \mu_{y}]$)	\\
	$k$ & Number of \emph{Chandra} observations being modeled\\
    $\Delta x_{K}$   &  Translational astrometric shift in $x$ ($K=[1, \dots,  k-1]$) \\
    $\Delta y_{K}$   &  Translational astrometric shift in $y$ \\
	$f_{BG}$	& Total count ratio between a given background component and the point source components	\\
	$M_{j}$ & Given model being analyzed by \BAYMAX{} \\
	$\theta_{j}$    & Parameter vector for $M_{j}$, i.e. [$\mu$, $f$, $f_{BF}$, $\Delta x_{K}$, $\Delta y_{K}$] . \\
	\hline 
\end{tabular*}
\end{center}
Note. -- Columns: (1) symbols used throughout the text; (2) definitions. 
\end{table*}
%%%%%%%%%%%%%%%%%%

\subsection{Bayesian Inference}
\par In order to determine the likelihood of a dual X-ray point source, \BAYMAX{} calculates the Bayes factor ($BF$). The $BF$ represents the ratio of the marginal probability density of the observed data $D$ under one model, to the marginal density under a second model. Here, each model is parameterized by a parameter vector, $\theta$. For our analyses on dual AGN candidates, the two models are a dual point source model ($M_{2}$) vs. a single point source model ($M_{1}$):
\begin{equation}
BF = \frac{\int P(D\mid\theta_{2},M_{2}) P(\theta_{2}\mid M_{2}) d\theta_{2}}{\int P(D\mid\theta_{1},M_{1}) P(\theta_{1}\mid M_{1}) d\theta_{1}}
\end{equation}
Because we are assuming that $M_{2}$ and $M_{1}$ are a priori equally probable, Bayes factor directly represents the posterior odds (see \citealt{Foord2019} for a more rigorous mathematical proof).  $BF$ values $>$1 or $<$1 signify whether $M_{2}$ or $M_{1}$, respectively, is more likely (however, see Section~\ref{sec:discussion} where we analyze false positive space to define a ``strong" $BF$). 
Below we go into brief detail regarding the steps required to calculate two main components of the Bayes Factor: the likelihood densities ($P(D\mid\theta_{j},M_{j})$) and the prior densities ($P(\theta_{j}\mid M_{j})$). In Table~\ref{tab:symbols} we list important symbols that will be referenced in the following Section.
\subsection{Modeling the PSF and Estimating the Likelihood Density}
BAYMAX{} compares calibrated events ($x_{i}$, $y_{i}$, $E_{i}$) from EDSER-reprocessed \emph{Chandra} observations to simulations of single and dual point sources that are based on the \emph{Chandra} PSF. We simulate and model the PSF for each observation individually using the Model of AXAF Response to X-rays ({\tt MARX}, \citealt{Davis2012}). {\tt MARX} simulates the \emph{Chandra} PSF of the optics from the High Resolution Mirror Assembly (HRMA), which is characterized by various parameters such as the source spectrum, the time of observation ({\tt TSTART}), the nominal position of the detector during the observation ({\tt RA\_Nom, Dec\_Nom, Roll\_Nom}), and the detector (ACIS-S). Thus, for any source with multiple observations (SDSS J0841+0101 and SDSS J1356+1026), the PSF is simulated and modeled for each observation individually.
\par For each observation, we use {\tt MARX} to simulate X-ray photons incident from a single point source centered on the observed central position of the AGN, $\mu_{\mathrm{obs}}$. Although the shape of the PSF is energy-dependent, the $x$, $y$ position of a photon with energy $E$ does not depend on the spectral shape of a given source.  Thus, our simulated PSF is independent of the spectral shape of our model. In order to robustly model the PSF, we generate $1\times10^{6}$ rays for each observation; here we exclude the simulated read-out strip provided by {\tt MARX} by setting the parameter {\tt ACIS\_Frame\_Transfer\_Time} to 0. The PSF is modeled as a summation of three circular concentric 2D Gaussians, where the amplitude and standard deviation of each Gaussian is energy-dependent. In past analyses, we have found that this model is a good approximation of the on-axis \emph{Chandra} PSF \citep{Foord2019}.  
\par Each photon is presumed to originate from (i) a point source or (ii) a background component. Regarding the single point source model: given a PSF centered at $\mu$, the probability that a photon is observed at sky coordinates $x_{i}$,$y_{i}$ with energy $E_{i}$ is $P(x_{i},y_{i}\mid \mu, E_{i})$. Similarly for the dual point source model, given the sky coordinates of a primary and secondary AGN ($\mu_{P}$ and $\mu_{S}$), the probability that a photon is observed at sky coordinates $x_{i}$,$y_{i}$ with energy $E_{i}$ is $P(x, y \mid \mu_{P},\mu_{S}, E, n_{S}/n_{P})$. Here $n_{S}/n_{P} = f$, which represents the ratio of the total counts between the secondary and primary AGN, where $0\le f\le1$ (see Table~\ref{tab:symbols}.)
\par There are several possible sources of X-ray contamination, including the Cosmic X-ray Background (CXB, which includes unresolved X-ray point sources such as background AGN), the non-X-ray background (NXB, caused by charged particles and $\gamma$-rays), and local, diffuse, X-ray emission.  There are many possible origins of local diffuse emission, which should be individually determined for a given system (see Section~\ref{J1356+2016}).  For the analysis presented in the paper, \BAYMAX{} fits for two different backgrounds: a lower count-rate component that represents the CXB and NXB, and a higher count-rate component that represents diffuse X-ray emission. This latter component is appropriate for merging systems, where extended gas is frequently detected in both simulations and observations (see, e.g., \citealt{Cox2006, Brassington2007, Sinha2009, Hopkins2013, Smith2018}), and is evident in the $HST$ observations of 2 of our 12 sources (see Fig.~\ref{fig:GalaxyImages}). We assume that photons originating from the background are uniformly distributed across a given region, such that the probability that a photon observed at location $x_{i}$,$y_{i}$ on the sky with energy $E_{i}$ is associated with a background component is $P(x_{i},y_{i}\mid f_{BG},E_{i})$. Here, $f_{BG}$ represents the ratio of counts between a given background component and the combined counts from all point source components.  Because we assume that each background component is uniformly distributed, $P(x_{i},y_{i}\mid f_{BG},E_{i})$ is always constant over a given region of interest.
\par As an example, given $n$ observed events, the likelihood density for the single point source model is:
\begin{equation}
    \begin{split}
    \mathcal{L}&=P(x,y\mid \mu, f_{BG}, E) = \prod_{i=1}^{n} P(x_{i},y_{i}\mid \mu,E_{i}) + P(x_{i},y_{i}\mid f_{BG},E_{i}) \\
    & = \prod_{i=1}^{n} \frac{ M_{1,i}(\theta_{1})^{D_{i}}}{D_{i}!} \exp(-M_{1,i}(\theta_{1})),
    \end{split}
\end{equation}
Here, the total probability is normalized by $f_{BG}$, such that the combined probability for each detected photon equals one. Given our PSF model, the probability for event $i$ is $M_{1, i}(\theta_{1})$, while $D_{i}$ is the event's data value. Due to \emph{Chandra} registering each event individually, we use Poisson likelihoods.

\subsection{Prior Distributions}
The parameter vectors for each model, for a given source, will depend on (i) the number of observations and (ii) the prior distributions for each parameter. Regarding point (i), 10/12 galaxies have $k=1$ observations, while SDSS J0854+0101 has $k=2$ and SDSS J1356+1026 has $k=3$. Thus, for the majority of our sample the parameter vectors for $M_{1}$ and $M_{2}$ are $\theta_{1}=[\mu, \log{f_{BG}}]$ and $\theta_{2}=[\mu_{{P}}$, $\mu_{{S}}, {\log{f}}, \log{f_{BG}}]$.  For SDSS J0854+0101 and SDSS J1356+1026, $\theta_{1}$ and $\theta_{2}$ also include $\Delta x_{K}, \Delta y_{K}$, which account for the translational components of the relative astrometric registration for the $K=[1, ..., k-1]$ observation (see Table~\ref{tab:symbols}).  
\par Regarding point (ii), any user-defined function can be used to describe the prior distributions for each parameter. We use continuous uniform distributions to describe the prior distributions of $\mu$, where the bounds of the distribution are determined by the spectroastrometric [\ion{O}{3}] $\lambda$5007 observations (see Section~\ref{sec:results}). The prior distribution for $\log{f_{BG}}$ is described by a Gaussian distribution, $N$($\mu_{BG}$,$\sigma^{2}_{BG}$), where $\mu_{BG}$ is estimated for each observation by selecting 10 random and source-free regions with a 2\arcsec~radius and within a 20\arcsec$\times$20\arcsec~ region centered on the AGN.  We set $\sigma_{BG}$ to 0.5, allowing for \BAYMAX{} to more easily move around parameter space.  
\par For $M_{2}$, the prior distribution for $\log{f}$ is described by a uniform distribution, bound between $-2$ and $0$. Regarding SDSS J0841+0101 and SDSS J1356+1026, the prior distributions of $\Delta x_{K}$ and $\Delta y_{K}$ are described by uniform distributions bound between $\delta \mu_{\mathrm{obs}}-3$ and $\delta \mu_{\mathrm{obs}}+3$, where $\delta \mu$ represents the difference between the observed central X-ray coordinates of the longest observation (Obs ID:18199 for SDSS J0841+0101 and Obs ID: 18826 for SDSS J1356+1026) and the $K=[1, ..., k-1]$ observation.

\subsection{Calculation of Bayes Factor}
In this section we briefly review how \BAYMAX{} implements model selection and parameter estimation, but refer the reader to \cite{Foord2019} for more details.
\par For model selection, \BAYMAX{} uses a sampling technique called nested sampling \citep{Skilling2004}, which efficiently samples through likelihood space to estimate the marginal likelihood, usually referred to as the Bayesian evidence and denoted by $Z$ (where $Z = \int P(D\mid\theta_{j},M_{j}) P(\theta_{j}\mid M_{j}) d\theta_{j}$; see \citealt{Skilling2004}, \citealt{Shaw2007}, \citealt{Feroz&Hobson2008}, and \citealt{Feroz2009} for more details.) In particular, \BAYMAX{} uses the {\tt PYTHON} package {\tt nestle},\footnote{https://github.com/kbarbary/nestle} which can estimate $Z$ on the order of minutes for low-count ($<100$) observations. For parameter estimation \BAYMAX{} uses {\tt PyMC3} \citep{Salvatier2016}, which uses a Hamiltonian Monte Carlo (HMC) sampling method to much more quickly converge than normal Metropolis-Hastings sampling.
\par The calculation of the $BF$ and the estimations of the posterior distributions are separated into two different processes, allowing the user flexibility to only estimate posteriors for sources of interest (i.e., that have $BF$ that favor the dual point source model). In general, nested sampling iterates through likelihood space in a coarser fashion and is a much faster calculation, as the maximum value in likelihood space only needs to be within the region where the points are sampled (as nested sampling is calculating an integral). While the $BF$ calculations take on the order of minutes, PyMC3 calculations (where the main goal is indeed to find the maximum likelihood) will take on the order of hours (for $>100$ counts). 
\par For unimodal posterior distributions, the posterior distributions returned by nestle and PyMC3 are generally consistent. In particular, for each source in our sample that we analyze with PyMC3, we find that nestle returns posteriors with consistent median values (at the 68\% confidence level), although the nestle posteriors are broader, a result of coarser sampling.
%

%%%%%%%%%%%%%%%%%%%%%%%%%%%%%%%%%%%%%
%%%%%%%%%%%%% RESULTS %%%%%%%%%%%%%%%
%%%%%%%%%%%%%%%%%%%%%%%%%%%%%%%%%%%%%
\section{Results}
\label{sec:results}

%%%%%% TABLE RESULTS %%%%%%
\begin{table}[t]
\begin{center}
\caption{Bayes Factor Results}
\label{tab:results}
\small
\begin{tabular*}{\columnwidth
}{ccc}
	\hline
	\hline
	\multicolumn{1}{c}{Galaxy Name} & \multicolumn{1}{c}{non-informative $\ln{BF}$} & \multicolumn{1}{c}{informative $\ln{BF}$}\\
	\multicolumn{1}{c}{(1)} & \multicolumn{1}{c}{(2)} & \multicolumn{1}{c}{(3)} \\
	\hline
	SDSS J0142$-$0050 & $-3.14\pm0.76$ & $-1.46\pm0.71$ \\
	SDSS J0752+2736 & 4.90$\pm$0.51 & 0.25$\pm$0.43 \\
	SDSS J0841+0101 & 9.97$\pm$0.75 & 5.91$\pm$0.78 \\
	SDSS J0854+5026 & 0.26$\pm$0.59 & 0.18$\pm$0.37 \\
	SDSS J0952+2552 & 0.52$\pm$0.38 & $-0.83\pm0.35$ \\
    SDSS J1006+4647 & 0.47$\pm$0.40 & 0.41$\pm$0.63 \\
    SDSS J1126+2944 & 1.50$\pm$0.41 & 3.54$\pm$0.43 \\
    SDSS J1239+5314 & $-3.36\pm0.85$ & $-3.43\pm0.50$ \\
    SDSS J1322+2631 & 0.36$\pm$0.62 & $-0.91\pm0.40$ \\
    SDSS J1356+1026 & 41.65$\pm$0.65 & 34.78$\pm$0.70 \\
    SDSS J1448+1825 & 1.43$\pm$0.55 & 2.95$\pm$0.52 \\
    SDSS J1604+5009 & $-0.45\pm0.50$ & $-0.83\pm0.49$ \\
	\hline 
	\hline
\end{tabular*}
\begin{tabular*}{\columnwidth
}{cccc}
	\multicolumn{1}{c}{Galaxy Name} & \multicolumn{1}{c}{$\ln{\frac{Z_{1,\mathrm{D}}}{Z_{1}}}$} & \multicolumn{1}{c}{$\ln{\frac{Z_{2,\mathrm{D}}}{Z_{2}}}$} 
	& \multicolumn{1}{c}{$\ln BF_{\mathrm{D}}$}\\
	\multicolumn{1}{c}{(1)} & \multicolumn{1}{c}{(2)} & \multicolumn{1}{c}{(3)} & \multicolumn{1}{c}{(4)} \\
	SDSS J0841+0101 & 139$\pm$0.75 & 148$\pm$0.71 & -2.62$\pm$0.65 \\
	SDSS J1356+1026 & 264$\pm$0.75 & 238$\pm$0.79 & 8.70$\pm$0.70 \\
	\hline
\end{tabular*}
\end{center}
Top. -- Columns: (1) SDSS galaxy designation; (2) $\ln{BF}$ values, defined as $Z_{2}/Z_{1}$, using non-informative priors on the location of $\mu$; (3) $\ln{BF}$ values, defined as $Z_{2}/Z_{1}$, using informative priors on the location of $\mu$. \\
Bottom. -- Columns: (1) SDSS galaxy designation; (2) $\ln{ \frac{ Z_{1,\mathrm{D}}}{Z_{1}}}$, using informative priors; (3) $\ln{ \frac{ Z_{2,\mathrm{D}}}{Z_{2}}}$, using informative priors; (4) $\ln{BF_\mathrm{D}}$, defined as $Z_{2,\mathrm{D}}/Z_{1,\mathrm{D}}$, using informative priors
\end{table}
%%%%%%%%%%%%%%%%%%

%%%%%%%%%%%%%%%%%%%%%%%%%%%%%%%%%%%%%%%%%%%%%
%%%%%%%%%%%%%% FIGURE %%%%%%%%%%%%%%%%%%%%%%%
%%% PYMC3 output from BAYMAX simulations %%%%
%%%%%%%%%%%%%%%%%%%%%%%%%%%%%%%%%%%%%%%%%%%%%
\begin{figure*}
\centering
    \begin{minipage}{0.27\linewidth}
    \includegraphics[width=\linewidth]{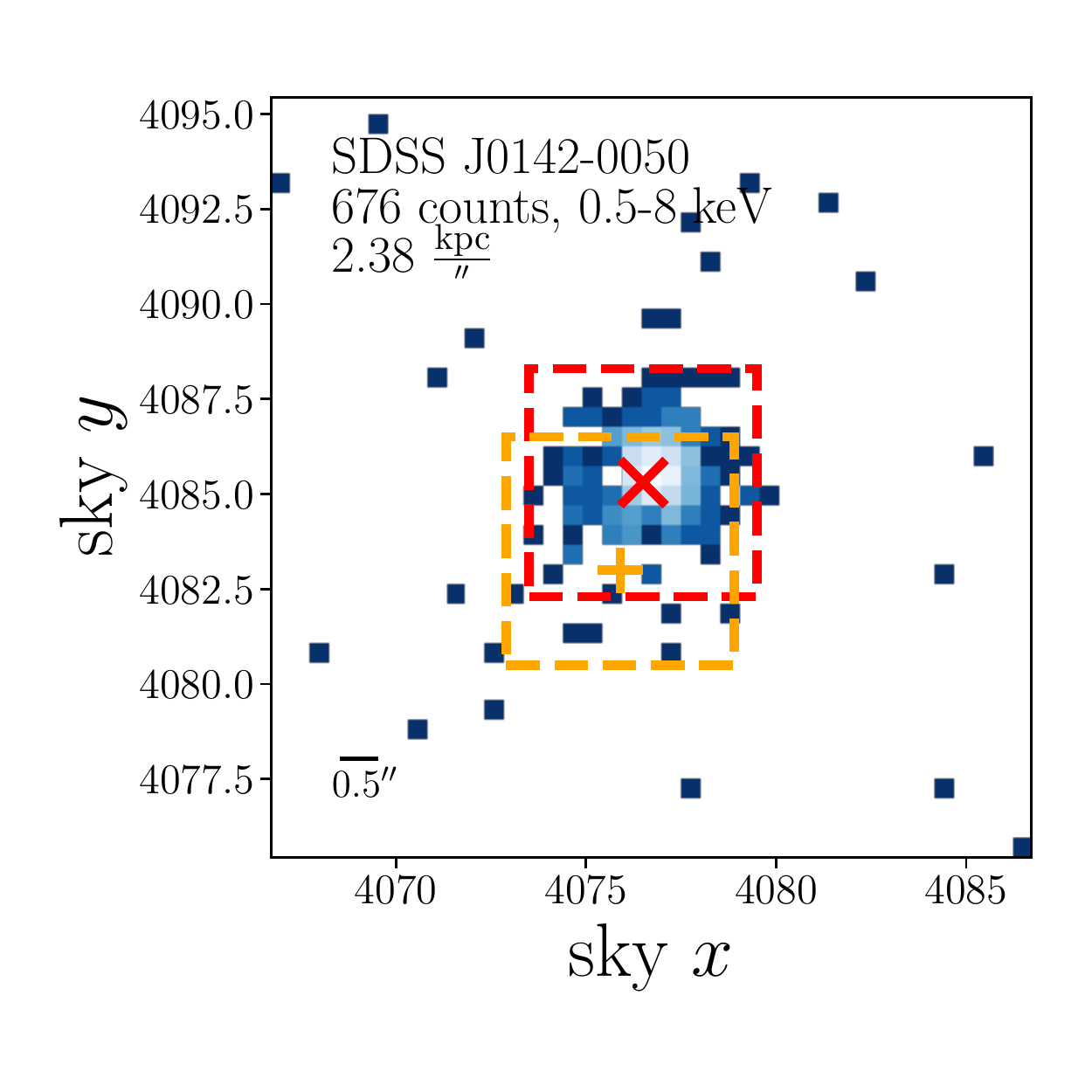}
    \end{minipage}
    \hspace{0.5cm}\vspace{-1cm}
    \begin{minipage}{0.23\linewidth}
    \includegraphics[width=\linewidth]{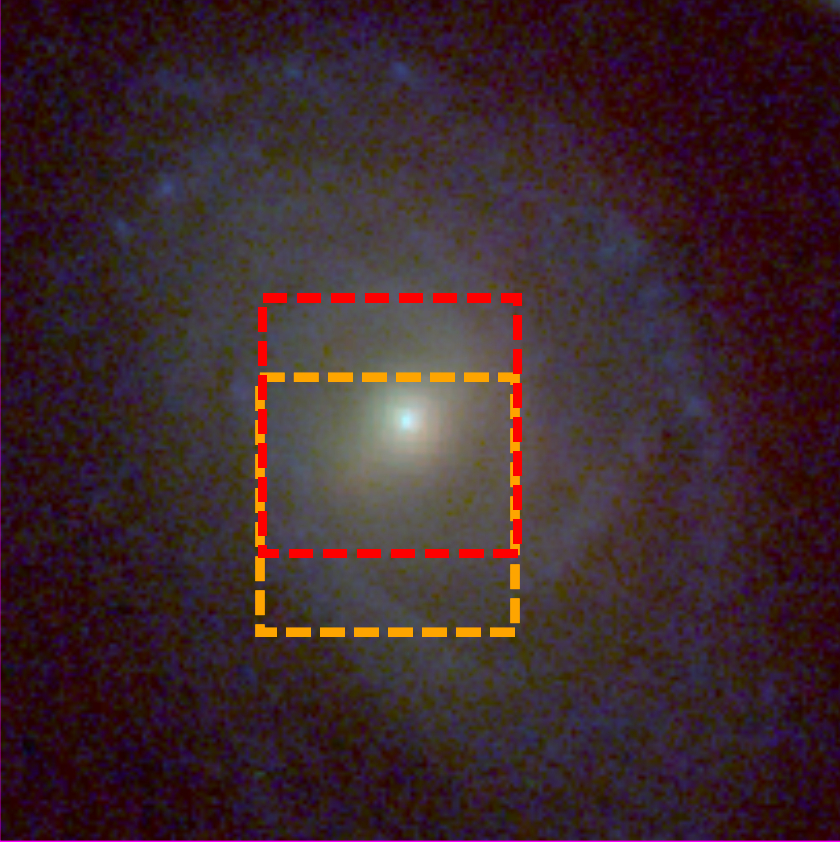}
    \end{minipage}
    \vspace{1cm}
 
     \begin{minipage}{0.27\linewidth}
    \includegraphics[width=\linewidth]{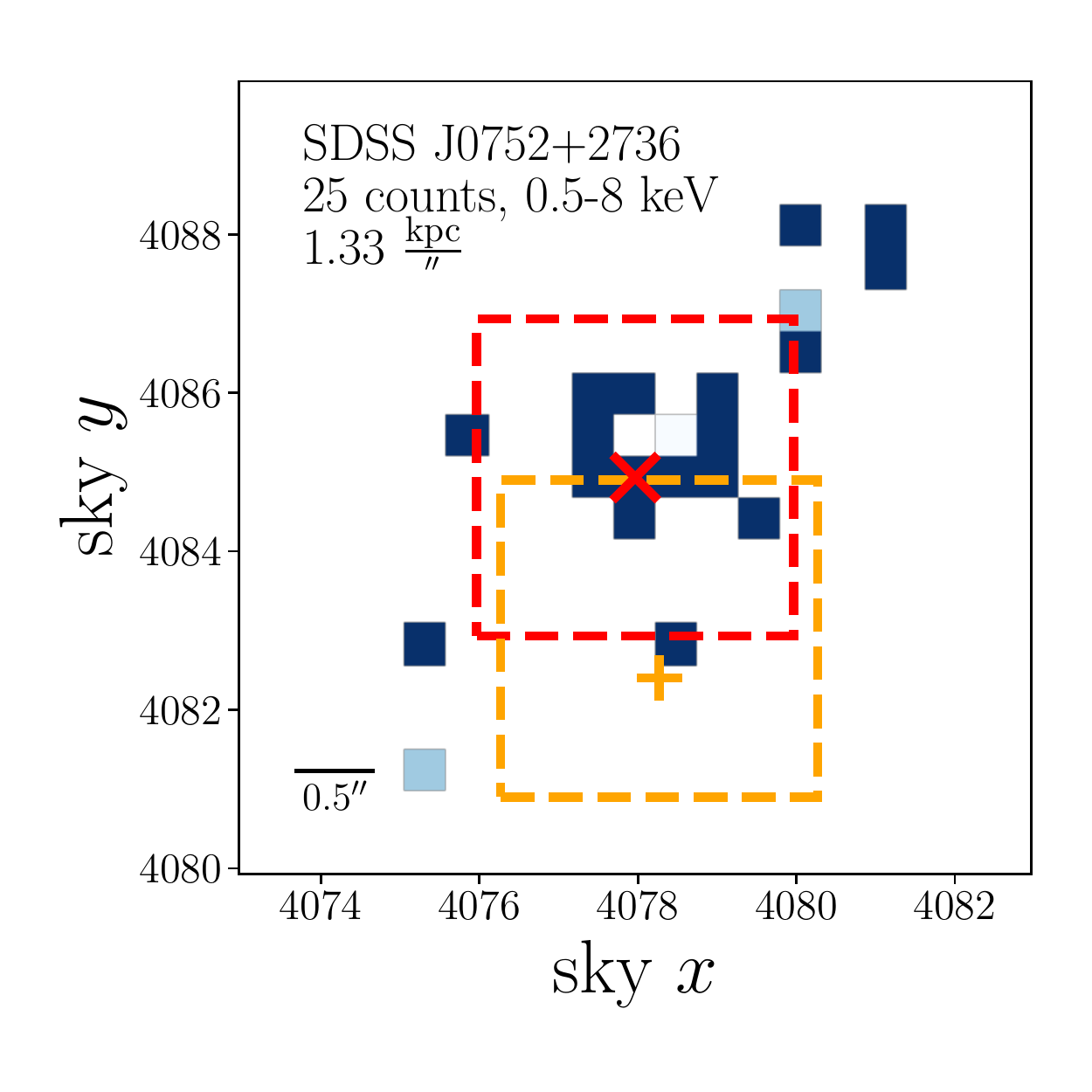}
    \end{minipage}
    \hspace{0.5cm}\vspace{-1cm}
    \begin{minipage}{0.23\linewidth}
    \includegraphics[width=\linewidth]{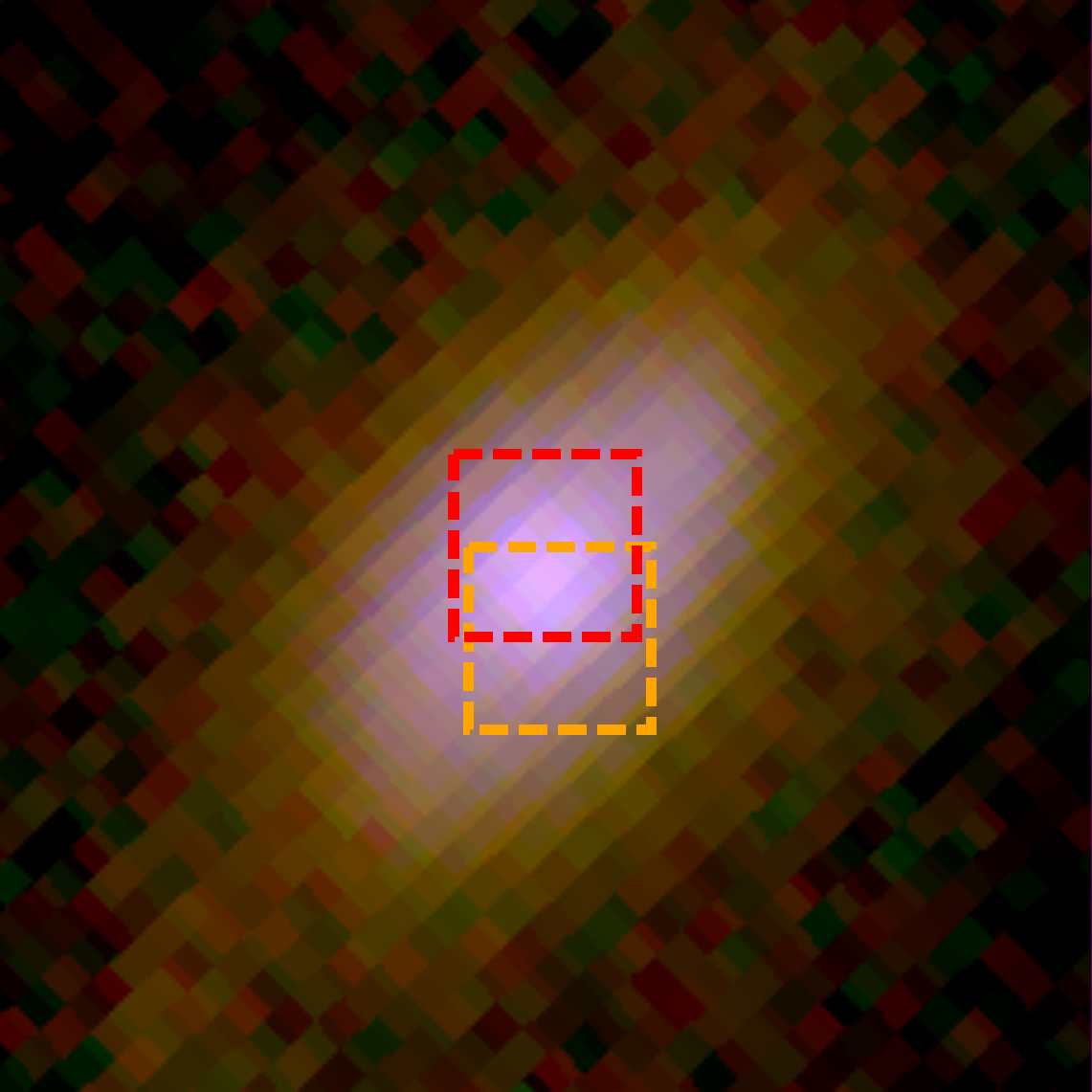}
    \end{minipage}
    \vspace{1cm}
 
      \begin{minipage}{0.27\linewidth}
    \includegraphics[width=\linewidth]{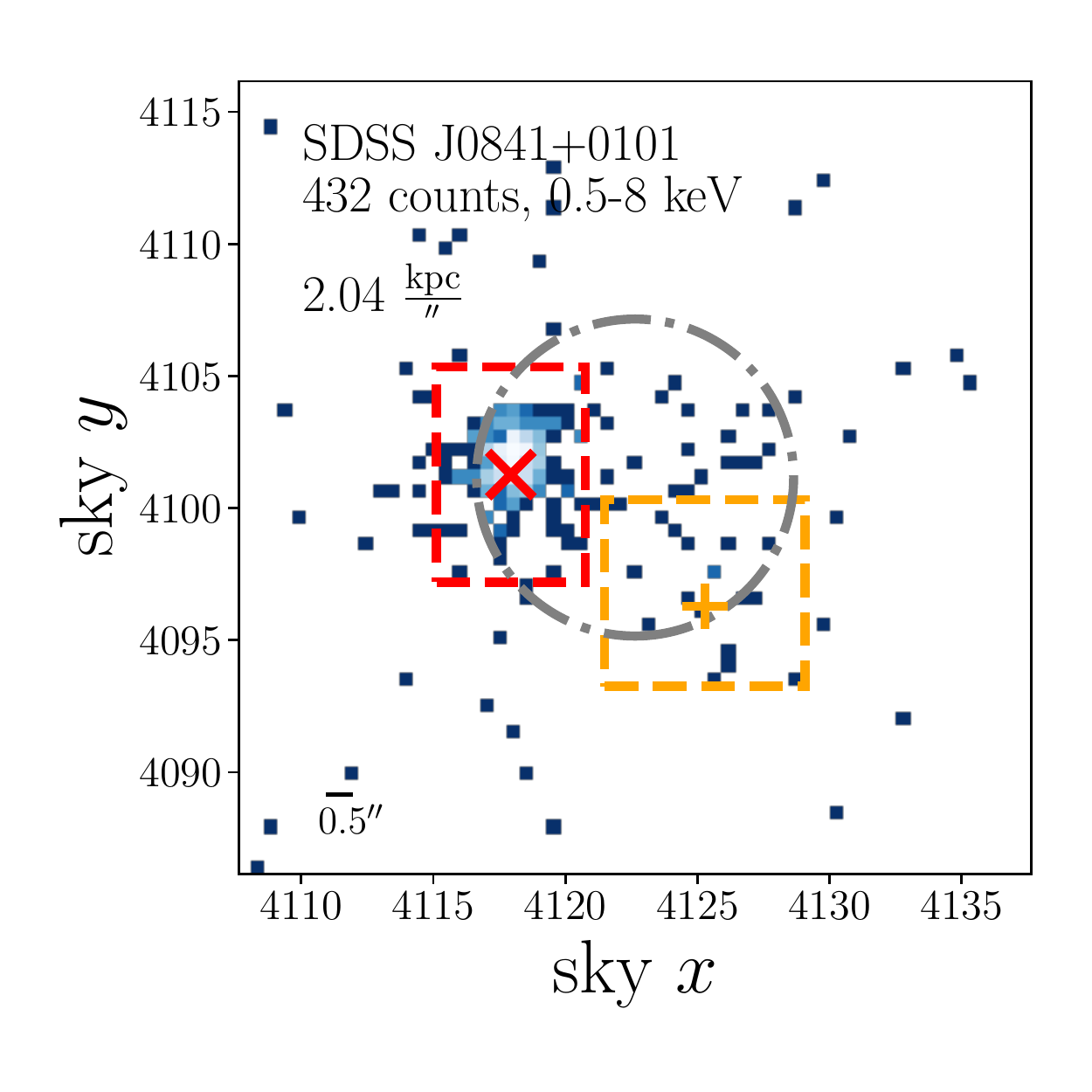}
    \end{minipage}
    \hspace{0.5cm}\vspace{-1cm}
    \begin{minipage}{0.23\linewidth}
    \includegraphics[width=\linewidth]{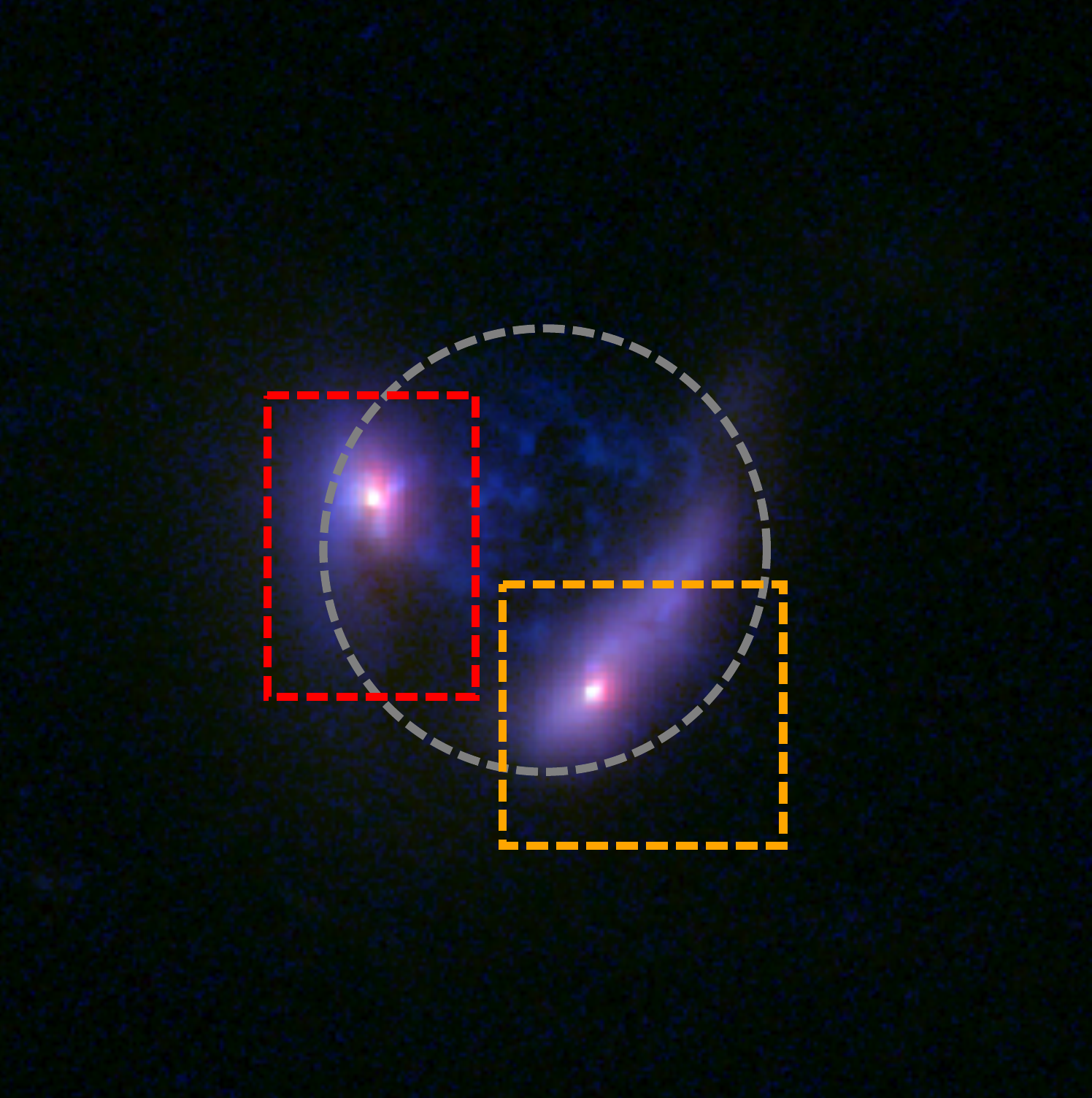}
    \end{minipage}
    \vspace{1cm}

 \begin{minipage}{0.27\linewidth}
    \includegraphics[width=\linewidth]{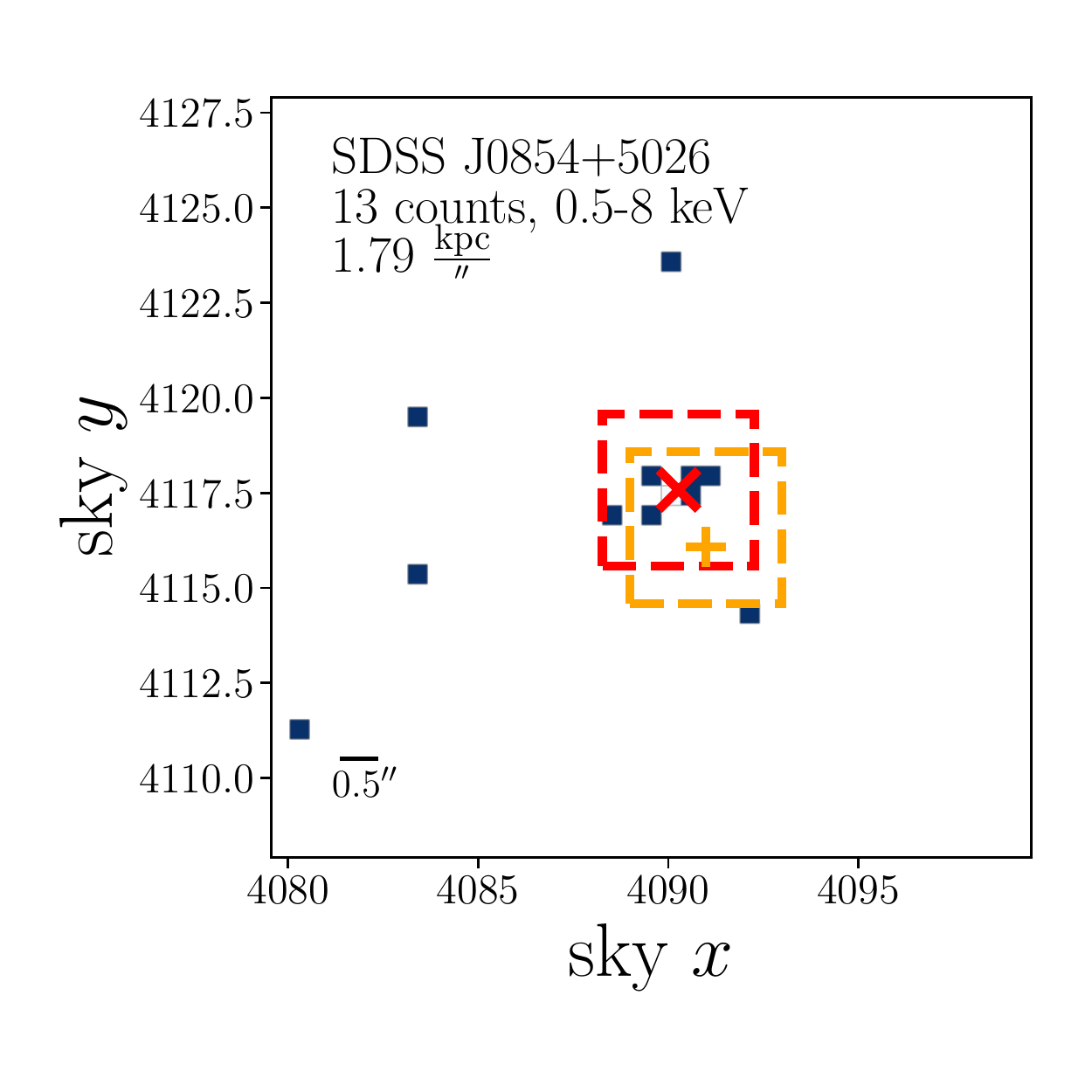}
    \end{minipage}
    \hspace{0.5cm}\vspace{-1cm}
 \begin{minipage}{0.23\linewidth}
    \includegraphics[width=\linewidth]{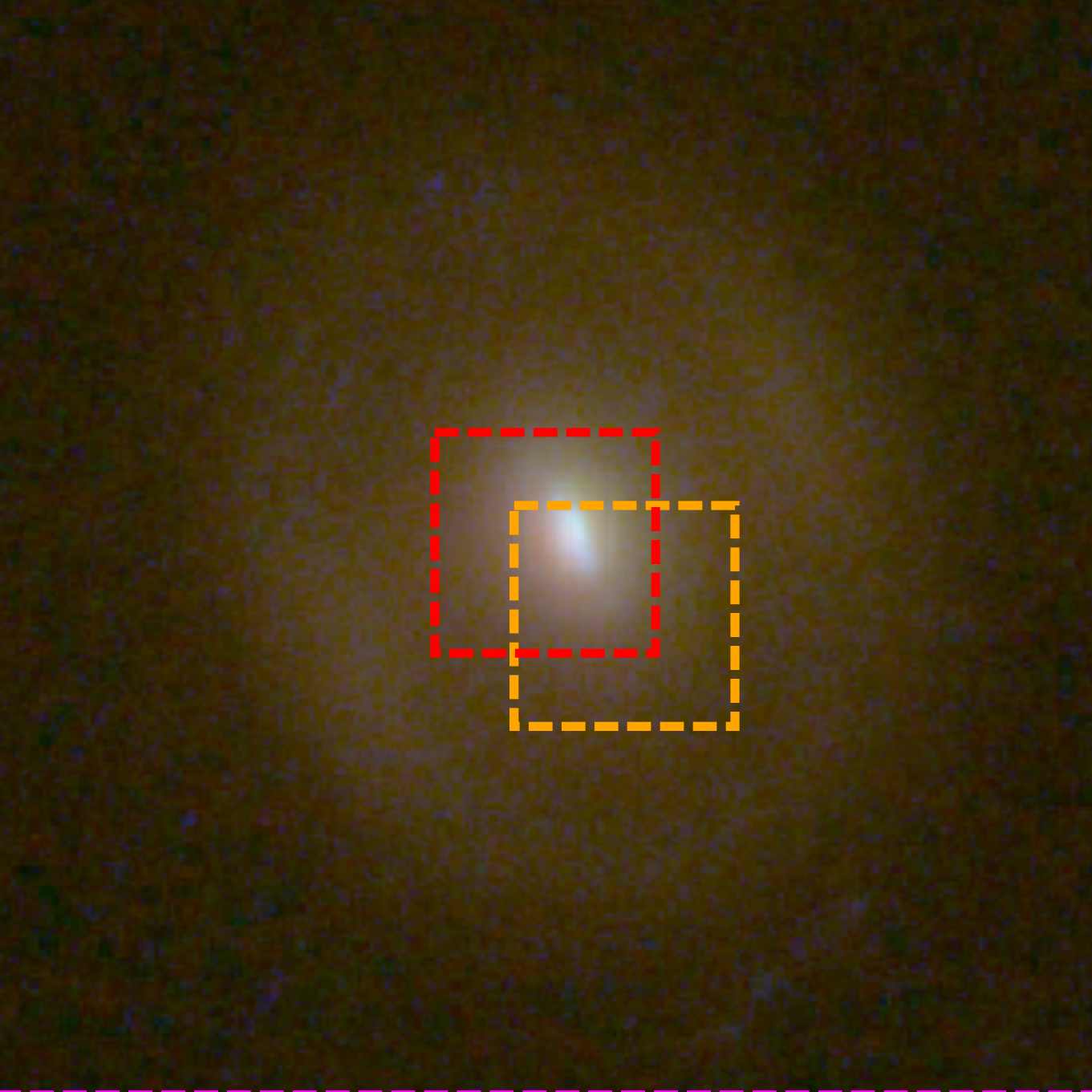}
    \end{minipage}
    \vspace{1cm}

\caption{\emph{Chandra} $0.5$--$8$ keV observations of each candidate dual AGN (left) and corresponding combined optical observations of the same field-of-view (right).  In each X-ray image we mark the location of each [\ion{O}{3}] $\lambda$5007 emission component with a red ``x" and an orange ``+".  We show the sky $x$, sky $y$ region, within which the informative priors for $\mu$ are constrained to in red and orange boxes.  When using non-informative priors, the central locations for the primary and secondary are allowed to be anywhere within the image. For SDSS J0841+0101 we denote the region within which the diffuse emission background component is restricted to with a gray box. Additionally, for SDSS J0841+0101 we show the combined X-ray emission for all \emph{Chandra} observations, where we use the best-fit astrometric shift values as found by \BAYMAX{}. The X-ray images have been binned to \emph{Chandra}'s native pixel resolution; all images are scaled in log-space with minimum and maximum counts/bin as follows: SDSS J0142$-$0050 (min=1, max=92), SDSS J0752+2736 (min=1, max=3), SDSS J0841+0101 (min=1, max=24), SDSS J0854+5026 (min=1, max=2). All the optical images are combined \emph{HST} images, with the exception of SDSS J0752+2736, which is an SDSS gri color composite image.  For the \emph{HST} images, we combine the F160W (red), F814W (green), and F438W (blue), with the exception of J1604+5009 (red: F105W; green: F621M; blue: F547M; GO 12521, PI: Liu).  In all panels, north is up and east is to the left, and a 0\farcs{5} bar is shown to scale.}
\label{fig:GalaxyImages}
\end{figure*}
%%%%%%%%%%%%%%%%%%%%%%%%%%%%%%%%%%%%%%%%%%%%%
%%%%%%%%%%%%%%%%%%%%%%%%%%%%%%%%%%%%%%%%%%%%%
\begin{figure*}
\centering
    \begin{minipage}{0.32\linewidth}
    \includegraphics[width=\linewidth]{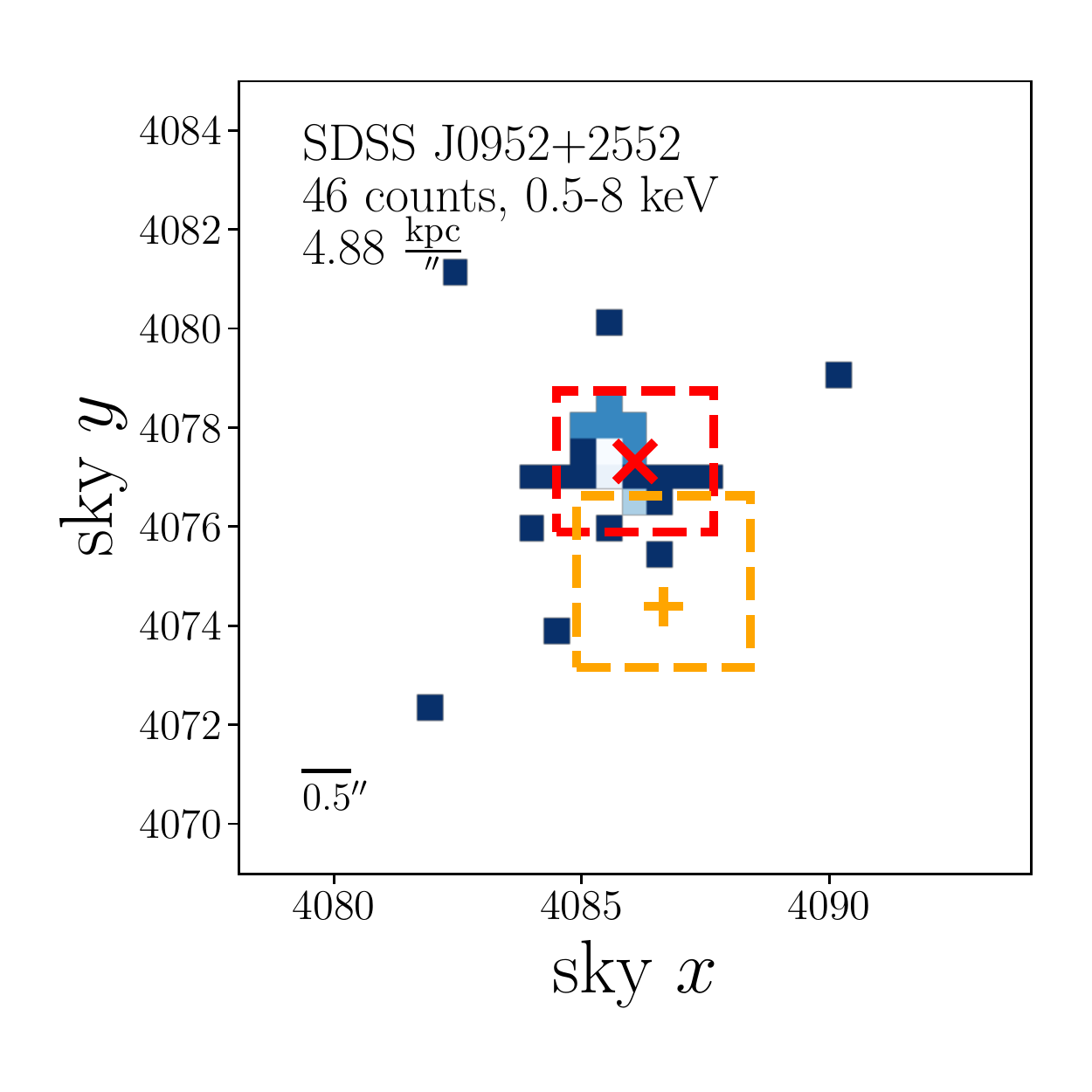}
    \end{minipage}
    \hspace{0.5cm}\vspace{-1cm}
    \begin{minipage}{0.32\linewidth}
    \includegraphics[width=0.85\linewidth]{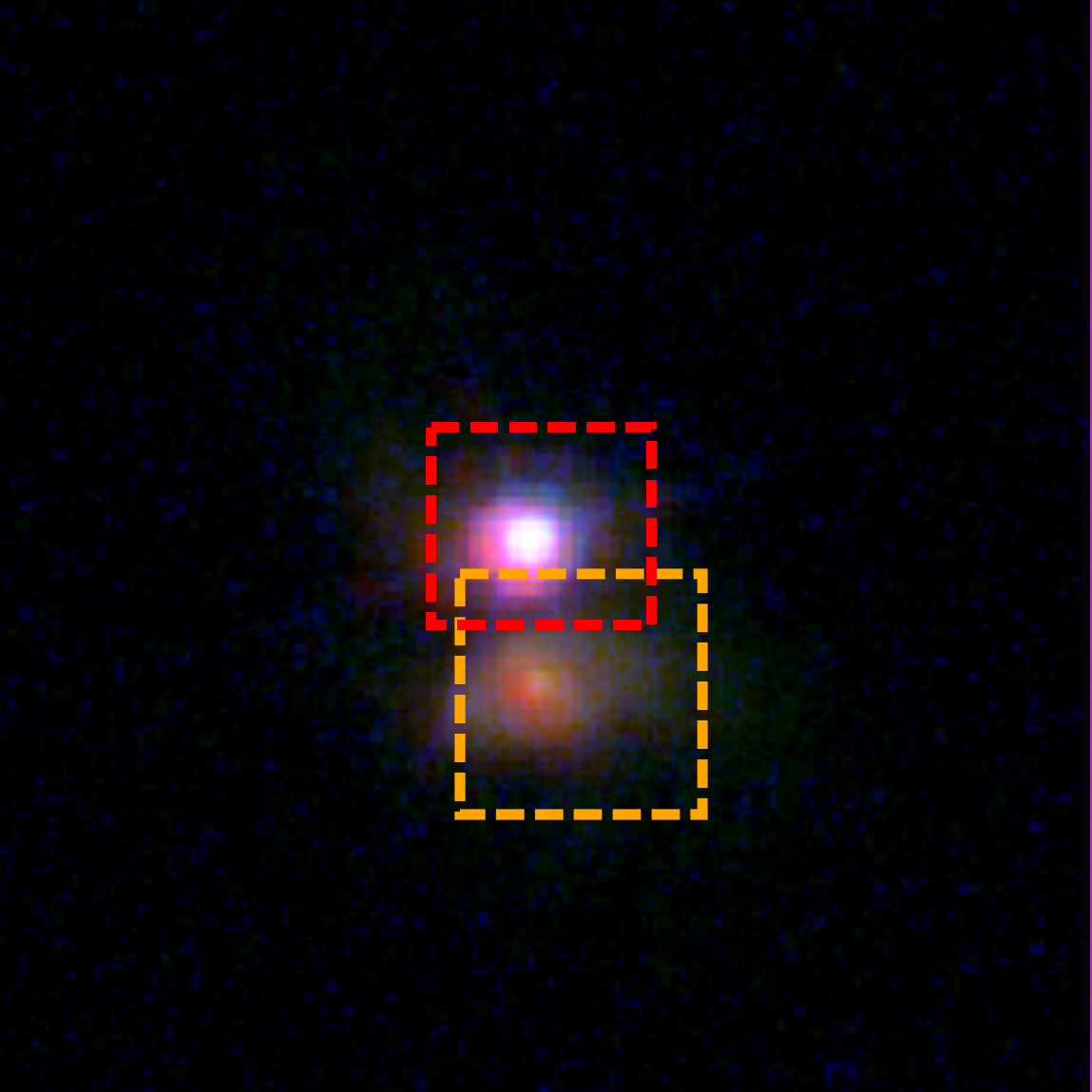}
    \end{minipage}
    \vspace{1cm}

    \begin{minipage}{0.32\linewidth}
    \includegraphics[width=\linewidth]{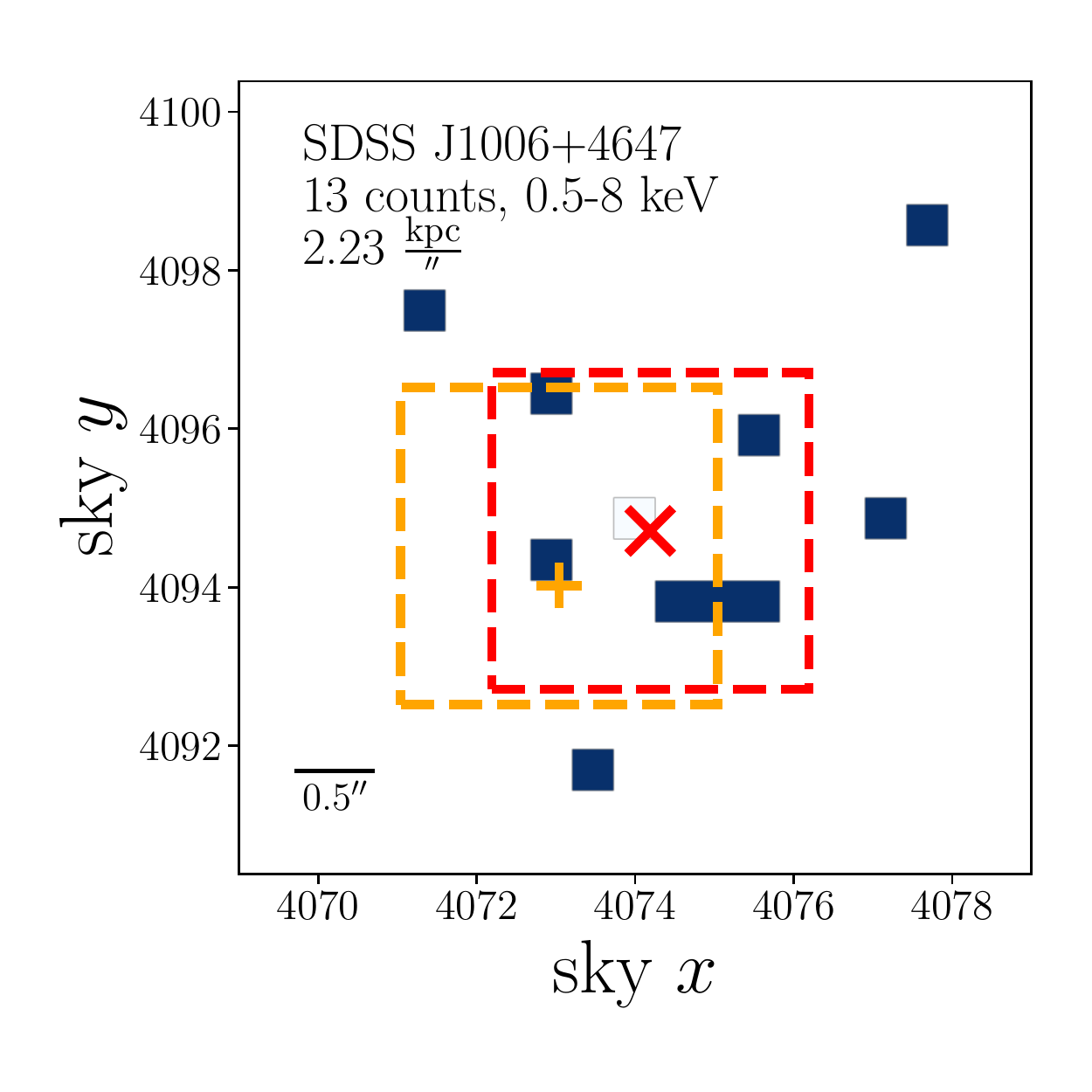}
    \end{minipage}
    \hspace{0.5cm}\vspace{-1cm}
    \begin{minipage}{0.32\linewidth}
    \includegraphics[width=0.85\linewidth]{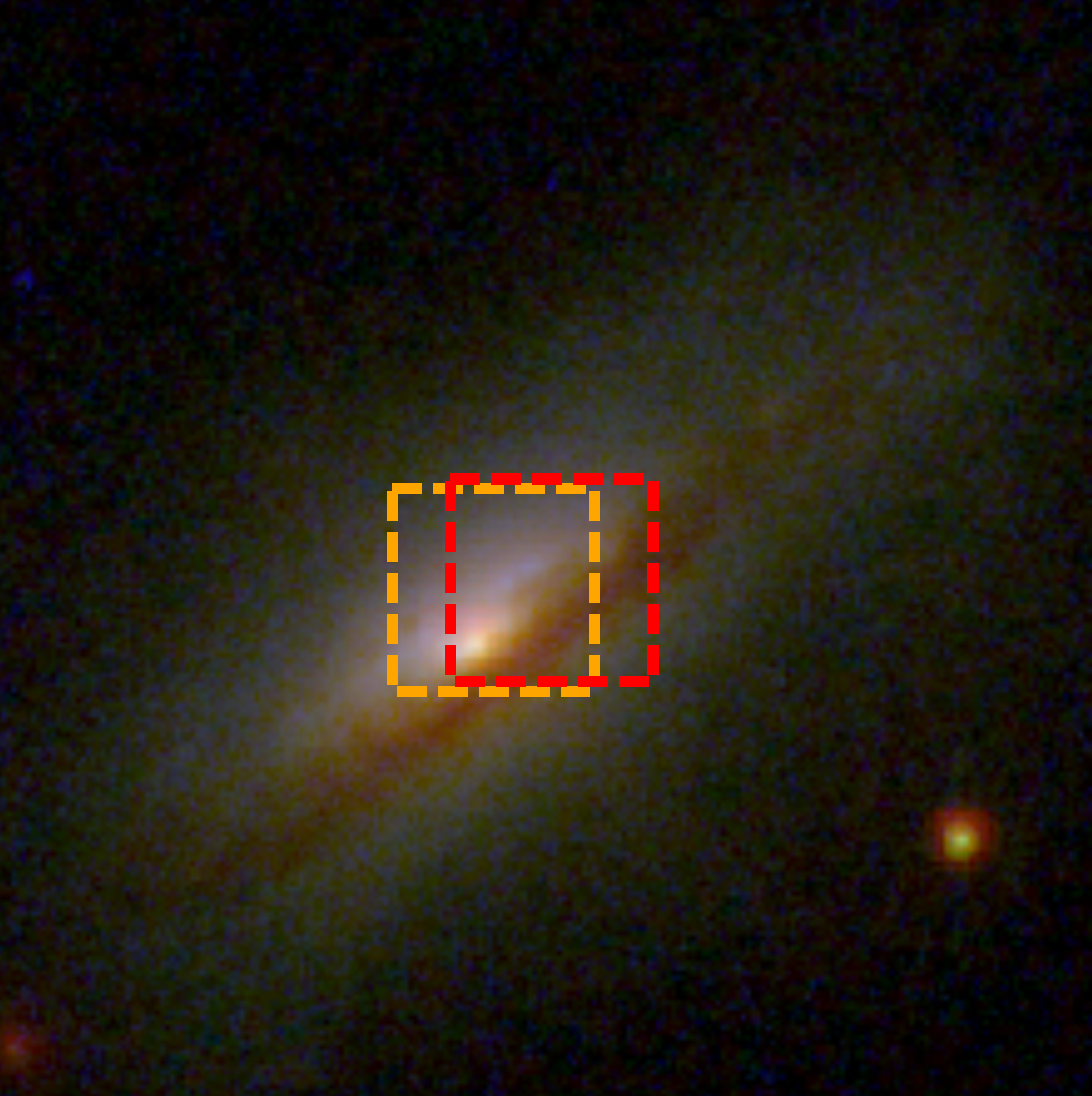}
    \end{minipage}
    \vspace{1cm}

    \begin{minipage}{0.32\linewidth}
    \includegraphics[width=\linewidth]{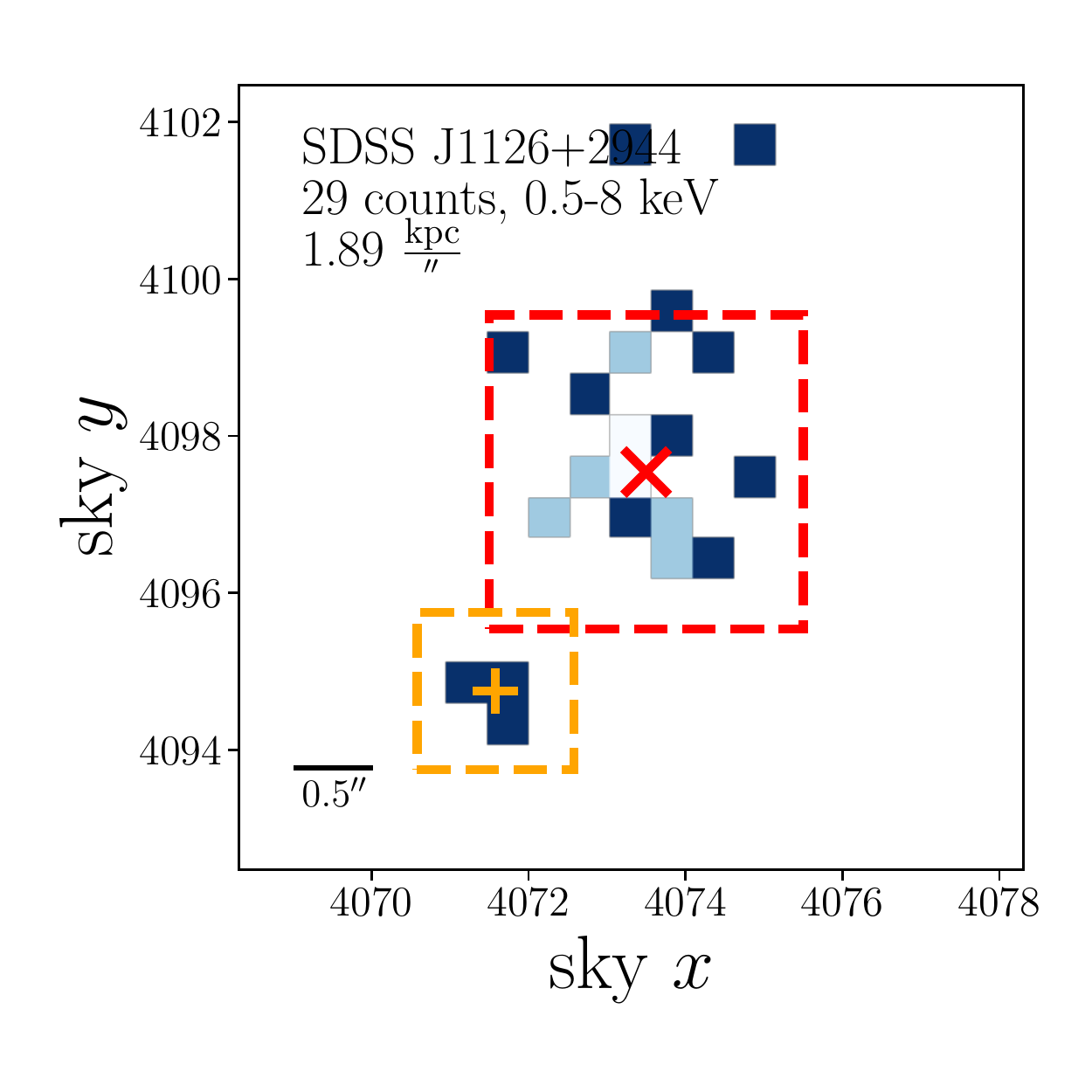}
    \end{minipage}
    \hspace{0.5cm}\vspace{-1cm}
    \begin{minipage}{0.32\linewidth}
    \includegraphics[width=0.85\linewidth]{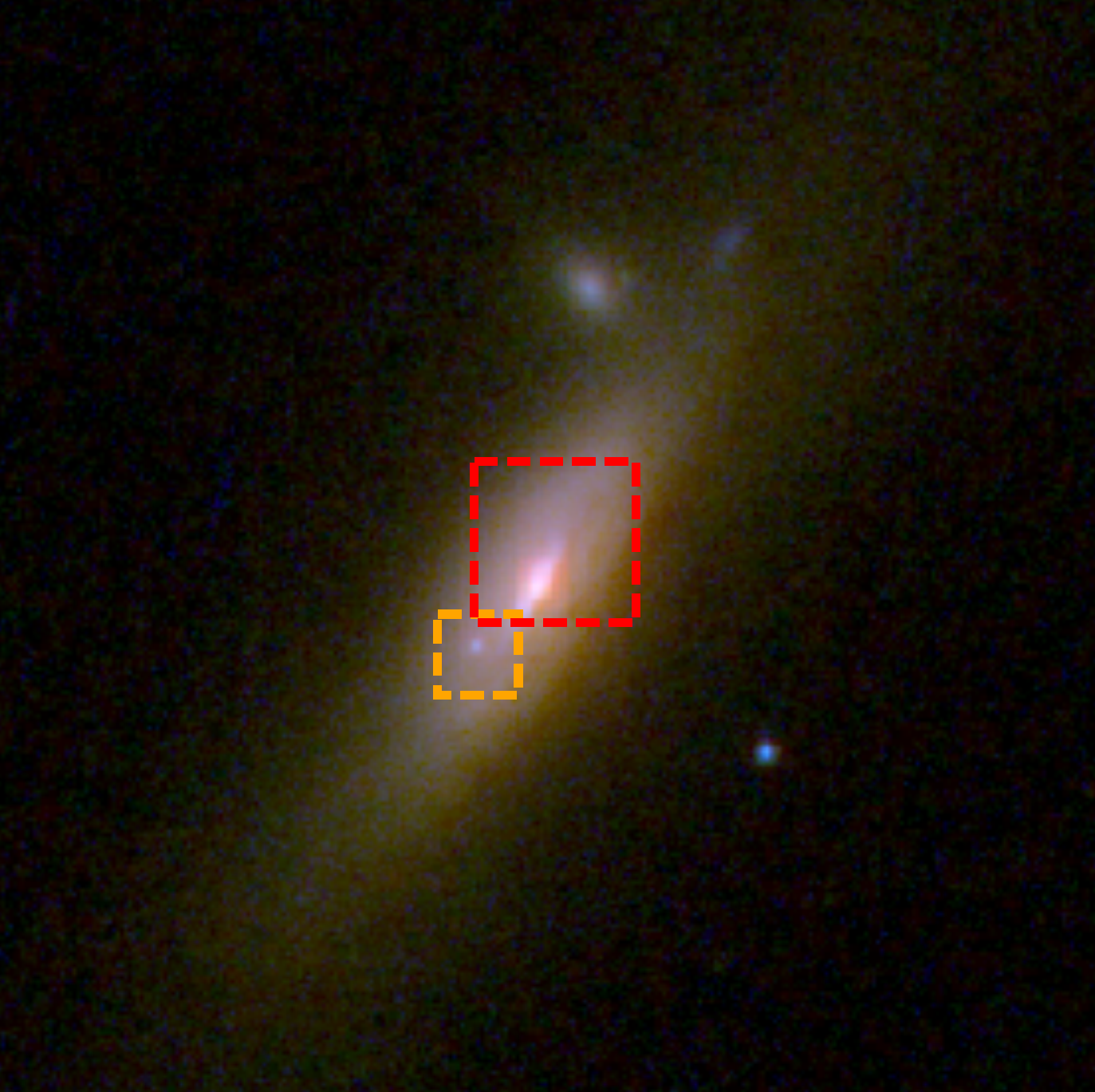}
    \end{minipage}
    \vspace{1cm}

    \begin{minipage}{0.32\linewidth}
    \includegraphics[width=\linewidth]{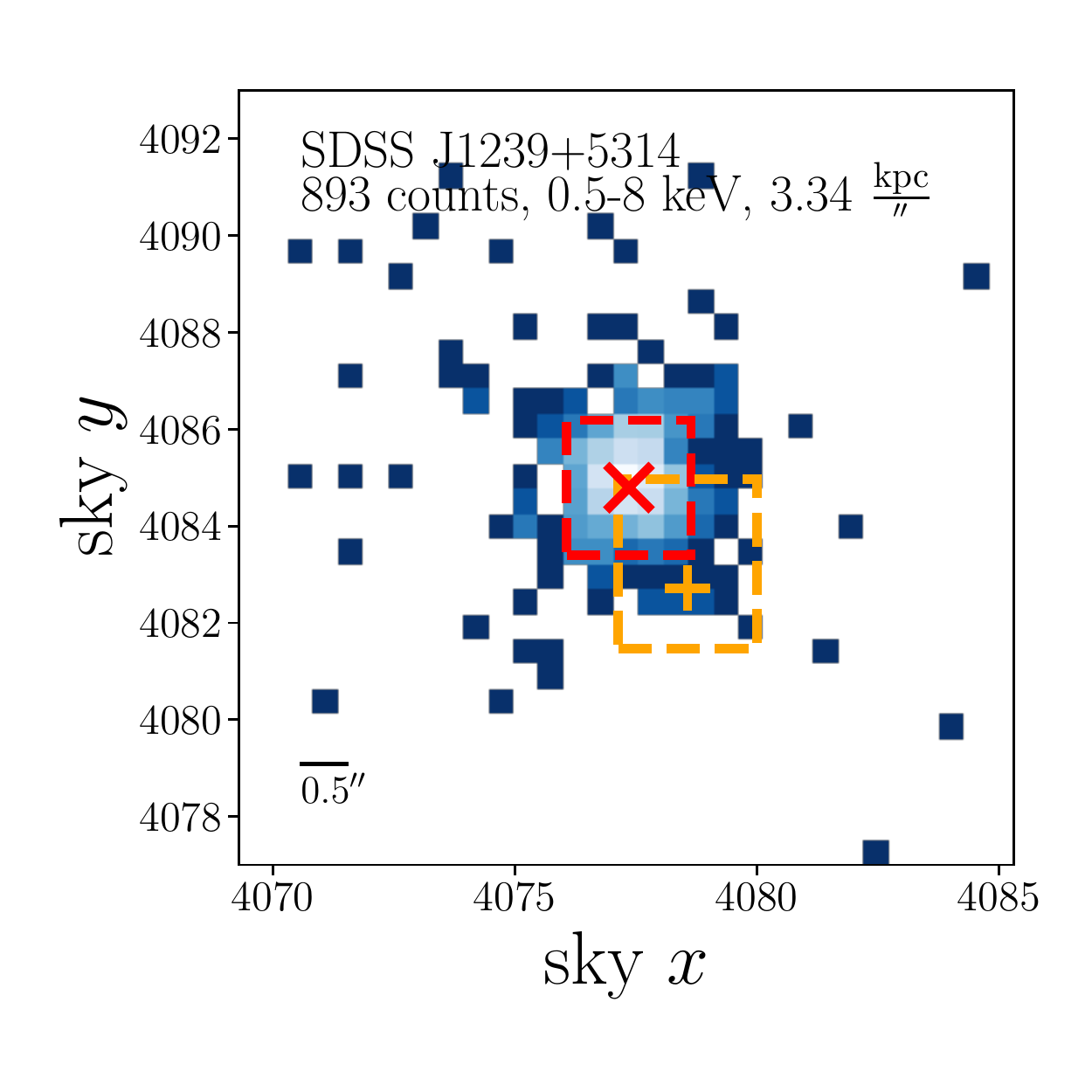}
    \end{minipage}
    \hspace{0.5cm}\vspace{-1cm}
    \begin{minipage}{0.32\linewidth}
    \includegraphics[width=0.85\linewidth]{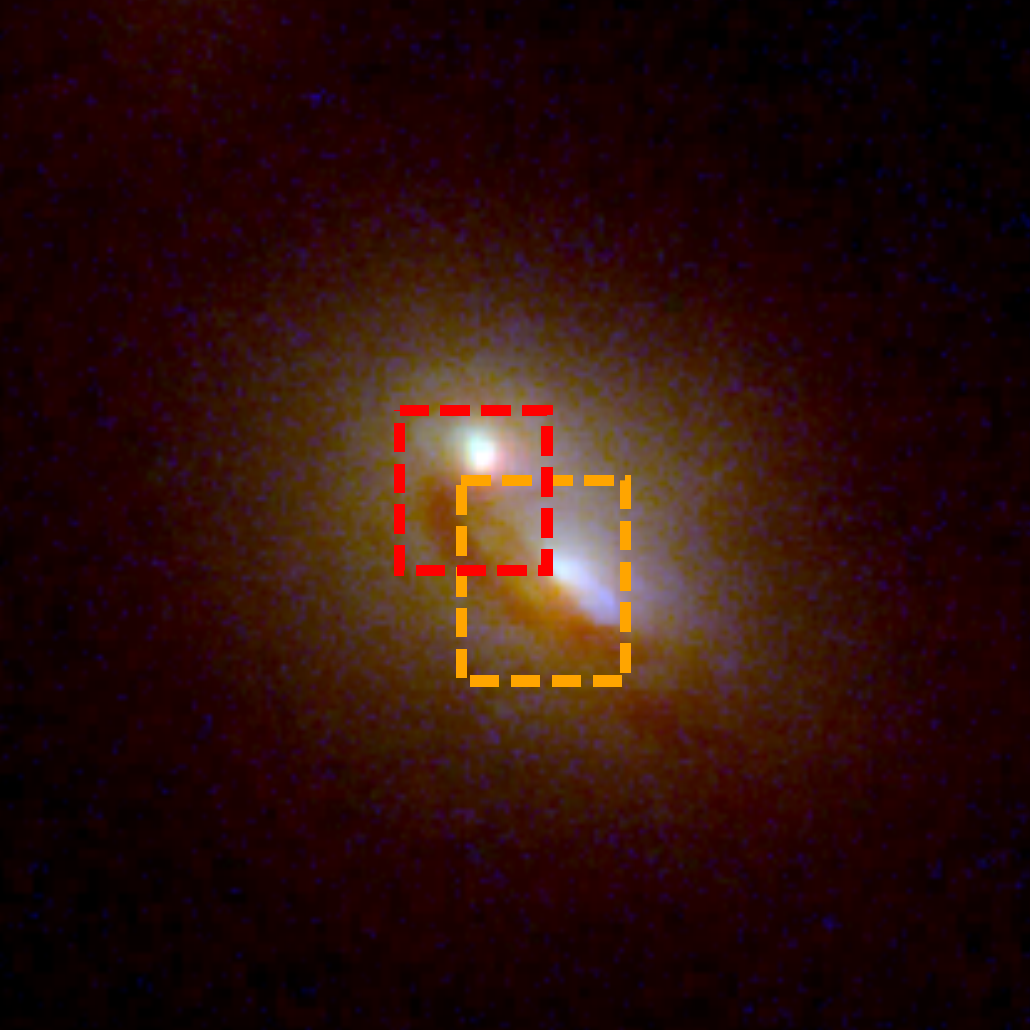}
    \end{minipage}
    \vspace{1cm}
\caption{All images are scaled in log-space with minimum and maximum counts/bin as follows: SDSS J0952+2552 (min=1, max=8), SDSS J1006+4647 (min=1, max=3), SDSS J1126+2944 (min=1, max=3), SDSS J1239+5314 (min=1, max=147).} 
\end{figure*}

%%%%%%%%%%%%%%%%%%%%%%%%%%%%%%%%%%%%%%%%%%%%%
%%%%%%%%%%%%%%%%%%%%%%%%%%%%%%%%%%%%%%%%%%%%%
%%%%%%%%%%%%%%%%%%%%%%%%%%%%%%%%%%%%%%%%%%%%%
%%%%%%%%%%%%%%%%%%%%%%%%%%%%%%%%%%%%%%%%%%%%%
\begin{figure*}
\centering
    \begin{minipage}{0.32\linewidth}
    \includegraphics[width=\linewidth]{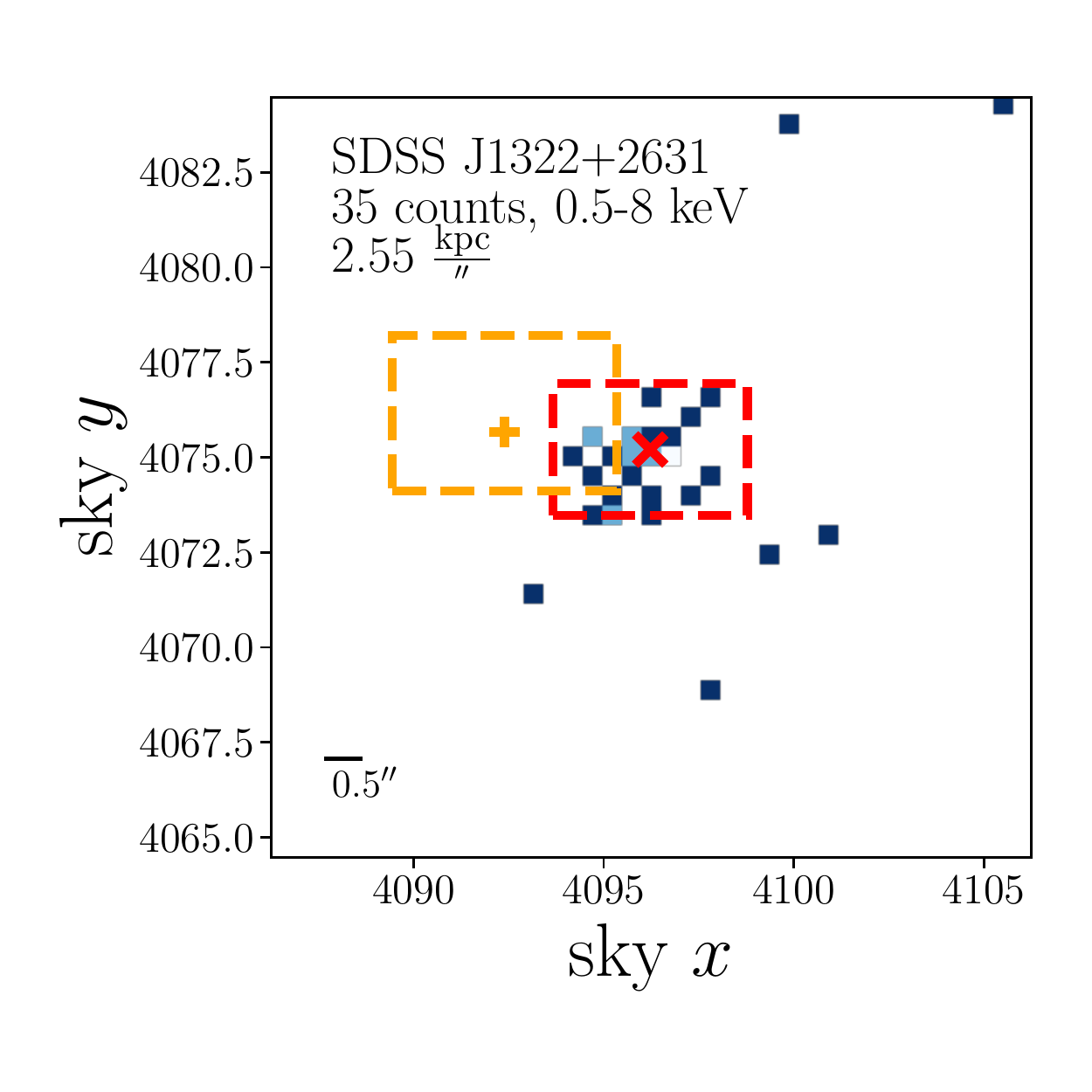}
    \end{minipage}
    \hspace{0.5cm}\vspace{-1cm}
    \begin{minipage}{0.32\linewidth}
    \includegraphics[width=0.85\linewidth]{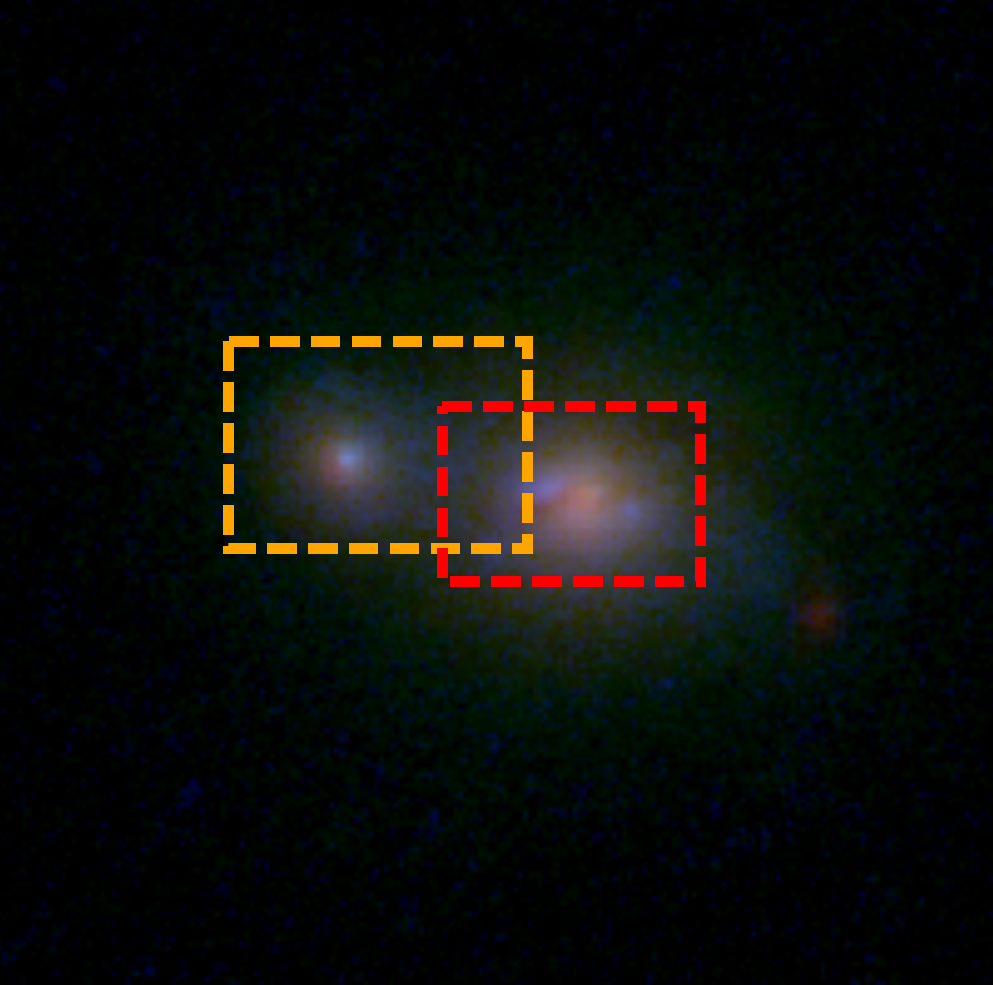}
    \end{minipage}
    \vspace{1cm}

    \begin{minipage}{0.32\linewidth}
    \includegraphics[width=\linewidth]{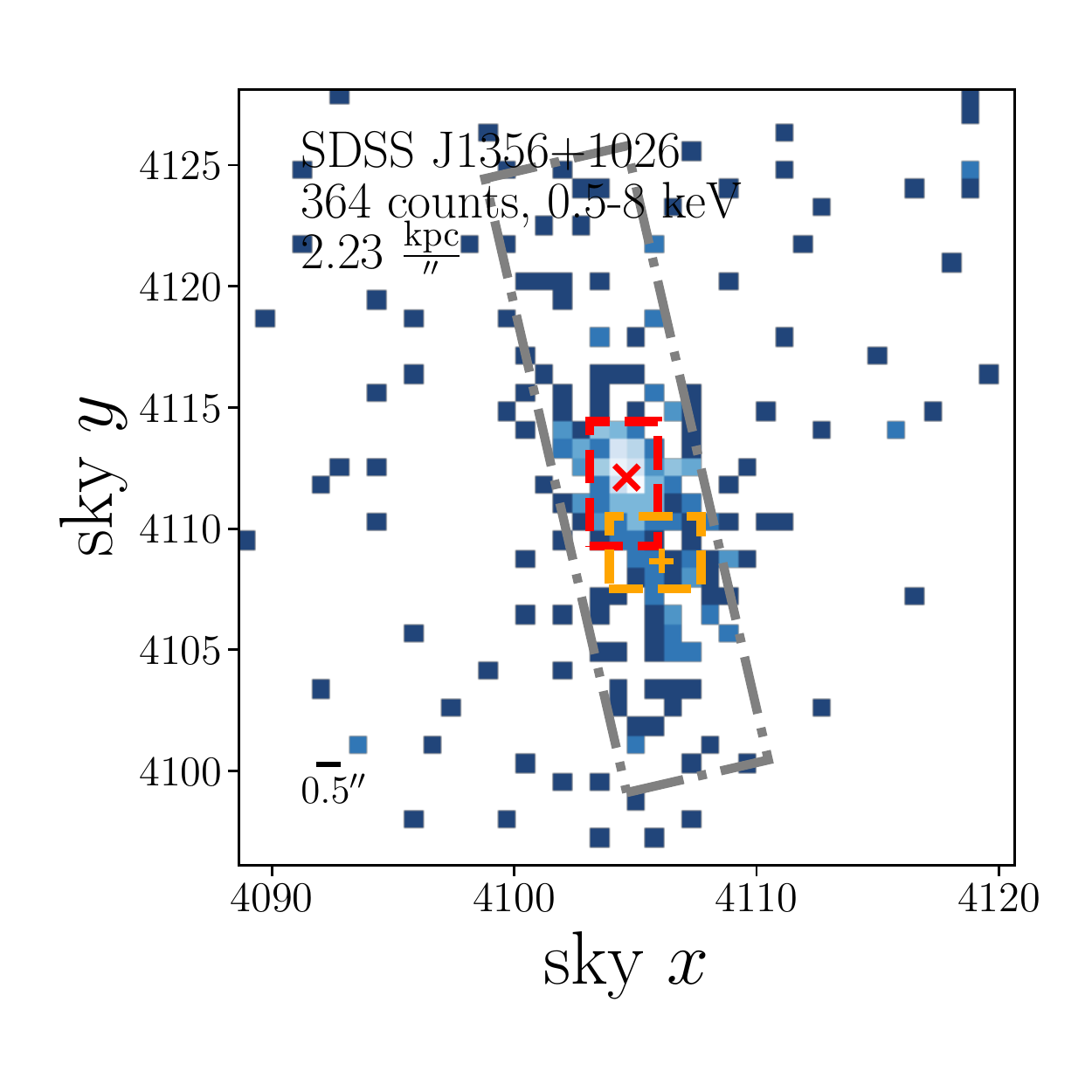}
    \end{minipage}
    \hspace{0.5cm}\vspace{-1cm}
    \begin{minipage}{0.32\linewidth}
    \includegraphics[width=0.85\linewidth]{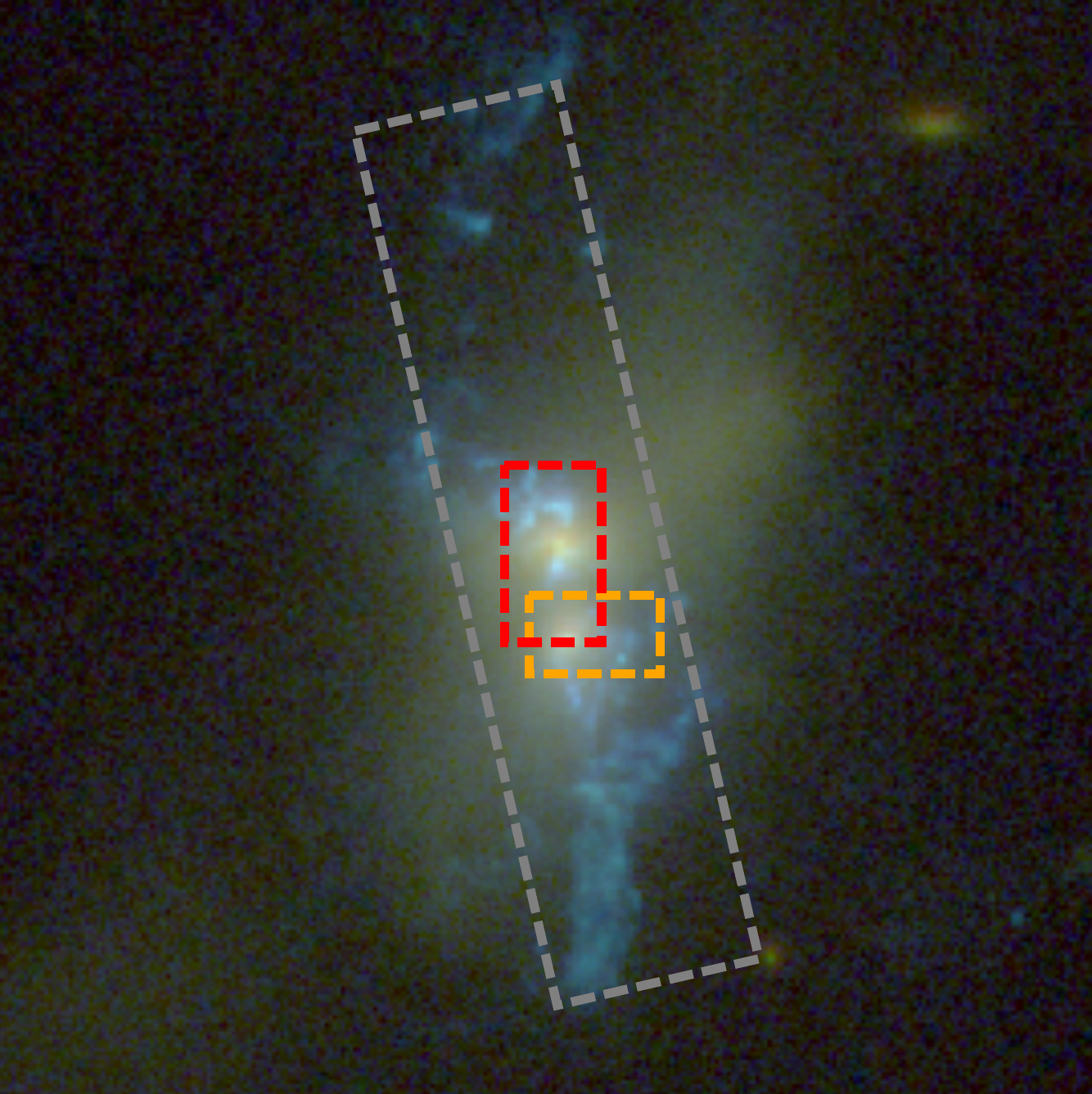}
    \end{minipage}
    \vspace{1cm}

    \begin{minipage}{0.32\linewidth}
    \includegraphics[width=\linewidth]{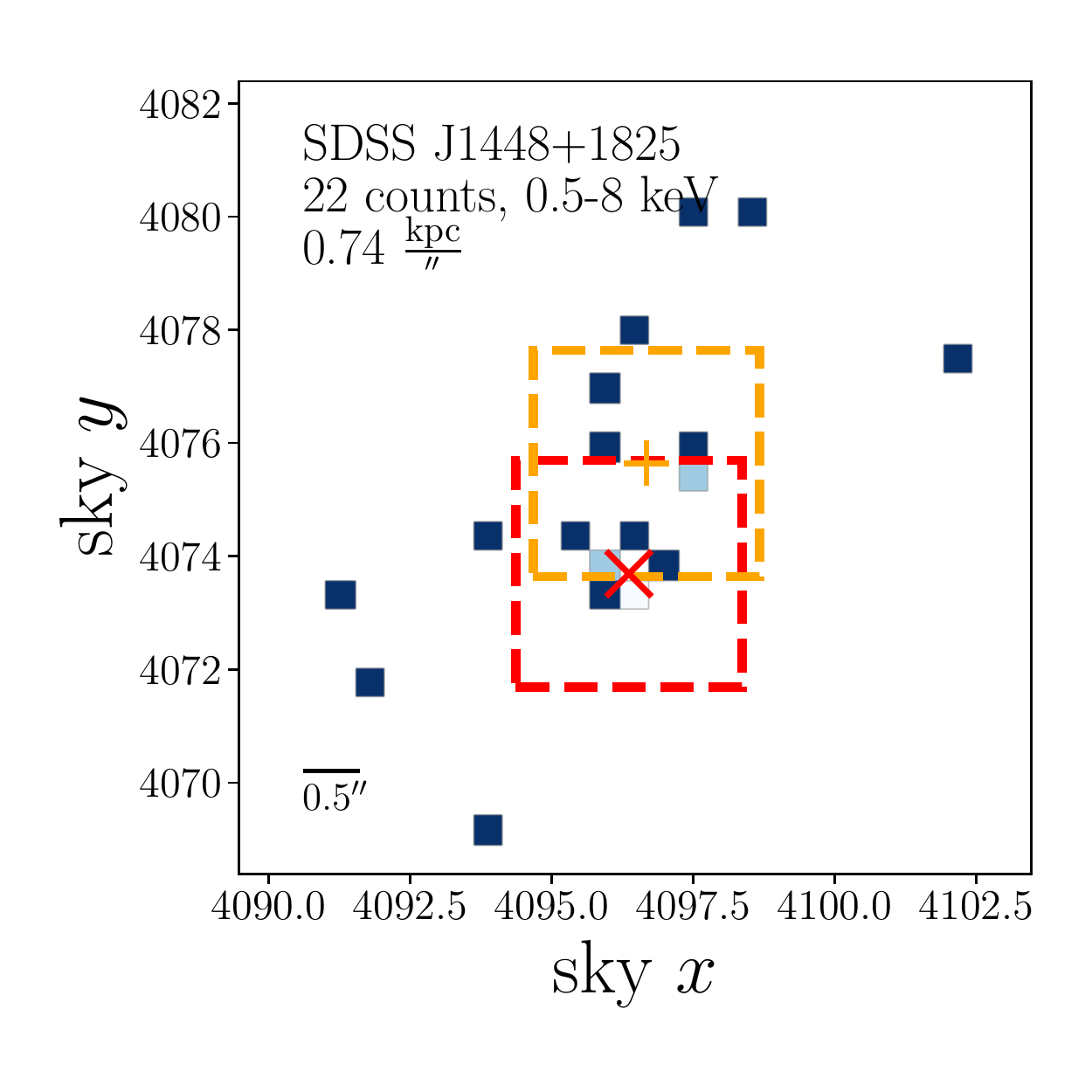}
    \end{minipage}
    \hspace{0.5cm}\vspace{-1cm}
    \begin{minipage}{0.32\linewidth}
    \includegraphics[width=0.85\linewidth]{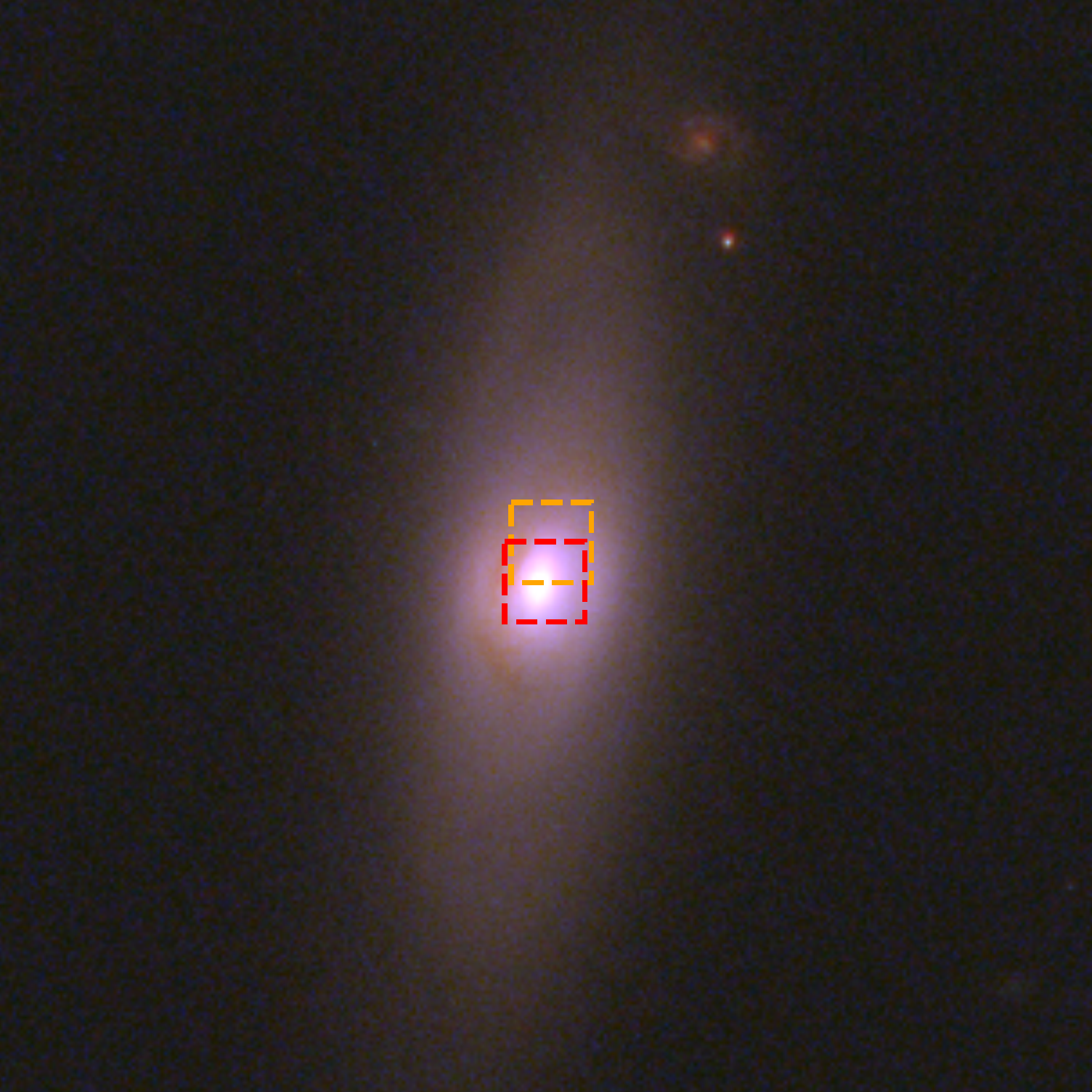}
    \end{minipage}
    \vspace{1cm}

    \begin{minipage}{0.32\linewidth}
    \includegraphics[width=\linewidth]{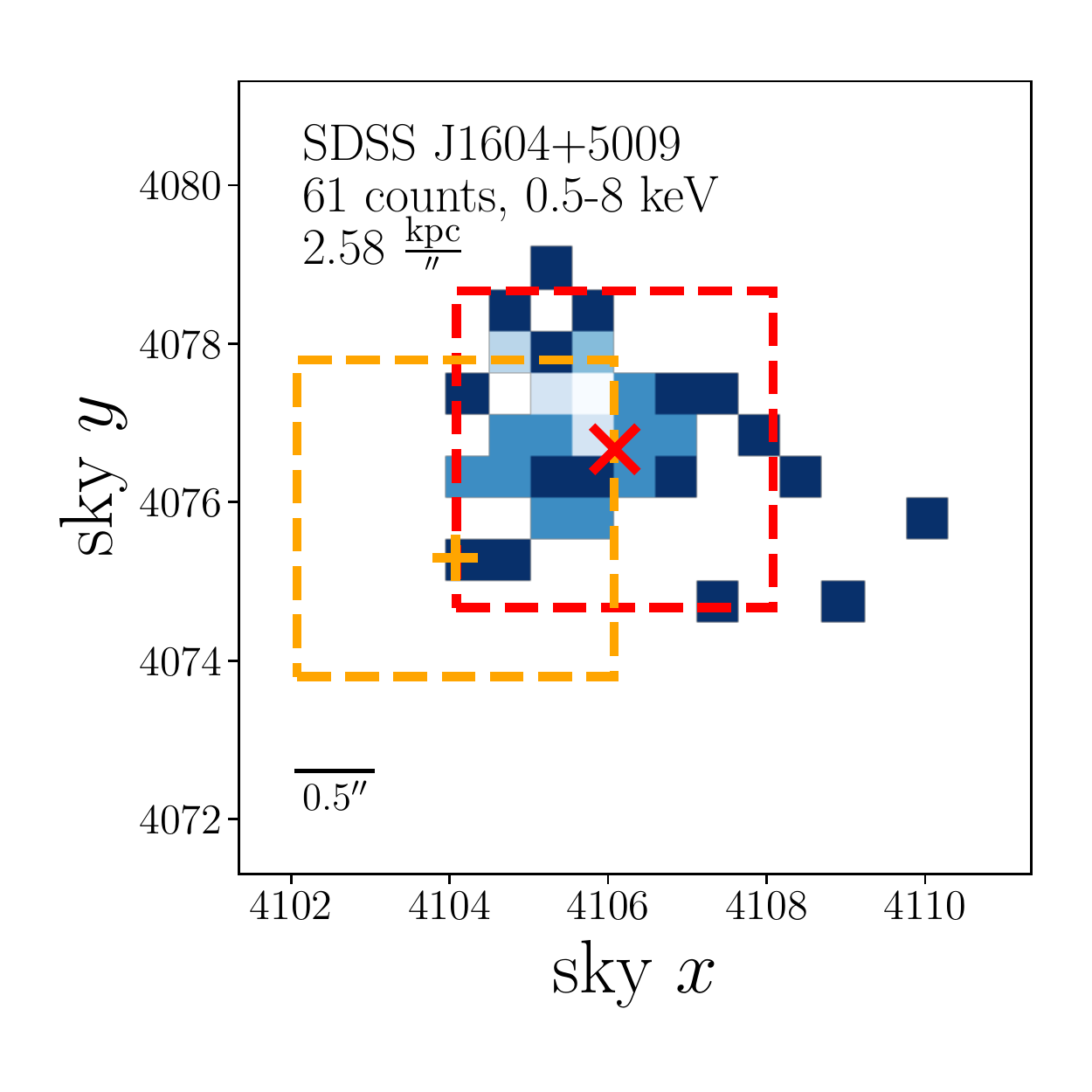}
    \end{minipage}
    \hspace{0.5cm}\vspace{-1cm}
    \begin{minipage}{0.32\linewidth}
    \includegraphics[width=0.85\linewidth]{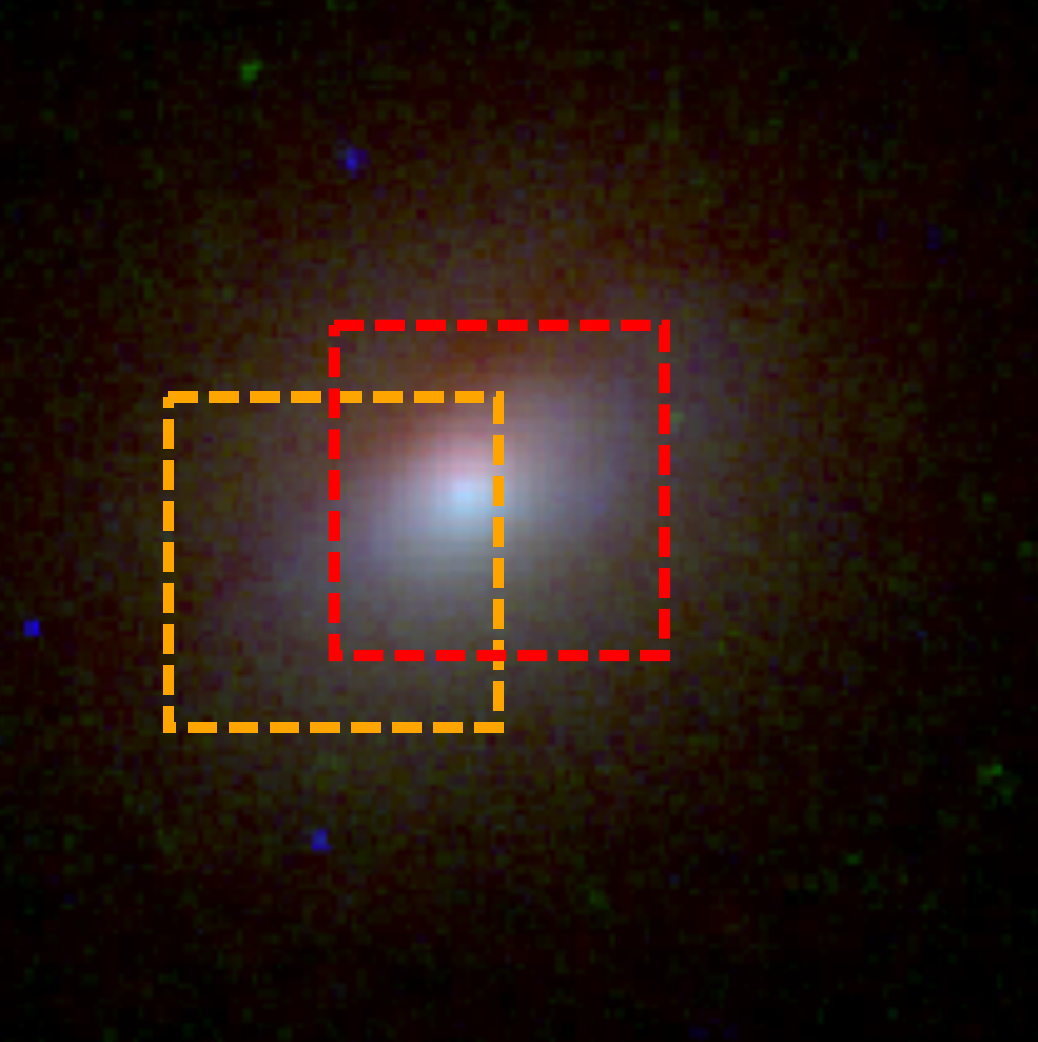}
    \end{minipage}
        \vspace{1cm}
\caption{Similar to SDSS J0841+0101, for SDSS J1356+1026 we denote the region within which the diffuse emission background component is restricted to with a gray box. Additionally, we show the combined X-ray emission for all \emph{Chandra} observations, where we use the best-fit astrometric shift values as found by \BAYMAX{}. All images are scaled in log-space with minimum and maximum counts/bin as follows: SDSS J1322+2631 (min=1, max=4), SDSS J1356+1026 (min=1, max=24), SDSS J1448+1825 (min=1, max=3), SDSS J1604+5009 (min=1, max=6).} 
\end{figure*}

%%%%%%%%%%%%%%%%%%%%%%%%%%%%%%%%%%%%%%%%%%%%%
%%%%%%%%%%%%%%%%%%%%%%%%%%%%%%%%%%%%%%%%%%%%%

For each observation, we restrict our analysis to photons with energies between $0.5$--$8$ keV.  We analyze the photons contained within square regions that are centered on the nominal X-ray coordinates of the AGN, $\mu_{\mathrm{obs}}$.  The length of each square is defined as $l_{\mathrm{box}}$, where $l_{\mathrm{box}}$ varies between 10 and 32 sky-pixels for each observation (4.95\arcsec~ and 15.84\arcsec~, respectively. See Figure~\ref{fig:GalaxyImages}). The known asymmetric \emph{Chandra} PSF feature is within this extraction region \citep{Juda&Karovska2010}, and sits approximately 0\farcs7 from the center of the AGNs.  Because our PSF model does not take into account this asymmetry, we mask the feature in all exposures before running \BAYMAX{}.
\subsection{Bayes Factor Results}
\par For each galaxy, we first run \BAYMAX{} using one background component (which accounts for the emission associated with the CXB and unresolved X-ray point sources) and non-informative priors, e.g., the prior distributions for $\mu$ are uniform distributions bound between $\mu_{\mathrm{obs}}-\frac{l_{\mathrm{box}}}{2}$ and $\mu_{\mathrm{obs}}+\frac{l_{\mathrm{box}}}{2}$. We then run \BAYMAX{} using informative priors, where the distributions for $\mu$ are constrained by and centered on the spatial position of the [\ion{O}{3}] $\lambda$5007 components (see Figure~\ref{fig:GalaxyImages}).  Here, the sky $x$ and sky $y$ limits of each prior distribution were determined by visually identifying where one may expect a galactic nucleus via the optical observations. Lastly, we note that our prior distributions for $\mu$ are wide enough to account for the relative astrometric shifts between the \emph{Chandra} and optical observations ($>1\arcsec$, see \citealt{Comerford2015}). 
\par To test the impact of the MCMC nature of nested sampling, we run \BAYMAX{} 100 times on each dataset.  The spread of the $\ln{BF}$ values are well-described by a Gaussian, and error bars are defined by the best-fit standard deviation.  In Table~\ref{tab:results} we list the various $\ln{BF}$ values for each of the 12 systems.  Here $\ln{BF}$ is defined as the logarithm ratio of the evidence for the dual point source model to the single point source model (i.e., $Z_{2}/Z_{1}$). Thus, values that are less than 0 are systems that are better described by the single point source model.
\subsection{Adding an Extended Background Emission Component}
\par We more closely analyze the 5 systems that have $\ln{BF}$ greater than 0 in favor of the dual point source model. Two of these galaxies, SDSS J0841+0101 and SDSS J1356+1026, show evidence for extended emission in the $HST$ F438W filter.  Because our background model is spatially uniform, we are assuming a constant background rate across the entire image. With this current model, it is possible that a region of background with a higher count-rate can be mistaken for a resolved point source sitting among a background with a lower count rate.  Although multiple analyses on SDSS J0841+0101 have concluded that the emission is consistent with two point sources (\citealt{Comerford2015, Pfeifle2019}, in addition to our analysis of two point sources and uniform background without additional components), contamination from extended diffuse emission better explains why \BAYMAX{} favors a dual point source more strongly using non-informative priors for SDSS J0841+0101.  The ``secondary" is most likely sitting in a region of X-ray emitting diffuse gas that is inconsistent with the spatial position of the nucleus of the merging galaxy. Similarly for SDSS J1356+1026, \BAYMAX{} favors a dual point source more strongly using non-informative priors.  The true nature of the extended X-ray emission has been studied extensively in the past \citep{Greene2009, Greene2012, Greene2014} and was found to most likely arise from photoionization and/or shocks from a quasar-driven superwind.
\par Thus, for SDSS J0841+0101 and SDSS J1356+1026 we add an additional background component to our model. In Figure~\ref{fig:GalaxyImages} we show these additional regions of background components in gray dash-dotted regions, where the position and size of these regions are visually determined from the $HST$ images. Within these regions, \BAYMAX{} fits for a different background fraction, $f_{BG}$, than for outside these regions.  We include the diffuse component when it is statistically favored, as determined by \BAYMAX{}.  In particular, for both the single and dual point source models, we compare the evidence of the original models ($Z_{1}$ and $Z_{2}$) to evidence the models that include a diffuse emission component ($Z_{1,\mathrm{D}}$ and $Z_{2,\mathrm{D}}$). We use informative priors for the locations of $\mu$, as shown in Figure~\ref{fig:GalaxyImages}. For both SDSS J0841+0101 and SDSS J1356+1026, we find that including a diffuse emission component is strongly preferred for both the single and dual point source models ($\ln{\frac{Z_{1, \mathrm{D}}}{Z_{1}}}, \ln{\frac{Z_{2, \mathrm{D}}}{Z_{2
}}} \gg 0$, see Table~\ref{tab:results}). With our updated model, SDSS J0841+0101 is no longer consistent with emission from two resolved point sources, and is instead better described by one point source with two background components ($\ln{BF_{\mathrm{D}}}<0$). However, SDSS J1356+1026 remains better described by two point sources. 
\par We analyze how the $BF$ determined by \BAYMAX{} depends on the shape and size of the additional background component. Specifically, because the diffuse emission surrounding SDSS J1356+1026 has an extreme spatial extension ($\approx$ 20 kpc) and is potentially driven by a superwind, the spatial distribution of extended gas is likely to be non-uniform within our square region ($l_{\mathrm{box}}=32$ sky-pixels or $\approx$35 kpc at $z=0.123$).  However, we find that that our results do not change when constraining our analysis to counts within a smaller $l_{\mathrm{box}}$ values (i.e, a physically smaller area over which the diffuse emission is more accurately modeled as spatially uniform); given the low number of counts available, we conclude that the X-ray emission of the diffuse background component can be appropriately modeled with a spatially uniform distribution.
\par Similarly for SDSS J0841+0101, under the assumption that the region dominated by extended diffuse emission surrounds both optical nuclei, the model favored by \BAYMAX{} remains a single point source, regardless of the shape. Naturally, as the size of the diffuse emission background component increases (and the size of the X-ray background region decreases) $M_{j,\mathrm{D}}$ begins to resemble $M_{j}$, with one (dominant) background component. However, as long as the diffuse emission region is constrained to overlap with emission seen in the $HST$ F438W filter (which represents a more informative model), the models favored by \BAYMAX{} remain a single point source for SDSS J0841+0101 and a dual point source for SDSS J1356+1026.  We conclude that SDSS J0841+0101 is most likely a single resolved point source, surrounded by extended diffuse X-ray emission while SDSS J1356+1026 is most likely a dual point source system, also surrounded by extended diffuse gas (for more details on the origin of this emission, we refer the reader to Section~\ref{J1356+2016}). 
\par In general, the user should test various models that are considered appropriate for a given observation. For the 3 other sources in which BAYMAX favored the dual point source model (SDSS J0752+2736, SDSS J1126+2944, and SDSS J1448+1825), we do not see any evidence of an additional high-count background, in either the X-ray or complementary optical observations (and, on average, these observations had a low number of total counts), and thus we do not test for the significance of including additional high-count background for these observations. One may ask whether the emission is better described by (i) two point sources ($M_{2}$) or (ii) a single point source plus a compact region of diffuse emission (sitting at the location of the secondary; $M_{\mathrm{1+diff}}$). As an example, we can compare the $BF$ between these two models for SDSS J1126+2944. Similar to $M_{\mathrm{two}}$, for $M_{\mathrm{1+diff}}$ we parametrize the diffuse emission component by fitting for the count ratio between its emission and the emission of the primary ($f$). We use the same informative priors as shown in Fig.~\ref{fig:GalaxyImages}. We find a $BF$ = 2.72 ($\pm$ 1.55), in favor of $M_{\mathrm{1+diff}}$. The larger error bars are reasonable, given that there are only $\sim$3 counts associated with either the secondary point source / diffuse emission component. We stress, however, that a compact uniformly emitting region in this case is contrived and not physical; in such a case, we would assign prior odds that take this into account, keeping the $BF$ in realistic territory. When doing similar tests using high-count simulations of dual point sources (where each point source is contributing $>$50 counts), the $BF$ in favor of $M_{\mathrm{2}}$ exceeds $10^{20}$, a reflection of our robust PSF models.
\par Thus, we update our list of dual AGN candidates to four systems: SDSS J0752+2736, SDSS J1126+2944, SDSS J1356+1026, and SDSS J1448+1825. SDSS J0752+2736 has a $BF$ value in favor of the dual point source model only when using non-informative priors, while the remaining three systems have $BF$ values in favor of the dual point source model when using both informative and non-informative priors.
\subsection{Strength of the Bayes Factor}
For each dual point source system, we analyze the strength of the $BF$. In the historical interpretation of the strength of the $BF$(see \citealt{Jeffreys1935} and \citealt{Kass&Raftery1995}), values between $\approx3 -10$ were defined as ``substantial", while values $>10$ were defined as ``strong".  However, these $BF$ value bins were arbitrarily defined; of course, the interpretation of a ``strong" $BF$ value depends on the context.  For each dual point source system we run false-positive tests to better, and uniquely, define a ``strong" $BF$.
\par The false-positive tests are set-up as follows: we create single point source simulations based on each observation in {\tt MARX}.  We constrain our analysis to the counts contained within the same sky  coordinates and energy cuts as the observations, use the same informative priors (or, non-informative in the case of SDSS J0752+2736), and add a uniform background contribution with a similar background fraction as each observation.  This results in simulations with a similar fraction of background counts as well as total number of counts as the observation. For SDSS J1356+1026, we also add a synthetic diffuse emission component (or, a background component with a higher count-rate) that is constrained within the same region as shown in Fig.~\ref{fig:GalaxyImages}. For each system, we run \BAYMAX{} on 1000 simulations and calculate what fraction have $BF$ values \emph{in favor} of a dual point source. Besides defining a ``strong" $BF$ value for each source, this technique also allows us to measure the probability that each system is more likely two point sources versus one point source. 
\par For the false-positive runs based on SDSS J0752+2736, SDSS J1356+1026, and SDSS J1448+1825, 99\% of the $\ln{BF}$ values are $<0.92$ in favor of a dual point source system; while for SDSS J1126+2944 99\% of the $\ln{BF}$ values are $<1.80$ in favor of a dual point source system.  Additionally, none of the 1000 simulations have $\ln{BF}$ values in favor of a dual point source model greater than what we measure (i.e., there is $<$99.9\% chance that a single point source with a comparable number of counts would return a $BF$ value, in favor of the dual point source model, greater than what we measure). Thus, we classify each Bayes factor value as ``strong" in favor of the dual point source model.

\section{Nature of the Dual point source Systems}
\label{sec:Nature}
\par We find that 4 of the 12 galaxies have strong $BF$ values in favor of the dual point source model: SDSS J0752+2736, SDSS J1126+2944, SDSS J1356+1026, and SDSS J1448+1825.  Generally, we find that the locations for a primary and secondary X-ray point source for  SDSS J1126+2944, SDSS J1356+1026, and SDSS J1448+1825 using non-informative priors are consistent, at the 68\% C.L, with those found using informative priors (albeit, with larger relative uncertainties). Further, because our informative priors are based on the locations of the spatially resolved \ion{O}{3} emission components, as presented in \cite{Comerford2015}, which were found to be consistent with the locations of the galactic nuclei, the best-fit \BAYMAX{}-derived separations for SDSS J1126+2944, SDSS J1356+1026, and SDSS J1448+1825 are, by nature, consistent with the separations between the optical nuclei. The remaining 8 galaxies have $\ln{BF}$ that favor a single point source, or are consistent with 0 at the 99.7\% confidence level (see Table~\ref{tab:results}). 
\par Before we investigate the nature of each dual point source system, it is important to note the specific differences in our analysis versus the original analysis presented in \cite{Comerford2015}: (i) in the original analysis the X-ray model contained two sources with a separation and orientation on the sky that were fixed at the measured separation and position angle of the two [\ion{O}{3}] $\lambda$5007 emission components; and (ii) the significance of each of the two sources in the model were estimated individually, such that each system could be categorized into three groups: no point source, 1 point source, or 2 point sources.  
\par Regarding (i), because we run \BAYMAX{} using both informative and non-informative priors, we are sensitive to detecting emission from a point source anywhere in the image.  Regarding (ii), because \BAYMAX{} is a \emph{comparative} analysis, we can only conclude that each system is either better explained by a single or dual point source.  Although the 8 systems with $\ln{BF}$ values $<0$ are better explained by a single point source versus a dual point source, they require a specified model for comparison in order to better understand their true nature. For example, one could compare a single point source to a uniform background, in order to analyze whether the emission is consistent with a compact object versus the CXB.  However, we note that the true origin of the X-ray emission of these 8 systems is outside the scope of this paper. 
\par In the following section, we aim to better understand the true nature of the 4 X-ray dual point source systems.  In order to better determine the likelihood that each dual point source system is actually composed of two AGN, we analyze the posterior distributions and X-ray spectra.  For each system, we determine the best-fit values of each fit parameter using the median values of their posterior distributions, which is appropriate given their unimodal nature.   In Table~\ref{tab:pymc3} we list the best-fit values for $r$, $\log{f}$ and $\log{f_{BG}}$. 
\subsection{X-ray Spectral Analysis of Individual point source Components}
The spectral fits and flux values are determined using XSPEC, version 12.9.0 \citep{Arnaud1996}. For each point source component (2 per system, the primary and secondary point source), we create 1000 spectral realizations by probabilistically sampling from the full distribution of counts.  Each spectral realization uses $\theta_{2}$ values that are drawn from the posterior distributions as determined by \BAYMAX{}.  For each iteration, \BAYMAX{} assigns each count to a specific model component (i.e, the primary, secondary, or background), based on the relative probabilities of being associated with each component. We then fit the spectra of the counts associated with the primary and secondary (and thus, they are background-subtracted spectra), and create distributions of spectral parameters and flux values based on the best-fit values of each fit. This allows, for the first time, a spectral analysis of \emph{individual} point source components in candidate dual AGN systems that are closely separated.  This type of analysis is useful for measuring the fluxes of each source, as well as better constraining the flux ratio between the secondary and primary.  Specifically, fitting 1000 spectral realizations for each point source allows for estimations on the flux ratio, whereas \BAYMAX{} calculates the likely count ratio.
\par Each point source component is modeled as either a simple absorbed power-law ({\tt phabs}$\times${\tt zphabs}$\times${\tt pow}; hereafter $m_{\mathrm{spec,1}}$) or an absorbed power-law with Compton scattering ({\tt phabs}$\times$({\tt pow} + {\tt zphabs}$\times${\tt pow}); $m_{\mathrm{spec,2}}$), where the power-law indices are tied to one-another.  This latter model has been found to accurately describe the spectra of AGNs in merger-environments (see, e.g., \citealt{Pfeifle2019}). Although the Compton scattering component can be fit using a physically motivated model (such as BNTorus; \citealt{BrightmanNandra2011}), doing so with a high statistical significance requires more counts than the observations contain. When using phenomenological models to describe our low-count spectra, the {\tt zphabs} component in $m_{\mathrm{spec,2}}$ effectively accounts for the Compton scattering.  We implement the Cash statistic ({\tt cstat}; \citealt{Cash1979}) in order to best assess the quality of our model fits.  Specifically, the latter model is used if it results in a statistically significant improvement in the fit, such that $\Delta C_{\mathrm{stat}}>$ 2.71 (see, e.g.,\citealt{Tozzi2006, Brightman2012}), corresponding to a fit improvement with 90\% confidence (however, this is only valid if $C_{\mathrm{stat}}$/dof$\approx$1, see \citealt{Brightman2014} and references therein).  
\par It has been found that the constraint on the power-law spectral index, $\Gamma$, is poor for low-count ($<$500) \emph{Chandra} spectra (where the average uncertainty on $\Gamma$ is $> 0.5$, \citealt{Brightman2012}). However, the large uncertainties introduced into the spectral fit can be reduced by fixing $\Gamma$. Thus, for those sources with an average of $<$10 counts (i.e., most of the secondary point sources), we assume the simpler spectral model, $m_{spec, 1}$, and fix $\Gamma$ to a value of 1.8 \citep{Corral2011,Yan2015}.  For the primary point sources (where the average number of $0.5$--$8.0$ keV counts ranges from $15$--$177$), if the best-fitting model where $\Gamma$ is free is a significantly better fit than the best-fitting model where $\Gamma$ is fixed (using the same criterion of $\Delta C_{\mathrm{stat}} > 2.71$) we choose the model with $\Gamma$ as free as the best-fitting model.  The exception to this is if nonphysical values were pegged for $\Gamma$ (i.e., values greater than 3 or less than 1; see \citealt{Ishibashi2010}) or if the extragalactic column density $N_{H}$ was pegged to values $>10^{24}$ cm$^{-2}$. Only one primary point source, SDSS J1356+1026, met this criterion.  For each model, we fix the column density to the Galactic value \citep{Kalberla2005} as well as the redshift to that of the host galaxy.  
\par We use the criterion of $L_{2-7~\mathrm{keV, unabs}} > 10^{40}$ erg s$^{-1}$, as a first pass, to rule out possible non-AGN contributions.  At X-ray luminositinies below this threshold, there are a handful of different possible sources of contamination, including a high-mass X-ray binary (HMXB) or an ultraluminous X-ray source (ULX).  Although the most luminous ULXs may contain a black hole of intermediate ($>100$ $M_{\odot}$) mass, the compact object is still thought to be accreting matter from a massive donor star.  Thus, these systems can be viewed as HMXBs in a broader sense.  The majority of the high-mass X-ray binary population has $2$--$7$ keV X-ray luminosities between 10$^{38}$--10$^{39}$ erg s$^{-1}$, while the ULX population dominates at the highest luminosities, with $L_{2-7~\mathrm{keV}}>$ 10$^{39}$ erg s$^{-1}$ (e.g., \citealt{Swartz2011, Walton2011}). The overall X-ray luminosity function of HMXBs and ULXs indicates a general cutoff at $L_{2-7~\mathrm{keV}}=$10$^{40}$ erg s$^{-1}$ (e.g., \citealt{Mineo2012, Sazonov2017, Lehmer2019}), and previous studies on XRB contamination in both late- and early-type galaxies have concluded that the majority of nuclear (within $2\arcsec$ of the galactic nucleus) X-ray point sources with $L_{2-7~\mathrm{keV}}>$ 10$^{40}$ erg s$^{-1}$ are highly unlikely to be emission associated with accretion onto XRBs \citep{Foord2017, Lehmer2019}.  We note, however, that such studies have yet to be carried out for a sample of merging systems, where merger-induced shocks and starbursts can amplify the surrounding X-ray emission. Particularly in the case of SDSS J1356+1026, which visibly has more complicated surroundings, we look at additional environmental aspects (see Section~\ref{J1356+2016}) before classifying the likely nature of the X-ray emission. 
\par In addition to $L_{2-7~\mathrm{keV,~unabs}}$, we analyze the hardness ratio of each, $HR$, defined as $HR = (H − S)/(H + S)$. Here, $H$ and $S$ are the number of hard and soft X-ray counts, where the threshold between the two is set to $2$ keV.  We list the best-fit values for each spectral parameter, $F_{0.5-8}$, $L_{2-7~\mathrm{keV,~unabs}}$, and $HR$, in Table~\ref{tab:spectra} (we denote the values for the primary and secondary with subscripts $p$ and $s$). For SDSS J1126+2944$_{p}$, SDSS J1356+1026$_{p}$, and SDSS J1448+1825$_{p}$ we quote the unabsorbed $2$--$7$ keV luminosities from $m_{\mathrm{spec,2}}$. However, because the best-fit extragalactic column density will be systematically lower for the simpler model ($m_{\mathrm{spec,1}}$), we also list the best-fit parameters for $m_{\mathrm{spec,1}}$ for the purposes of comparison between the primary and secondary in a given system. 
\begin{figure*}
\centering
    \includegraphics[width=0.42\linewidth]{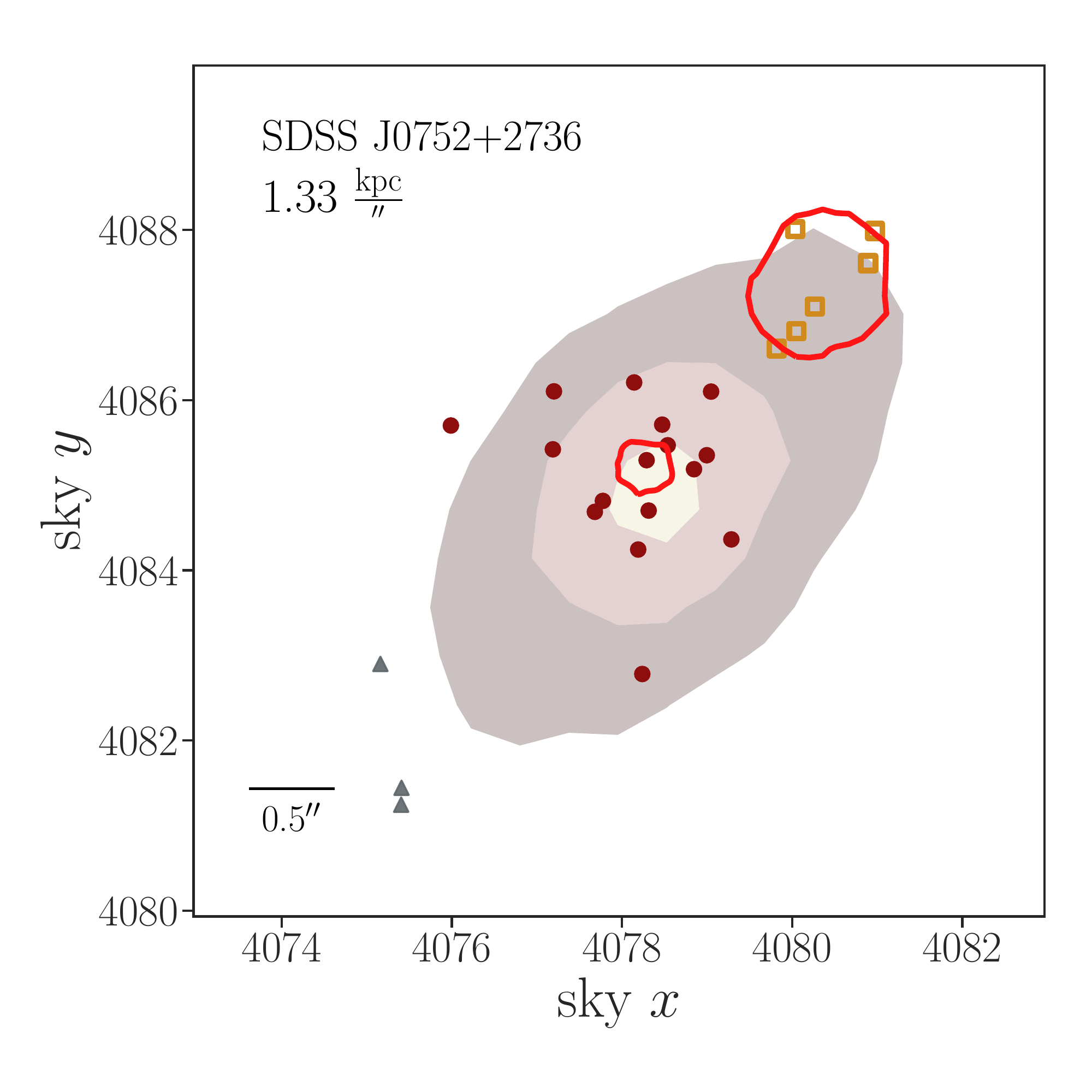}{\hskip 0pt plus 0.3fil minus 0pt}
    \includegraphics[width=0.45\linewidth]{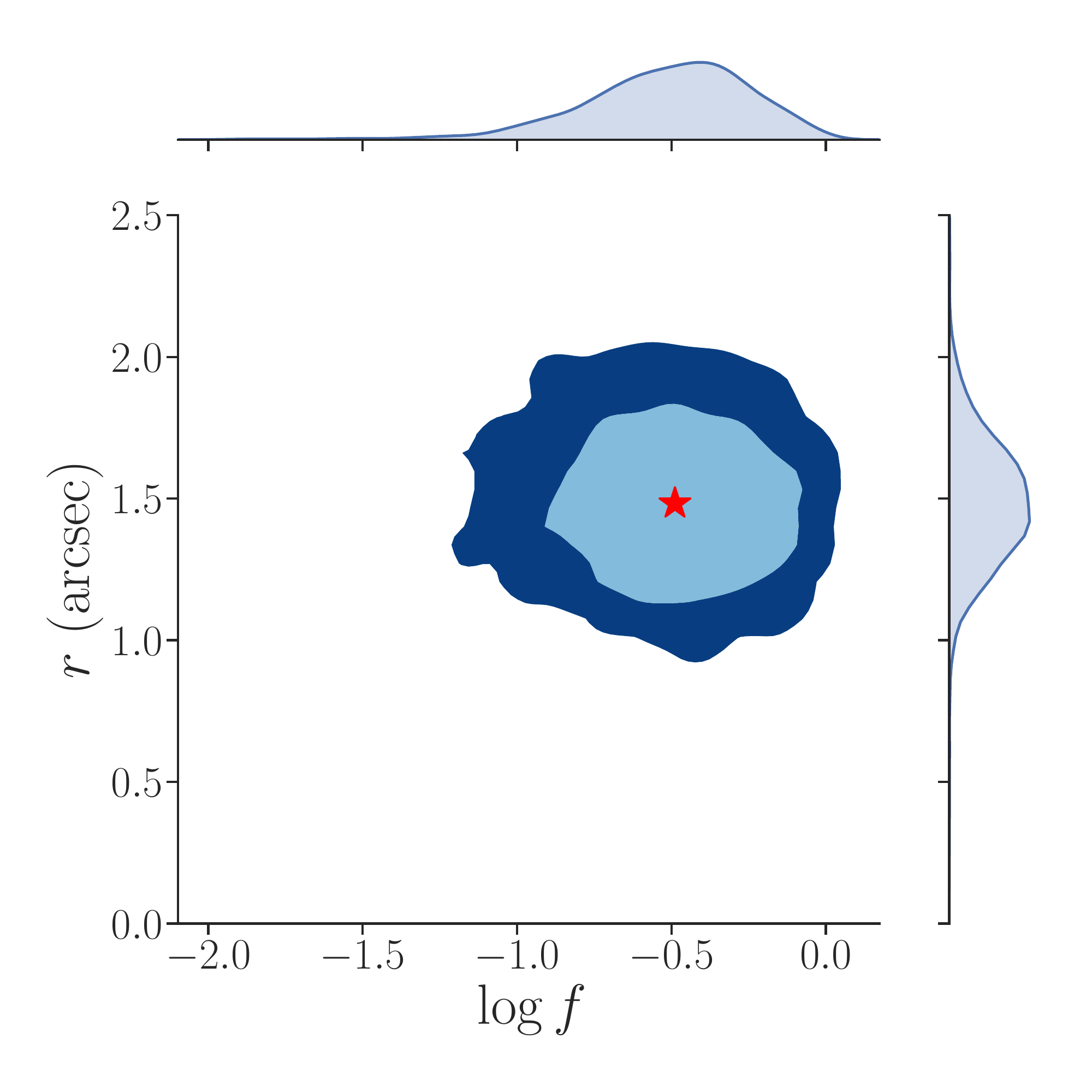}{\vskip 0.09cm plus 1fill}
    
    \includegraphics[width=0.42\linewidth]{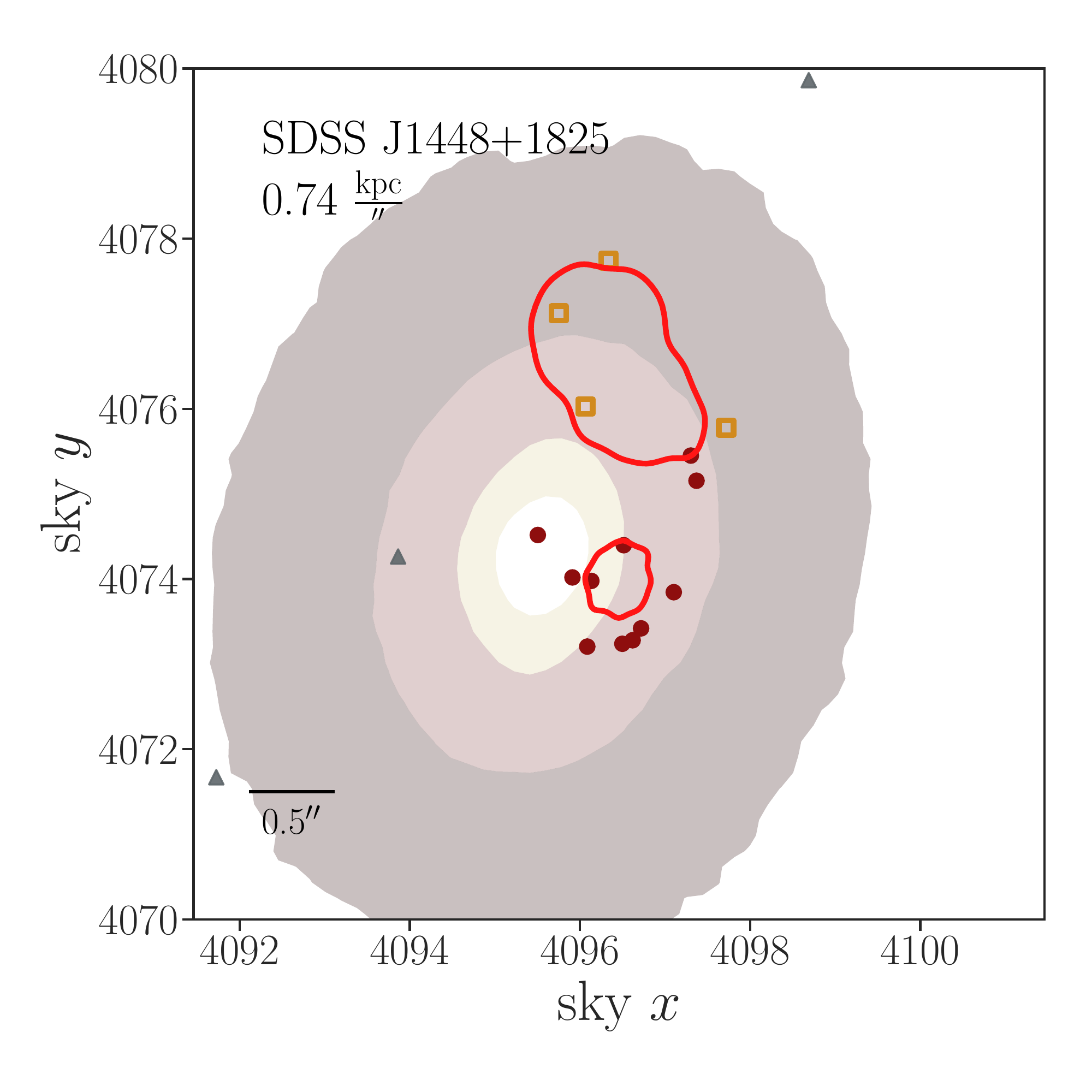}{\hskip 0pt plus 0.3fil minus 0pt}
    \includegraphics[width=0.45\linewidth]{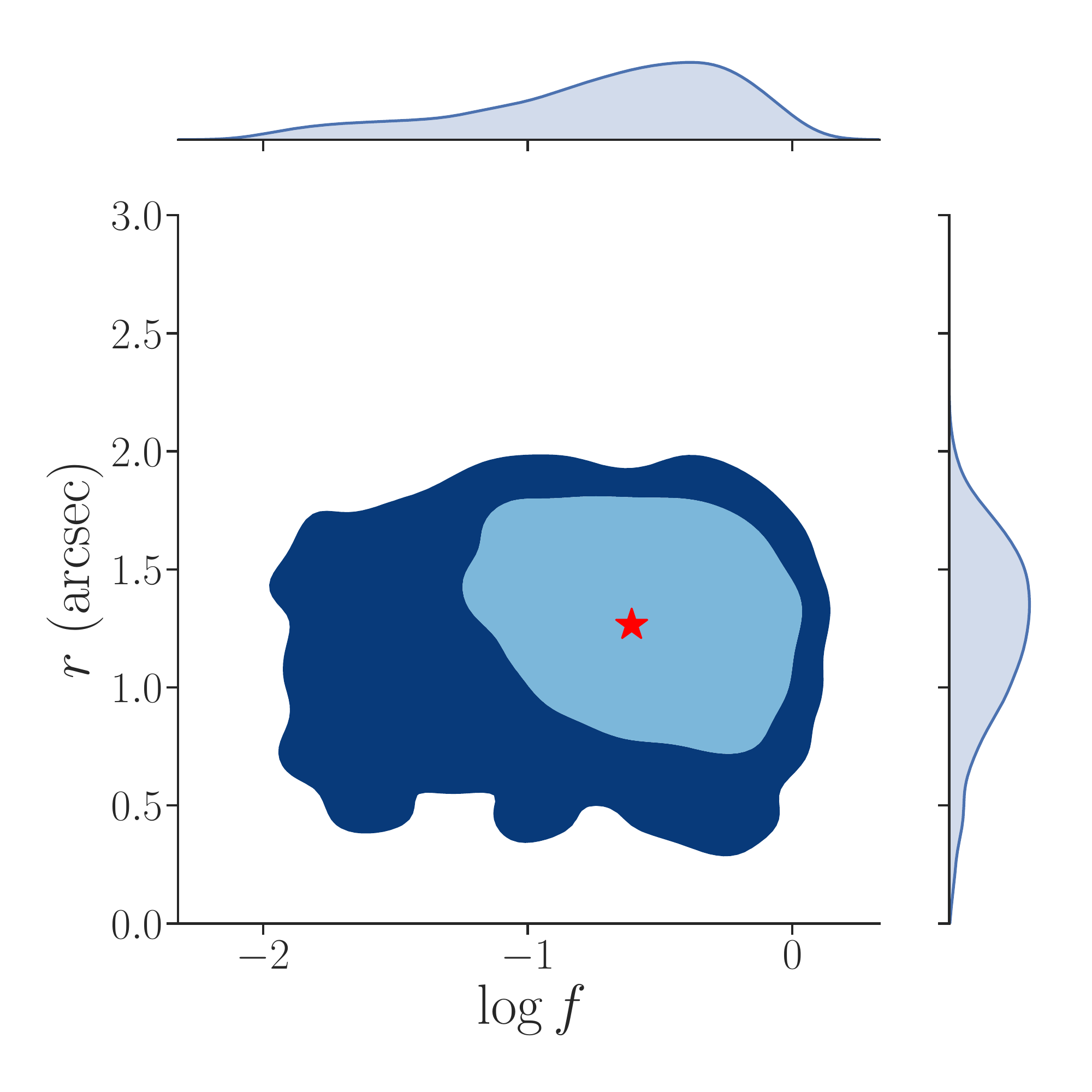}{\vskip 0.09cm plus 1fill}
    \caption{\emph{Left:} The $0.5$--$8$ keV datasets for the two dual AGN candidates SDSS J0752+2736 and SDSS J1448+1825.  We plot the 68\% confidence intervals (red lines) for the best-fit sky $x$ and sky $y$ positions for a primary and secondary.  Here, counts most likely associated with the primary are denoted by circles, counts most likely associated with the secondary are denoted by squares, counts most likely associated with background are shown as faded triangles.  In order to more clearly see the results, we do not bin the data. Contours of the $HST$ F160W observations of the host galaxies are overplotted (with the exception of SDSS J0752+2736, which are contours of the SDSS i-band observation).~\emph{Right}: Joint posterior distribution for the separation $r$ (in arcseconds) and the count ratio (in units of $\log{f}$), with the marginal distributions shown along the border. The 68\%, and 95\% confidence intervals are shown in blue contours. We denote the location of the median of the posterior distributions with a red star.}
\label{fig:PyMC3_1}
\end{figure*}

\begin{figure*}
\centering

    \includegraphics[width=0.42\linewidth]{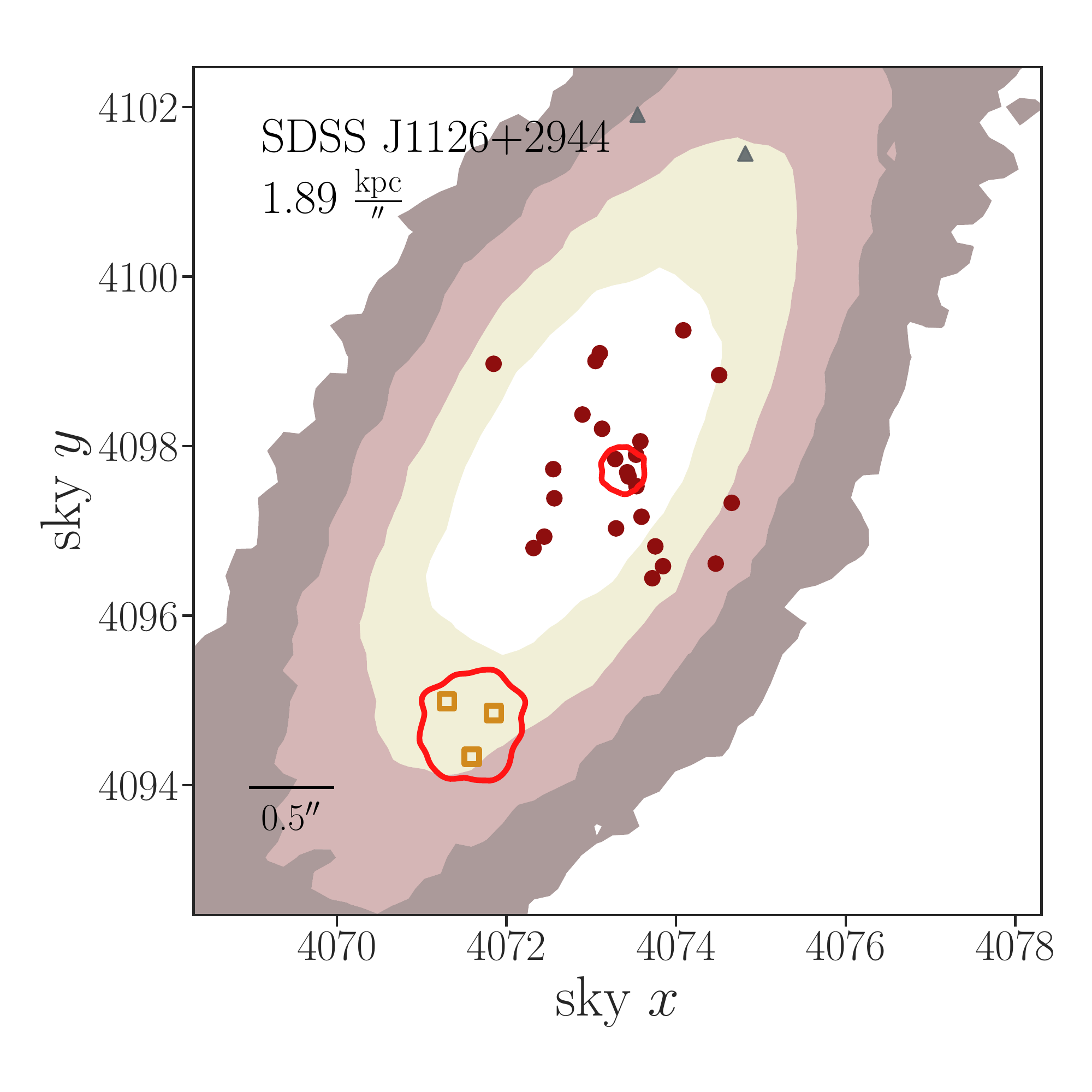}{\hskip 0pt plus 0.3fil minus 0pt}
    \includegraphics[width=0.45\linewidth]{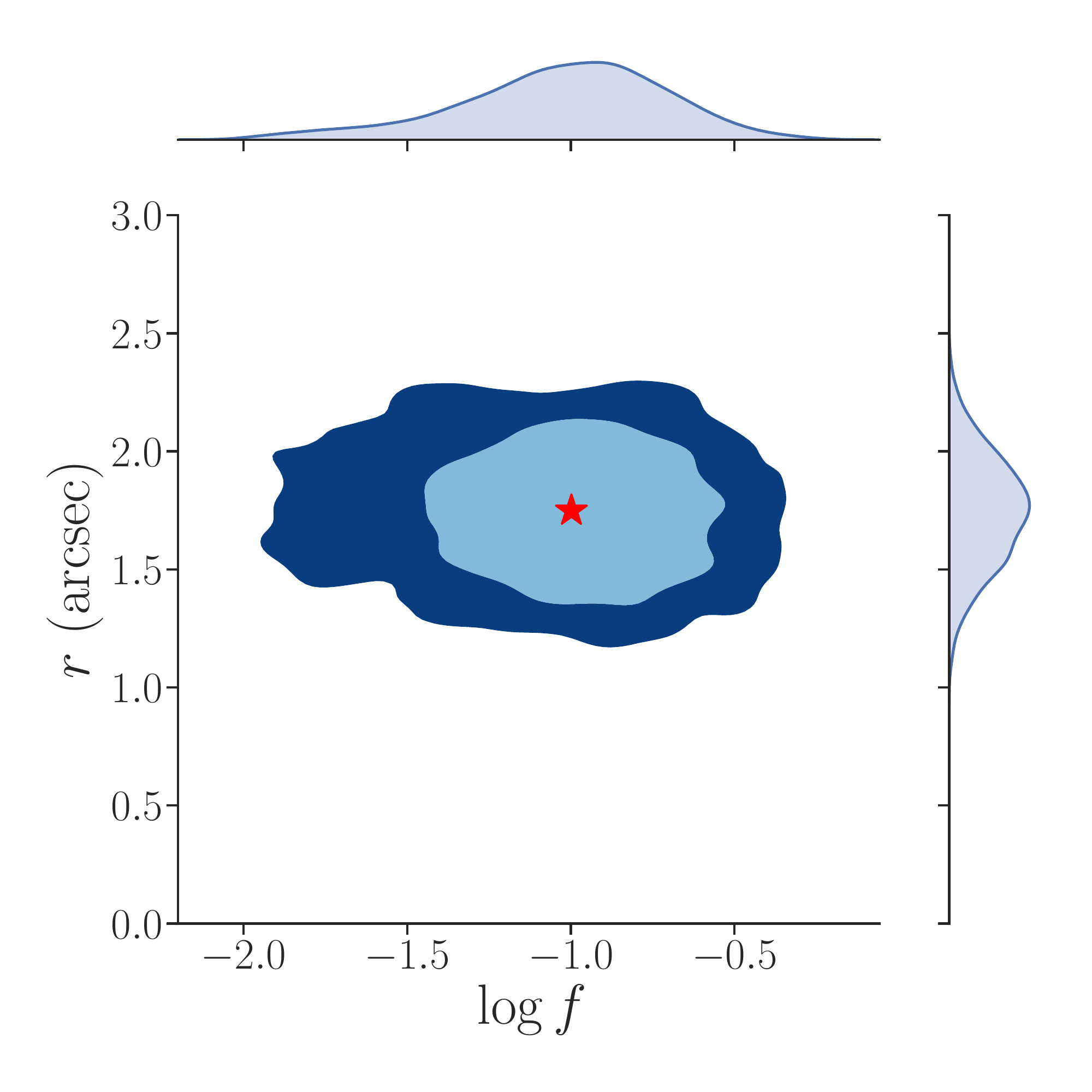}{\vskip 0.09cm plus 1fill}

    \includegraphics[width=0.42\linewidth]{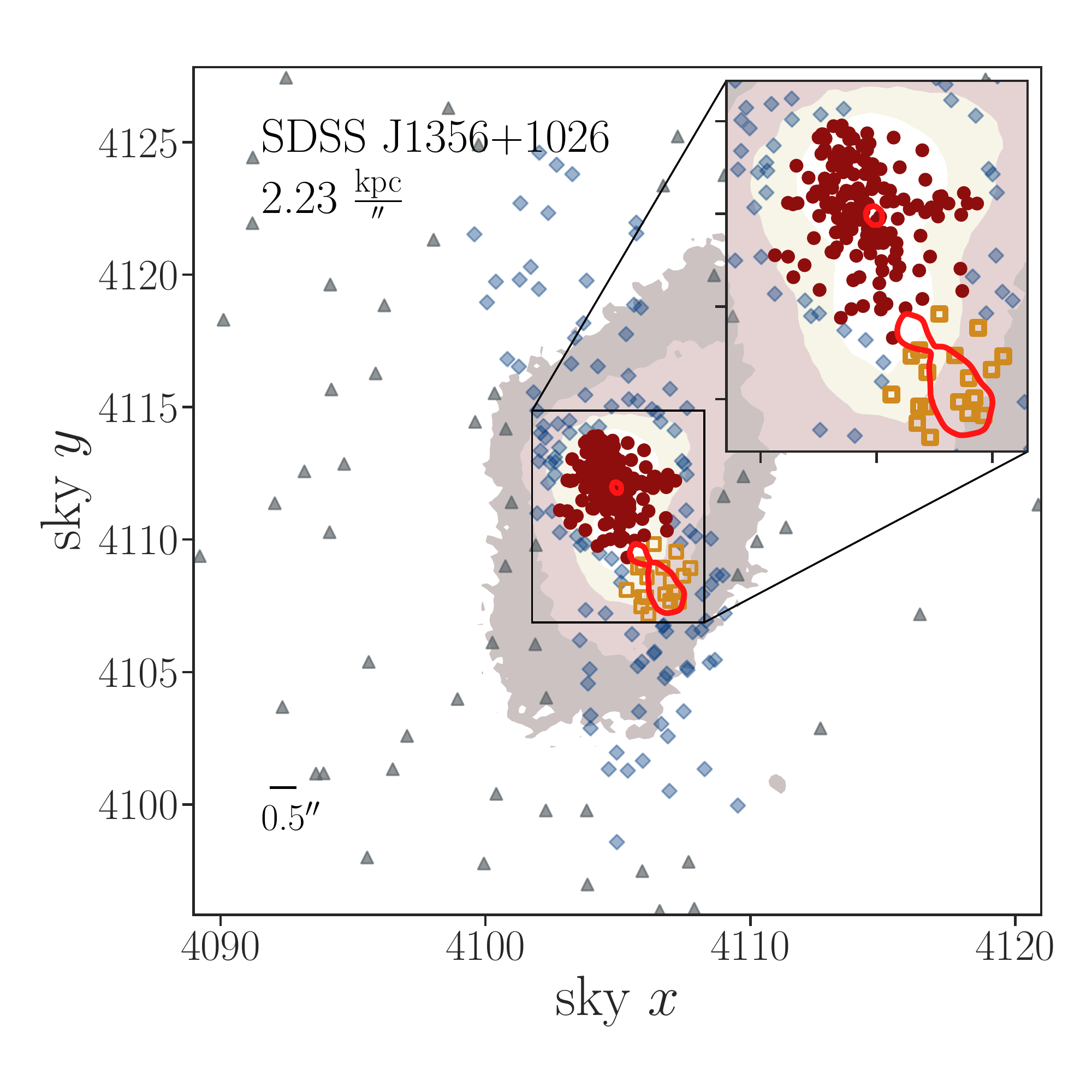}{\hskip 0pt plus 0.3fil minus 0pt}
    \includegraphics[width=0.45\linewidth]{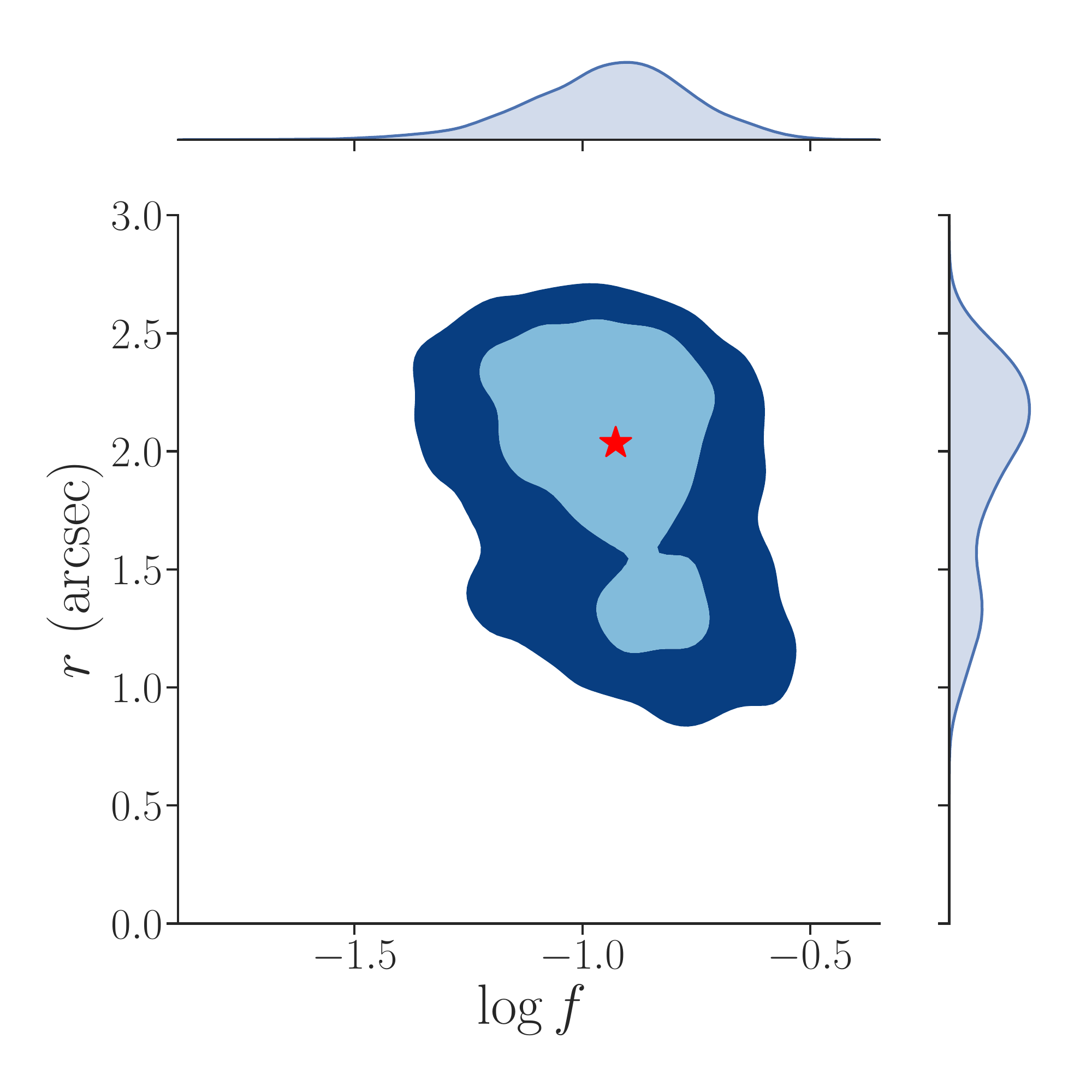}{\vskip 0.09cm plus 1fill}
    \caption{The $0.5$--$8$ keV datasets for the two dual AGN candidates whose primary and secondary X-ray point sources meet our AGN luminosity criterion (left) and the joint posterior-distribution for the separation $r$ and the count ratio (right).  Symbols and contours follow the same guidelines as Fig.~\ref{fig:PyMC3_1}. For SDSS J1356+1026, we denote diffuse emission background with faded diamonds.}
\label{fig:PyMC3_2}
\end{figure*}

%%%%%% TABLE PYMC3 RESULTS %%%%%%
\begin{table*}[t]
\begin{center}
\caption{Posterior Results for $\theta_{2}$}
\label{tab:pymc3}
\small
\begin{tabular*}{0.95\textwidth
}{cccccccc}
	\hline
	\hline
	\multicolumn{1}{c}{Galaxy Name} &
	\multicolumn{1}{c}{$\alpha_{p}$} &
	\multicolumn{1}{c}{$\delta_{p}$} &
	\multicolumn{1}{c}{$\alpha_{s}$} &
	\multicolumn{1}{c}{$\delta_{s}$} &
	\multicolumn{1}{c}{$r$ (arcsec)} & 
	\multicolumn{1}{c}{$\log{f}$} & \multicolumn{1}{c}{$\log{f_{bkg}}$} \\
	\multicolumn{1}{c}{(1)} & \multicolumn{1}{c}{(2)} & \multicolumn{1}{c}{(3)} & \multicolumn{1}{c}{(4)}  & \multicolumn{1}{c}{(5)}  & \multicolumn{1}{c}{(6)} &
	 \multicolumn{1}{c}{(7)}  & \multicolumn{1}{c}{(8)}\\
	\hline
	SDSS J0752+2736 & 7:52:23.341 & +27:36:43.516 & 7:52:23.266 & +27:36:44.562 & 1.50$\pm$0.30 & $-0.47\pm$0.36 & $-0.72\pm$0.40 \\
    SDSS J1126+2944 & 11:26:59.534 & +29:44:42.573 & 11:26:59.602 & +29:44:41.101 & 1.74$\pm$0.33 & $-1.00\pm$0.44 & $-0.97\pm$0.40 \\
    SDSS J1356+1026 & 13:56:46.123 & +10:26:09.321& 13:56:46.067 & +10:26:07.502 & 2.00$\pm$0.62 & $-0.92\pm$0.23 & $-0.16\pm$0.10 \\
    SDSS J1448+1825 & 14:48:04.174 & +18:25:37.925 & 14:48:04.177 & +18:25:39.115 & $1.29\pm0.52$ & $-0.45\pm$0.80 & $-0.40\pm$0.34 \\
	\hline 
	\hline
\end{tabular*}
\end{center}
Note. -- Columns: (1) SDSS galaxy designation; (2) the central R.A. of the primary X-ray source; (3) the central declination of the primary X-ray source; (4) the central R.A. of the secondary X-ray source; (6) the central declination of the secondary X-ray source; (6) the separation between the two point sources in arcseconds; (7) the log of the count ratio between the secondary and primary; (8) the log of the count ratio between the background contribution. For SDSS J1356+1026, the background component is defined as the diffuse emission component.  Each value is the best-fit value from the posterior distributions, defined as the median of the distribution. All posteriors distributions are unimodal, and thus the median is a good representation of the value with the highest likelihood (with the exception of $r$ for J1356+1026, see Fig.~\ref{fig:PyMC3_2}). Error bars represent the 68\% confidence level of each distribution.
\end{table*}
%%%%%% TABLE PYMC3 RESULTS %%%%%%

%%%%%% TABLE SPECTRAL RESULTS %%%%%%
\begin{table*}
\caption{Best-fit Spectral Parameters}
\label{tab:spectra}
\small
\begin{center}
\begin{tabular*}{0.8\linewidth
}{lcccccr}
	\hline
	\hline
	\multicolumn{1}{c}{Galaxy Name} &
	\multicolumn{1}{c}{$m_{\mathrm{spec, x}}$} & \multicolumn{1}{c}{$N_{H}$ (10$^{22}$ cm$^{-2}$)} & \multicolumn{1}{c}{$\Gamma$} &
	\multicolumn{1}{c}{$F_{0.5-8~\mathrm{keV}}$} & 
	\multicolumn{1}{c}{$L_{2-7~\mathrm{keV,~unabs}}$} &
	\multicolumn{1}{c}{HR}\\
	\multicolumn{1}{c}{(1)} & \multicolumn{1}{c}{(2)} & \multicolumn{1}{c}{(3)} & \multicolumn{1}{c}{(4)} & \multicolumn{1}{c}{(5)} & \multicolumn{1}{c}{(6)} & \multicolumn{1}{c}{(7)}
	\\
	\hline
    SDSS J0752+2736$_{p}$ & 1 & $<10^{-2}$ & 1.8 & 2.74$\pm$0.59 & 1.67$\pm$0.36 & $-$0.24$\pm$0.1\\
	SDSS J0752+2736$_{s}$ & 1 & $<10^{-2}$ & 1.8 & 0.46$\pm$0.10 & 0.71$\pm$0.16 & $-$0.53$\pm$0.40\\
	\hline
    SDSS J1126+2944$_{p}$ & 2 & 34.20$\pm$2.00 & 1.8 & 33.50$\pm$2.90 & 284.50$\pm$55.90 & 0.32$\pm$.10\\
                  & 1 & 0.23$\pm$0.1 & 1.8 & 11.80$\pm$0.90 & 19.00$\pm$1.70 \\
    SDSS J1126+2944$_{s}$ & 1 &  14.30$\pm$7.70 & 1.8 & 4.64$\pm$2.00  & 28.80$\pm$15.80 & 0.30$\pm$0.2\\
    \hline
    SDSS J1356+1026$_{p}$ & 2 & 41.10$\pm$14.50 & 2.54$\pm$0.27
    & 26.80$\pm$4.50 & 3.40$\pm$1.60$\times10^{2}$ & 0.30$\pm$0.10\\
            & 1 & $<10^{-2}$ & 1.8 & 17.55$\pm$3.2 & 35.50$\pm$6.60 \\
    SDSS J1356+1026$_{s}$ & 1 & $<10^{-2}$ & 1.8 & 0.90$\pm$0.41 & 1.80$\pm$0.80 & $-$0.30$\pm$0.29 \\
    \hline
    SDSS J1448+1825$_{p}$ & 2 & 56.30$\pm$14.5 & 1.8 & 12.80$\pm$5.20 & 17.50$\pm$10.00 & $-0.1\pm0.20$\\
                    & 1 & $<10^{-2}$ & 1.8 & 4.20$\pm$1.00 & 0.75$\pm$0.2 \\
    SDSS J1448+1825$_{s}$ & 1 & $<10^{-2}$ & 1.8 & 0.56$\pm$0.47 & 0.11$\pm$0.097 & $-$0.98$\pm$0.1 \\
	\hline 
	\hline
\end{tabular*}
\end{center}
Note. -- Columns: (1) SDSS galaxy designation, we denote the primary and secondary with subscripts $p$ and $s$; (2) the spectral model used; (3) the best-fit extragalactic column density; (4) the assumed or best-fit spectral index; (5); the measured $0.5$--$8$ keV flux, in units of 10$^{-15}$ erg s$^{-1}$ cm$^{-2}$; (6) the rest-frame, unabsorbed, $2$--$7$ keV luminosity in units of 10$^{40}$ erg s$^{-1}$; (7) the hardness ratio, defined as $HR= (H-S) / (H+S)$.  Each best-fit value is defined as the median of the full distribution. Error bars represent the 1$\sigma$ confidence level of each distribution.
\end{table*}
%%%%%% TABLE SPECTRAL RESULTS %%%%%%
\begin{figure*}
\centering
    \includegraphics[width=0.45\linewidth]{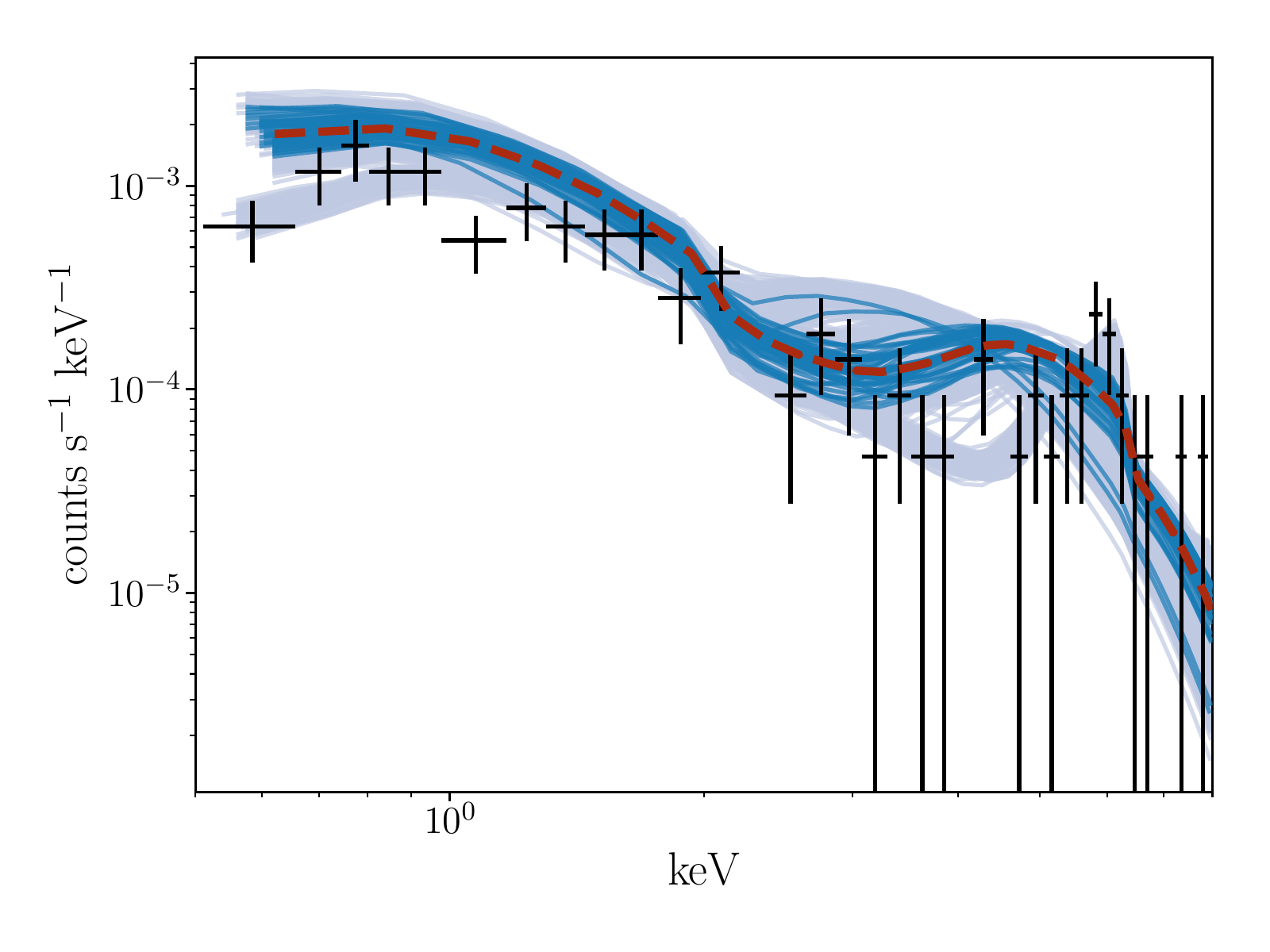}{\hskip 0pt plus 0.3fil minus 0pt}
    \includegraphics[width=0.45\linewidth]{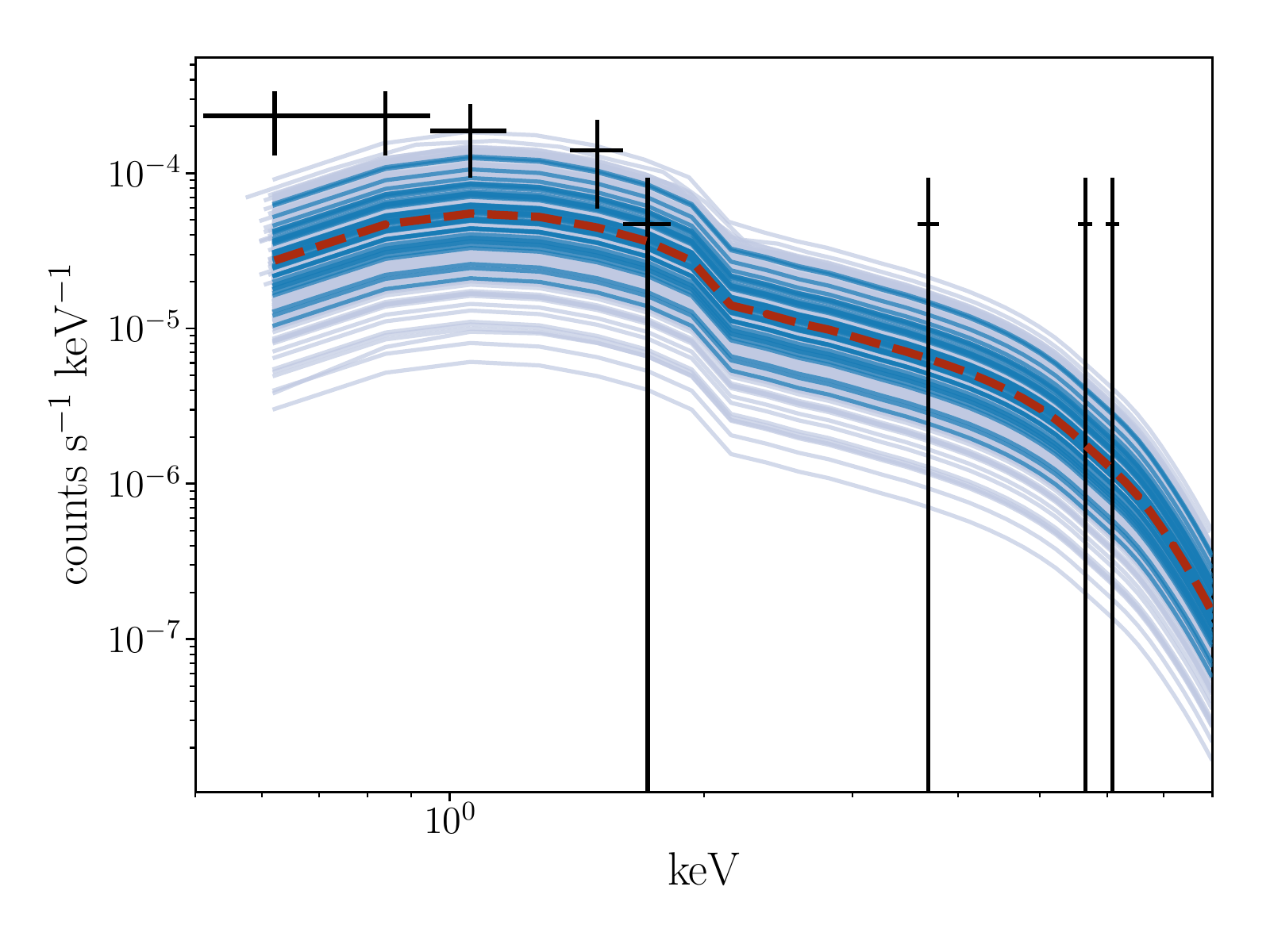}{\vskip 0.09cm plus 1fill}
 \caption{\emph{Chandra} spectral fits for 1000 realizations for J1356+1026 (left: primary point source, where the median number of counts is 177; right: secondary point source, where the median number of counts is 20), where the data have been folded through the instrument response. We overplot one of the spectral realizations with black points and plot the median spectral fit in a red dashed line}. We randomly select 50 of the 1000 spectral fits and plot them in dark blue to better highlight the density distribution of the lines. The spectra have been rebinned for plotting purposes. We fit J1356+1026$_{p}$ with the model {\tt phabs}$\times$({\tt pow} + {\tt phabs}$\times${\tt zphabs}$\times${\tt pow}), while we fit J1356+1026$_{s}$ with the model {\tt phabs}$\times${\tt zphabs}$\times${\tt pow}.  For J1356+1026$_{p}$, $\Gamma$ is allowed to vary, while for J1356+1026$_{s}$ we fix $\Gamma$ to a value of 1.8. We investigate whether the emission of the secondary is consistent with the emission of the diffuse background component by allowing $\Gamma$ vary. While $L_{\mathrm{2-7,~keV}}$ remains $>10^{40}$ erg s$^{-1}$, we can not differentiate this spectrum, at a statistical confidence level, from the diffuse emission component. We list the best-fit values for each model in Table~\ref{tab:spectra}, defined as the median of distribution of the best-fit values from the 1000 realizations.
\label{fig:J1356spec}
\end{figure*}

\subsection{SDSS J0752+2736 and SDSS J1448+1825: A High Probability of Contamination from XRBs}
\subsubsection{SDSS J0752+2736}
In the original analysis of SDSS J0752+2736, neither a primary or secondary point source were found to be statistically significant at the locations of each [\ion{O}{3}] $\lambda$5007 emission component \citep{Comerford2015}.  Our analysis with informative priors does not refute this conclusion, as the Bayes factor disfavors the dual point source model.  However, the dual point source model becomes favored when using non-informative priors, with $\ln{BF}=4.90\pm0.51$.  
\par Running \BAYMAX{} using non-informative priors, we analyze the posterior distributions for the locations of the primary and secondary ($\mu$), the count ratio ($f$), and the background fraction ($f_{BG}$). The best-fit position of each point source is shown in Figure~\ref{fig:PyMC3_1}, where the secondary appears to align with the position angle of the galaxy (see Fig.~\ref{fig:GalaxyImages}).  We also show the joint posterior distribution for the separation between the two point sources, $r$, and the logarithm of the count ratio, ($\log{f}$).  There are no $HST$ data for this system, and thus it is not possible to resolve potential optical cores or smaller galactic disturbances on the same scale as the estimated separation ($<2\arcsec$). We find the best-fit values for $r$ and $\log{f}$ to be $1.5\arcsec\pm0.30\arcsec$ and $-0.47\pm0.36$, respectively. At the 95\% C.L., the separation between the two point sources is greater than 0\farcs{5}.
\par We run our spectral analysis on SDSS J0752+2736, and find that, on average, the primary and secondary have 15 and 6 counts, respectively. We fit both the primary and secondary point source with a simple absorbed power-law ({\tt phabs}$\times${\tt zphabs}$\times${\tt pow}), with $\Gamma$ fixed to a value of 1.8.  For the primary, we calculate a total observed $0.5$--$8$ keV flux of ($2.74\pm0.59) \times 10^{-15}$ erg s$^{-1}$ cm$^{-2}$, while for the secondary we calculate a total observed $0.5$--$8$ keV flux of ($4.60\pm1.00) \times 10^{-16}$ erg s$^{-1}$ cm$^{-2}$ s$^{-1}$.  This corresponds to a rest-frame $2$--$7$ keV luminosity of ($1.67\pm0.36) \times 10^{40}$ erg s$^{-1}$ and ($7.1\pm1.60) \times 10^{39}$ erg s$^{-1}$ at $z=0.069$. Since we have fixed both point sources to have the same spectral shape, the count ratio that we calculate with \BAYMAX{}, should represent the flux ratio between the two sources. We find that the flux ratio we calculate via {\tt XSPEC} ($\approx$0.43) is consistent within the 68\% error interval of the $\log{f}$ posterior (where the median value is $\approx$0.34).
\par Although the primary point source X-ray luminosity meets our $L_{2-7~\mathrm{keV,~unabs}}$ criterion, the secondary does not. With an X-ray luminosity below 10$^{40}$ erg s$^{-1}$, we can not rule out contamination from possible XRBs or ULXs. Generally, the X-ray to optical flux ratio of most ULXs is very high \citep{Tao2011}, which is similar to low mass X-ray binaries and suggests that the optical emission arises from an accretion flow. This is consistent with the observation, as the secondary's position does not coincide with the measured bright [\ion{O}{3}] $\lambda$5007 emission component, and has no obvious optical counterpart in the SDSS image.
\par Given the point-like emission, spatial position, and $L_{2-7~\mathrm{keV,~unabs}}$ value, the emission of the secondary point source in SDSS J0752+2736 is consistent with what is expected from a ULX.  Thus, while the X-ray data are strongly indicative of a secondary point source, we can not conclude with a high certainty that SDSS J0752+2736 is a dual AGN system.

\subsubsection{SDSS J1448+1825}
\par In the original analysis of SDSS J1448+1825, neither a primary or secondary point source were found to be statistically significant at the locations of each [\ion{O}{3}] $\lambda$5007 emission components \citep{Comerford2015}.  However, we find that the dual point source model is favored when using both non-informative and informative priors, with a $\ln{BF}=1.43\pm0.55$ and $\ln BF=2.95\pm0.52$.  When using informative priors, we find the best-fit values for $r$ and $\log{f}$ to be $1.29\arcsec\pm52\arcsec
$ and $-0.45\pm0.80$, respectively.  However, at the 95\% C.L., the separation between the two point sources is $<0\farcs{5}$. 
\par We run our spectral analysis on SDSS J1448+1825, and find that the primary and secondary have, on average, 14 and 3 counts. We fit the primary AGN with $m_{spec, 2}$, fixing $\Gamma$ to a value of 1.8.  Here, we find that $\Delta C_{\mathrm{stat}}=5.7$, such that the more complicated spectral model is a statistically better fit.  For the primary, we calculate a total observed $0.5$--$8$ keV flux of ($12.80\pm5.20) \times 10^{-15}$ erg s$^{-1}$ cm$^{-2}$, while for the secondary we calculate a total observed $0.5$--$8$ keV flux of ($5.60\pm4.70) \times 10^{-16}$ erg s$^{-1}$ cm$^{-2}$ s$^{-1}$.  This corresponds to a rest-frame $2$--$7$ keV luminosity of ($17.50\pm10.00) \times 10^{40}$ erg s$^{-1}$ and ($1.10\pm0.97) \times 10^{39}$ erg s$^{-1}$ at $z=0.038$. 
\par Although \BAYMAX{} favors the dual point source model for J1448+1825, and the secondary's position is consistent with the secondary [\ion{O}{3}] $\lambda$5007 emission component, the rest-frame unabsorbed X-ray luminosity is below our criterion and the X-ray spectrum of the secondary is very soft ($HR \approx -1$). Based on our MC spectral analysis, there is $>$50\% chance that the count rate above 2 keV is 0.  Given the average X-ray count-rate ($\approx$1$\times$10$^{-4}$ cps) and assumed spectral shape ($\Gamma=1.8$), this is consistent with what is expected from a possible low-luminosity AGN ($\approx$1 count between $2$--$8$ keV), however we conservatively do not classify SDSS J1448+1825 as a dual AGN system.  Deeper observations of SDSS J1448+1825 will allow for better constraints on the spectral shape.
\subsection{SDSS J1126+2944: A dual AGN system with an ultra-compact dwarf galaxy candidate}
SDSS J1126+2944 was the only confirmed dual AGN candidate found in the analysis of \cite{Comerford2015}.  Specifically, both the primary and secondary X-ray point sources were found to be statistically significant at the locations of each [\ion{O}{3}] $\lambda$5007 emission component at a $5\sigma$ and $2.3\sigma$ confidence level, respectively.  Our results agree with these conclusions, as we find $\ln{BF} = 3.54\pm0.43$ when using informative priors based on the positions of the [\ion{O}{3}] $\lambda$5007 emission components. 
\par We analyze the posterior distributions for $\mu$, $f$, and $f_{BG}$ when \BAYMAX{} is run using informative priors. We find the best-fit values for $r$ and $\log{f}$ to be $1.74\arcsec\pm0.33$ and $-1.00\pm0.44$. At the 95\% C.L., the separation between the two point sources is greater than 1\arcsec.  The best-fit locations of the primary and secondary are shown in Figure~\ref{fig:PyMC3_2}.  
\par We run our spectral analysis on SDSS J1126+2944, and find that primary and secondary, on average, have 25 and 3 counts.  We fit the primary AGN with $m_{\mathrm{spec,2}}$, and fix the photon index of the power-law, $\Gamma$, to a value to 1.8.  On average we find that $\Delta C_{\mathrm{stat}}=39.6$, such that the more complicated spectral model is a statistically better fit.  For the primary, we calculate a total observed $0.5$--$8$ keV flux of ($3.35\pm0.29) \times 10^{-14}$ erg s$^{-1}$ cm$^{-2}$, while for the secondary we calculate a total observed $0.5$--$8$ keV flux of ($4.64\pm2.00) \times 10^{-15}$ erg s$^{-1}$ cm$^{-2}$ s$^{-1}$.  This corresponds to a rest-frame $2$--$7$ keV luminosity of ($2.84\pm0.55) \times 10^{42}$ erg s$^{-1}$ and ($2.90\pm1.58) \times 10^{41}$ erg s$^{-1}$ at $z=0.102$. 
\par We confirm that the location of the secondary coincides spatially with a faint point-like source discovered in the $HST$ F160W, F814W, and F438W images (hereafter  SDSS J1126+2944 SE ; \citealt{Comerford2015}). The merger ratio of the main host galaxy to SDSS J1126+2944 SE was found to be $\approx$ 460:1, and thus SDSS J1126+2944 SE was classified as a potential ultra-compact dwarf galaxy \citep{Comerford2015}.  Indeed, the estimated upper-limit of the half-light radius of 280 pc agrees with other ultra-compact dwarf galaxies that host a supermassive black hole (e.g., M60-UCD1; see \citealt{Seth2014}).  Specifically, M60-UCD1 is thought to be the remnant of a galaxy that was once more massive, but underwent tidal stripping via an encounter with the galaxy M60.  Due to signs that SDSS J1126+2944 underwent some kind of tidal disruption itself in the $HST$ images, this is a possible scenario.
\par As an additional step in the analysis of the true nature of the secondary point source, we compare the hard X-ray luminosity to the total, expected X-ray luminosity due to XRBs. Following a similar analysis to \cite{Foord2017}, we adopt an updated analytic prescription by \cite{Lehmer2019}, to estimate the total, $2$--$7$ keV luminosity expected from XRBs ($L^{\mathrm{gal}}_{\mathrm{XRB}}$). In particular, for a given stellar mass ($M_{\ast}$, which scales with the LMXB population; \citealt{Gilfanov2004}) and star formation rate (SFR, which scales with the HMXB population; \citealt{Mineo2012}) the total, $2$--$7$ keV luminosity from XRBs can be estimated \citep{Lehmer2019}. We use the values for $M_{\ast}$ and the SFR as estimated in \cite{Barrows2017b}, where they fit galaxy and AGN templates to the broadband photometric SEDs of a sample of dual AGN candidates, including SDSS J1126+2944 and SDSS J1356+1026. Since the values for $M_{\ast}$ and the SFR have only been measured for the primary galaxies, we are assuming that the primary and secondary AGNs are in galaxies with similar SFRs and stellar masses (and for SDSS J1126+2944 this is a conservative assumption, given the large mass ratio estimated by \citealt{Comerford2015}). We estimate a total $2$--$7$ keV luminosity expected from XRBs $L^{\mathrm{gal}}_{\mathrm{XRB}}=1.8^{+6.56}_{-1.58}\times10^{40}$~erg s$^{-1}$, over a factor of 10 less than the measured X-ray luminosity of the secondary point source.

\subsection{SDSS J1356+1026: A candidate dual AGN system among warm photoionized gas}
\label{J1356+2016}
Although SDSS J1356+1026 was originally found to have both a primary and secondary point source at a statistically significant confidence level ($5\sigma$ and $4.4\sigma$, respectively), it was conservatively categorized as a single AGN \citep{Comerford2015}, as the soft X-rays associated with an outflowing bubble \citep{Greene2012, Greene2014} complicated the identification of a possible dual AGN. We take into account possible contamination from diffuse gas associated with photoionization by including an additional background component to our model (see Section~\ref{sec:results}). Further, our model for the additional background component is uniform over energy-space (i.e., a 2 keV photon is just as likely as a 7 keV photon), and thus conservative, as we expect most of the diffuse emission, regardless of its physical origin, to be $<$3 keV. We find that (i) the results from \BAYMAX{} favor the model that includes the additional background component, for both the single and dual point source models, and (ii) the results from \BAYMAX{} remain in favor of the dual point source model, even when accounting for the diffuse emission. 
\par However, if the extended emission in SDSS J1356+1026 is a result of extreme photoionization via feedback of the primary AGN \citep{Greene2012, Greene2014}, there is a possibility that the secondary X-ray point-source is instead associated with a luminous [\ion{O}{3}] gas clump.  Thus, analyzing our best-fit parameters for the location, as well as carrying out an X-ray spectral analysis, are imperative for a better understanding of the most likely origin of emission.
\par We find the best-fit values are $r=2.00\arcsec\pm0.62\arcsec$ and $\log{f}=-0.92\pm0.23$. We note that the posterior distribution for $r$ has a slight bimodality, due to a bimodality in the $x$, $y$ position of the secondary X-ray point-source (see Fig.~\ref{fig:PyMC3_2}). This is likely due to the diffuse emission component contributing a large fraction of counts (69$\%$ of the counts emitted by both point sources, or $\approx$75\% of the counts emitted by the secondary). However, at the 95\% C.L., the separation between the two point sources remains $>0\farcs{5}$, and the location of the secondary is consistent, within the errors, of the merging galaxy's optical nucleus.
\par Running our spectral analysis, we find that the primary and secondary have, on average, 177 and 20 counts.  Here $\Delta C_{\mathrm{stat}}\approx8$, such that $m_{\mathrm{spec,2}}$ is favored for the primary point source at a significant level.  For the primary, we calculate a total observed $0.5$--$8$ keV flux of ($2.38\pm0.16) \times 10^{-14}$ erg s$^{-1}$ cm$^{-2}$, while for the secondary we calculate a total observed $0.5$--$8$ keV flux of ($9.00\pm4.10) \times 10^{-16}$ erg s$^{-1}$ cm$^{-2}$ s$^{-1}$.  This corresponds to a rest-frame $2$--$7$ keV luminosity of ($5.60\pm2.00) \times 10^{43}$ erg s$^{-1}$ and ($1.80\pm0.80) \times 10^{40}$ erg s$^{-1}$ at $z=0.123$. In Figure~\ref{fig:J1356spec}, we show the spectral fits for the 1000 realizations of both the primary and secondary point source.
\par Although we find that the position and luminosity of the secondary point source are consistent with what is expected from a central SMBH, there still exists the possibility of contamination from an [\ion{O}{3}] gas clump. Thus, we compare the spectrum of the counts associated with secondary point source to that of the counts associated with the diffuse emission.  When $\Gamma$ is allowed to vary, with unconstrained values, we find that the spectrum of the secondary point-source ($\Gamma \approx 3.0\pm0.64$) is consistent with spectrum of the diffuse emission ($\Gamma \approx 3.2\pm0.25$). We note that with the softer spectral fit, the total unabsorbed 2$-$7 keV luminosity of the secondary point-source still meets our AGN luminosity criterion ($L_\mathrm{2-7,~unabs} = 1.1\pm0.60\times$10$^{40}$ erg s$^{-1}$).  However, because we can not differentiate between the soft spectra of the secondary point source and diffuse emission at a statistical confidence level, we conservatively do not classify SDSS J1356+1026 as a dual AGN system. 
\par Similar to SDSS J1126+2944, we compare the hard X-ray luminosity of the secondary point source to the total, expected X-ray luminosity due to XRBs. Using $M_{\ast}$ and SFR values from \cite{Barrows2017b}, we find $L^{\mathrm{gal}}_{\mathrm{XRB}} = 8.92^{+3.37}_{-2.29}$, such that the total, expected X-ray luminosity for XRBs is greater than the measured X-ray luminosity measured for the secondary point source. Of course, this is not a perfect comparison, as we do not expect that all of the X-ray luminosity from the XRB population is contained within a compact 2\arcsec radius centered on the location of the secondary point source. However, it further exemplifies the complexities when attempting to classify the true nature of the secondary in SDSS J1356+1026.
%
%%%%%%%%%%%%%%%%%%%%%%%%%%%%%%%%%%%%%
%%%%%%%%%%%%% DISCUSSION %%%%%%%%%%%%%%%
%%%%%%%%%%%%%%%%%%%%%%%%%%%%%%%%%%%%%
\section{Discussion} 
\label{sec:discussion}
Using \BAYMAX{} on 12 dual AGN candidates, that were identified via [\ion{O}{3}] $\lambda$5007 emission, we have found that 4/12 have a $BF$ that favor the dual point source model, 2/12 have secondary point sources with X-ray luminosities consistent with an AGN, and 1/12 is likely true dual AGN system. Both SDSS J1126+2944 and SDSS J1356+1026 have strong $BF$ values in favor of a dual point source and have primary and secondary X-ray point sources with X-ray luminosities consistent with emission from AGN. However, due to the extreme feedback associated with SDSS J1356+1026$_{p}$ (seen in both \emph{Chandra} and \emph{HST} observations) there is a probability that the X-ray emission of SDSS J1356+1026$_{s}$ is instead due to a luminous [\ion{O}{3}] gas clump. Because we can not differentiate the spectrum of SDSS J1356+1026$_{s}$ from the background emission, we conservatively do not only classify SDSS J1356+1026 as a dual AGN system. SDSS J0752+2736 and SDSS J1448+1825 have $BF$ values that favor the dual point source model, however because of the large probability of contamination from XRBs, we additionally do not categorize them as dual AGNs. The remaining 8 galaxies (SDSS J0142$-$0050, SDSS J0841+0101, SDSS J0854+5026, SDSS J0952+2552, SDSS J1006+4647, SDSS J1239+5314, SDSS J1322+2631, and SDSS J1604+5009) have $BF$ values that do not favor the dual point source model.  In the following section, we aim to better understand \BAYMAX{}'s sensitivity across parameter space, as well as characterize all 12 galaxies via a multi-wavelength analysis.
\subsection{The Sensitivity of \BAYMAX{} Across Count Ratio Space}
We first discuss the significance of our results by analyzing \BAYMAX{}'s capabilities across a range of count ratio space for the dual point source model. In particular, we aim to understand where in parameter space \BAYMAX{} loses sensitivity for simulations with a comparable number of counts as the 8 systems in which the $BF$ value favored the single point source model.  This is done by running \BAYMAX{} on a {\tt MARX}-generated suite of simulated dual AGN systems that closely match the observed data and expected dual configurations.  The simulations have the same total number of counts between $0.5$--$8$ keV as each observation, with a primary and secondary AGN located at the spatial locations of the measured [\ion{O}{3}] $\lambda$5007 emission components. Further, each simulated AGN has the same $0.5$--$8$ keV spectrum as the observation, but with normalizations proportional to their count ratio.  We also add a spatially uniform background component to the simulations, where $f_{BG}$ is determined from the best-fits returned by \BAYMAX{}. For J0841+0101, we add an additional synthetic higher-count background to represent the diffuse emission component, which is constrained within the region shown in Fig.~\ref{fig:GalaxyImages}.
\par We simulate dual AGN systems with count ratios that range between $0.1$--$1.0$, with the exception of the highest-count observations (SDSS J0142$-$0050 and SDSS J1239+5314), where we can probe lower count ratios ($0.03$--$1.0$).  We analyze each simulation using the same informative priors as used for the real datasets.  For each $f$ value in parameter space, we analyze 100 simulations, and evaluate the mean $BF$ value.  We enforce a cut of $BF>3$, where only mean $BF$ values above this threshold are classified as strongly in favor of the dual point source model.
\par For SDSS J0142$-$0050 and SDSS J1239+5314 we find that \BAYMAX{} can correctly identify dual AGN systems with a strong $BF$ value for all count ratio values (down to $f=0.03$). This is not surprising, given that SDSS J0142$-$0050 and SDSS J1239+5314 have many counts ($>$ 600 counts between $0.5$--$8$ keV).  At the lower-end of the count ratios probed, the secondary is, on average, contributing $\geq$20 counts.  
\par For SDSS J1322+2631 and SDSS J0841+0101, \BAYMAX{} is able to correctly identify the systems as a dual point source for the entire count-ratio range analyzed ($f=0.1$--$1.0$). These results are not surprising, given that SDSS J1322+2631 has a large estimated separation between the [\ion{O}{3}] $\lambda$5007 emission components ($>2\arcsec$) and SDSS J0841+0101 has over 400 counts between 0.5--0.8 keV. For SDSS J0952+2552 and SDSS J1604+5009, \BAYMAX{} is able to correctly identify systems as a dual point source for $f=0.2$--$1.0$. Although these two systems have a comparable number of counts to SDSS J1322+2631 ($\approx$50), the projected separations between the [\ion{O}{3}] $\lambda$5007 emission components are smaller ($\approx$1\arcsec). We find that the strength of the $BF$ in favor of the dual point source model increases as a function of the count ratio in the simulations, where the $BF > 10^{2}$ for systems with $f\ge0.3$.
\par Regarding SDSS J0854+5026 and SDSS J1006+4647, given the small number of counts ($\approx$ 13 total counts between $0.5$--$8$ keV) as well as smaller estimated separations between the [\ion{O}{3}] $\lambda$5007 emission components ($<1\arcsec$), \BAYMAX{} is unable to favor the correct model, on average, for the entire range of $f$-values probed.  
\par Using the count ratio thresholds determined by \BAYMAX{}, we estimate the $2$-$7$ keV luminosities of possible secondary point sources that we are sensitive to. Assuming a power-law spectral shape with $\Gamma=1.8$, we find that \BAYMAX{} is capable of detecting secondary point sources with $L_{\mathrm{2-7~keV}} \ge 4\times 10^{40}$ at the lower-luminosity end (SDSS J0841+0101, where $z=0.096$) and $L_{\mathrm{2-7~keV}} \ge 6 \times 10^{41}$ at the higher-luminosity end (SDSS J0952+2552, where $z=0.339$).  More data on each of these sources, especially SDSS J0854+5026 and SDSS J1006+4647, will be necessary in order for a more thorough analysis of their true nature.

%%%%%%%%%%%%%%%%%%%%%%%%%%%%%%%%%%%%%%%%%%%%%%%%%%%%%
%%%%%%%%%%%%%%%%%%%% WISE FIGURE %%%%%%%%%%%%%%%%%%%%
%%%%%%%%%%%%%%%%%%%%%%%%%%%%%%%%%%%%%%%%%%%%%%%%%%%%%
\begin{figure}
\centering
\includegraphics[width=\linewidth]{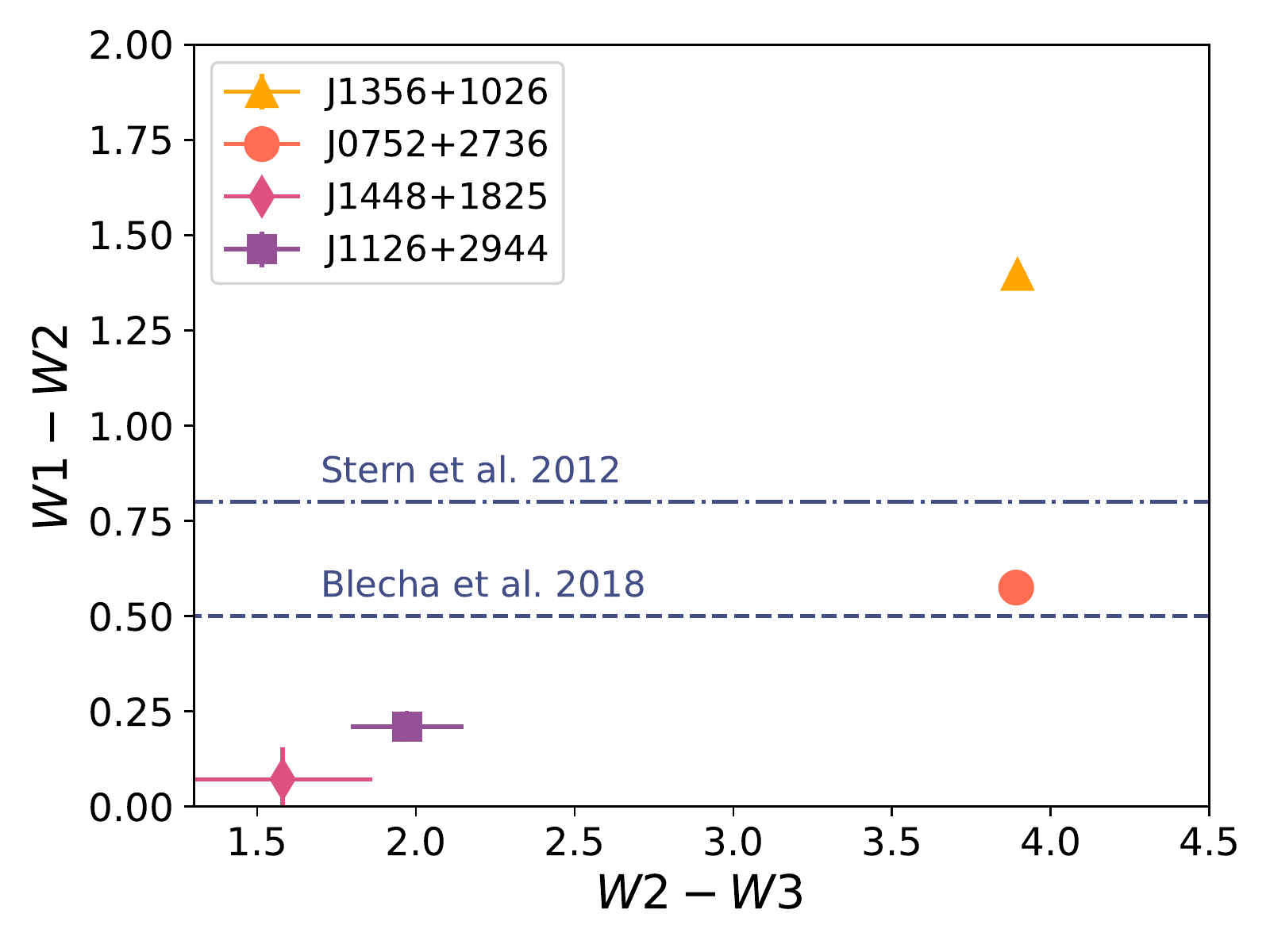}
\caption{$W1-W2$ vs.~$W2-W4$ color-color diagram for the four sources in the sample that have $BF$ values that favor the dual point source model.  We show various cuts, above which the majority of luminous AGNs \citep{Stern2012} and dual AGNs \citep{Blecha2013} should sit.  We find that one of the sources, SDSS J1356+1026, has an AGN-dominated infrared flux. This is not surprising, given the overall lower X-ray luminosities of these systems.}
\label{fig:WISEcolors}
\end{figure}

\subsection{Infrared Observations}
We re-plot the mid-infrared colors from the \emph{Wide-field Infrared Survey Explorer} (\emph{WISE}) for the subsample of 4 systems with $BF$ in favor of the dual point source model (which were previously examined for all 12 systems in \citealt{Comerford2015}). Here we incorporate recent results from simulations of merging galaxies \citep{Blecha2015}, where specific IR color-cuts in the \emph{WISE} bands were found to select both merger-triggered AGN and dual AGNs.
\par In general, IR colors are often used as a tool to identify AGNs \citep{Jarrett2011, Stern2012}, as mid-IR-selected AGNs are much less sensitive to attenuation by gas and dust than AGN selected in optical or soft X-ray bands.  The standard single-axis color-cut (above which, the source is likely an AGN) is $W1$-$W2$ $> 0.8$ \citep{Stern2012}, but multiple-axis cuts additionally using the $W3$ and $W4$ bands are used as well (see, e.g., \citealt{Jarrett2011}). However, such diagnostics are sensitive to only the most luminous AGNs, that are contributing a considerable fraction of the total bolometric luminosity \citep{Mateos2013}. At lower luminosities, the selection is largely incomplete and strongly biased against AGN residing in massive and/or star-forming hosts.
\par Similar conclusions have been reached for more recent studies looking at the mid-IR colors of merger-triggered AGN \citep{Blecha2015}, even in the late stages of gas-rich major mergers. More interestingly, however, \cite{Blecha2015} find that a less stringent single-color cut not only selects merger-triggered AGN with a much higher completeness, but selects virtually all bright dual AGNs (where each AGN have $L_{\mathrm{bol}} > 10^{44}$) throughout the merger.
\par Thus, we plot the $W1$--$W2$ vs.~$W2$--$W3$ colors of the four systems with $BF$ values in favor of the dual point source model in order to see if they lie in an interesting region of IR color-color space. We confirm the finding of \cite{Comerford2015} that only one of the four systems, SDSS J1356+1026, has an AGN-dominated mid-infrared flux. This is not surprising, given the overall lower luminosities of the sources (each point source has $L_{2-7~\mathrm{keV, unabs}} <10^{42}$ erg s$^{-1}$, besides SDSS J1356+1026).    Additionally, both SDSS J1356+1026 and SDSS J0752+2736 lie above the less stringent single-color cut found in \cite{Blecha2015}.  In the future, confirmation of dual-AGNs via IR colors may be achieved with AO imaging in the near-/mid-IR bands, where the primary and secondary X-ray point sources can be analyzed individually. 
\subsection{Optical Narrow-line Ratio Diagnostics}
\par We compare the classification of the central ionizing source via available optical spectroscopic data to the conclusions reached by our X-ray analysis.  In particular, we analyze how the [\ion{O}{3}]$/$H$\beta$ and [\ion{N}{2}]$/$H$\alpha$ ratios compare to the line ratio BPT diagram (``Baldwin, Phillips, \& Terlevich"; see \citealt{Baldwin1981, Kewley2006}) for the four systems with $BF$ values in favor of the dual point source model. These line ratio diagnostics can be used to classify the dominant energy source in emission-line galaxies.  
\par With available long-slit spectroscopic data, we are capable of extracting BPT diagnostics individually for the primary and secondary X-ray point-sources.  However, because the original long-slit spectroscopic analysis was designed to target the [\ion{O}{3}]$\lambda$5007 emission, we do not have information regarding the [N~II] or H$\alpha$ emission.  Thus, we compare these data-points to $\log{\mathrm{[O~III]/H\beta}}=1$, above which the line ratios are consistent with emission from an AGN, at all reasonable $\log{\mathrm{[N~II]/H\alpha}}$ values. Such an analysis will allow us to better understand the true nature of the 4 systems with a $BF$ that favor the dual point source model. In particular, because SDSS J1356+1026 is more complicated to classify, we are interested in whether this additional optical analysis classifies the secondary point source as an AGN.
\par
\begin{figure}
\centering
\includegraphics[width=0.5\textwidth]{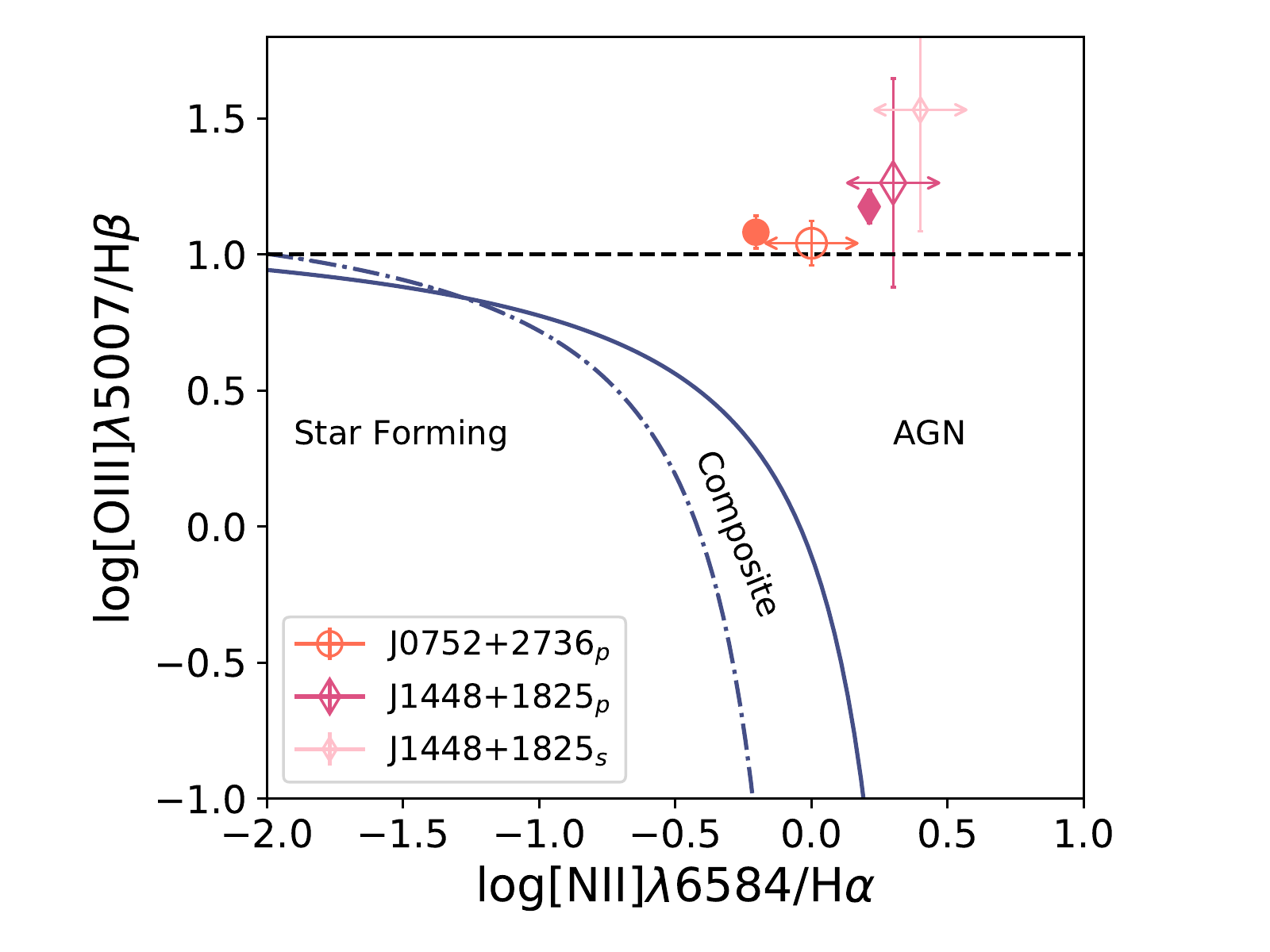}
\includegraphics[width=0.5\textwidth]{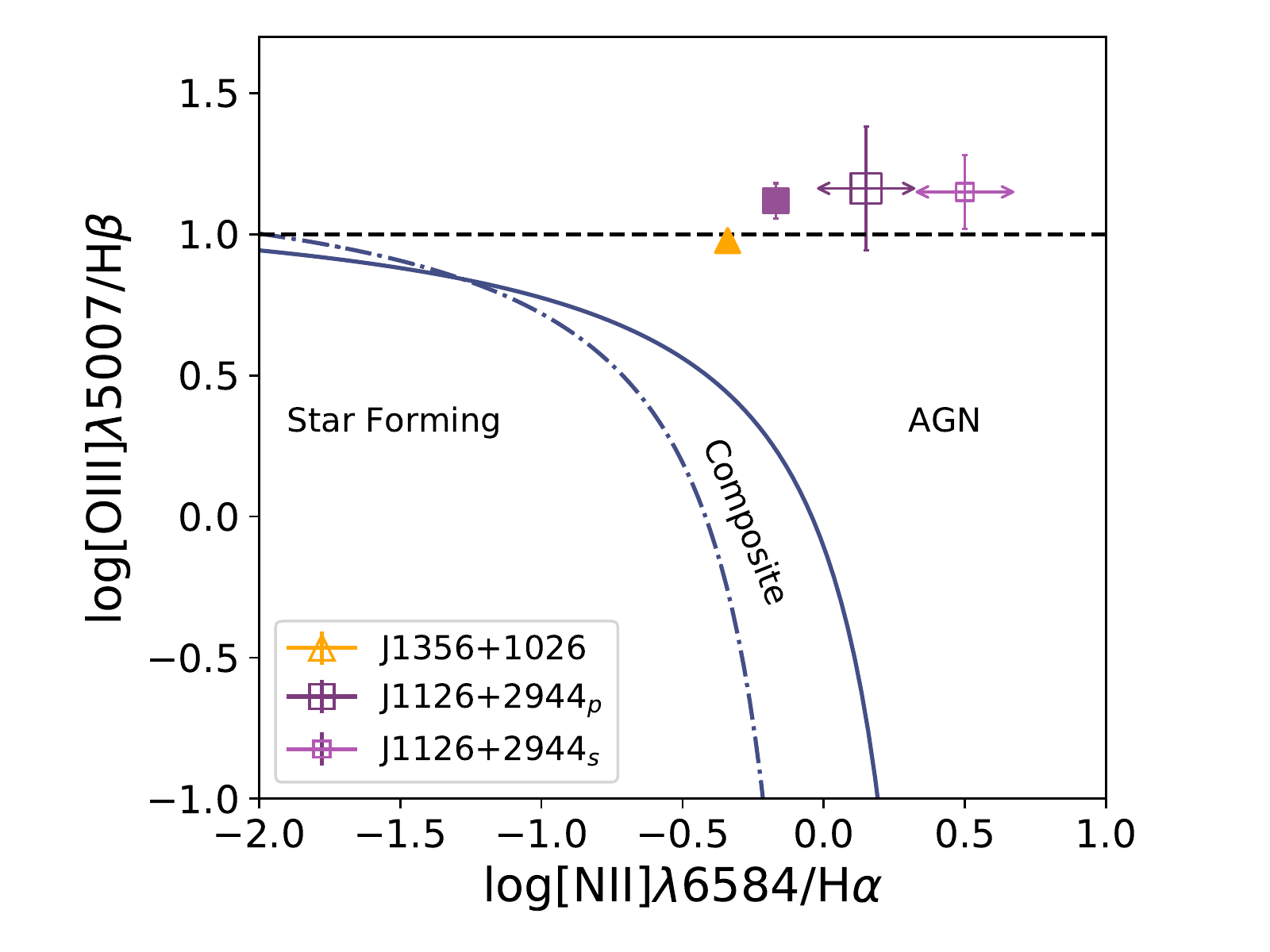}
\caption{BPT optical spectroscopic line ratio diagrams, based on the [\ion{O}{3}]$/$H$\beta$ to [N~II]/H$\alpha$ emission line ratio. The blue lines represent the \cite{Kewley2001} (solid) and \cite{Kauffmann2003} (dot-dashed) demarcations, which separate different sources of photoionizaiton. We plot the line ratios for SDSS J075+2736 and SDSS J1448+1825 in the top panel and those for SDSS J1356+1026 and SDSS J1126+2944 in the bottom panel. We show the average [\ion{O}{3}]$/$H$\beta$ line ratio values for the long-slit data with open markers, where we note that values $\log{\mathrm{[O~III]/H\beta}}>1$ (black dashed line) are consistent with AGN photoionization, at all reasonable $\log{\mathrm{[N~II]/H\alpha}}$ values.  Additionally, we show the [\ion{O}{3}]$/$H$\beta$ to [\ion{N}{2}]/H$\alpha$ ratios for each system using available Sloan spectra (filled markers).  For each marker we include 1$\sigma$ error bars.  For SDSS J0752+2736$_{p}$ and SDSS J1356+1026, we find that the line ratios of each system are consistent with AGN photoionization. For SDSS J1448+1825, we find that the line ratios of the primary and secondary X-ray point sources are consistent with AGN photoionization; although the X-ray luminosity of this source is below our AGN luminosity criterion, it's possible that the X-ray emission of the secondary point source is highly obscured.  For SDSS J1126+2944, we find that the line ratios of each point source are consistent with AGN photoionization, in agreement with our X-ray analysis. Because we have no \ion{N}{2}$/$H$\alpha$ for the long-slit data, we choose x-axis coordinates near the respective SDSS measurements.}
\label{fig:BPTdiagrams}
\end{figure}

 We use available long-slit optical spectroscopic data for SDSS J0752+2736, SDSS J1126+2944, and SDSS J1448+1285 \citep{Comerford2012}; while we use archival Sloan spectral data for SDSS J1356+1026. SDSS J0752+2736 and SDSS J1126+2944 were observed with the Blue Channel Spectrograph on the MMT 6.5 m telescope ($0\farcs29$/pixel, \citealt{Schmidt1989}), while SDSS J1448+1285 was observed with the Kast Spectrograph on the Lick 3 m telescope ($0\farcs78$/pixel).  In Figure~\ref{fig:BPTdiagrams} we plot pairs of line ratios for each system.  Each long-slit observation used a 1200 lines mm$^{-1}$ grating and was centered so that the wavelength range covered H$\beta$ and [\ion{O}{3}], given the various redshifts of each system.  Due to the larger diameter of SDSS optical fibers (3\arcsec), the line ratio values calculated for SDSS J1356+1026 represent the line ratios for the primary and secondary AGN combined. In general, however, we find that the line ratios estimated with Sloan spectra are consistent with those estimated from long-slit spectra (see Fig.~\ref{fig:BPTdiagrams}).  
\par Each system with long-slit spectroscopic data (MMT and Lick) was observed twice, with the slit oriented along two different position angles on the sky.  Here, the line ratios were estimated by collapsing the spectrum along the spatial direction (to increase the S/N), fitting the spectra with either one or two Gaussians, and averaging the line ratios between the two position angles.  For each system we identify whether the primary or secondary X-ray point source is spatially coincident with the red- or blue-shifted emission components using the spatial information provided by the long-slit observations.
\par Given that the locations of our informative priors for the primary and secondary X-ray point sources are based on the spatially resolved positions of red- and blue-shifted components of the [\ion{O}{3}] long-slit observations (as determined in \citealt{Comerford2012}), we assume that each red- and blue-shifted component of a given spectrum represent emission from the primary and secondary X-ray point source (with the exception of SDSS J0752+2736, where no X-ray point source was found at a position consistent with a peak in the [\ion{O}{3}] emission). 
\par Another limitation of the original analysis targetting the [\ion{O}{3}]$\lambda$5007 emission is that the H$\beta$ emission is generally quite faint relative to the [\ion{O}{3}]$\lambda$5007 emission lines.  For SDSS J0752+2736 we are unable to fit a Gaussian to the red-shifted component of the H$\beta$ emission line (which corresponds to the secondary X-ray point source) with any statistical confidence ($>$1$\sigma$). For SDSS J1448+1285 we are able to fit Gaussians to both the red- and blue-shifted H$\beta$ components, however we note that the estimated H$\beta$ flux values are statistically significant at $<3\sigma$, and should be interpreted with skepticism. For SDSS J1126+2944, we are able to cleanly decompose the two X-ray point source components in H$\beta$ velocity space (where the H$\beta$ flux values are statistically significant at $>3\sigma$); thus, we measure individual line ratios for the primary and secondary with high statistical confidence.
\par For SDSS J1126+2944, we find that the line ratios of each X-ray point source are consistent with AGN photoionization, in agreement with our X-ray analysis.  Additionally, the line ratios of SDSS J1356+1026, and the primary X-ray point source in SDSS J0752+2736, are consistent with AGN photoionization. Lastly, both the primary and secondary X-ray point sources in SDSS J1448+1825 have [\ion{O}{3}]$/$H$\beta$ ratios consistent with AGN photoionization.  This is surprising, given that the X-ray luminosity of the secondary X-ray point source is below our AGN luminosity criterion ($L_\mathrm{2-7,~unabs}<10^{40}$ erg s$^{-1}$).  If SDSS J1448+1825 is indeed a dual AGN, the low X-ray luminosity of the secondary AGN may be a result of the merger environment (see, e.g., \citealt{Liu2013}).  Here, the X-ray emission is more susceptible to obscuration by the excess of gas at the galaxy center than the optical flux (which is emitted on larger physical scales than the X-ray flux).  Indeed, in our spectral analysis we find that SDSS J1448+1825 has one of the largest extragalactic column densities ($>50 \times 10^{22}$ cm$^{-2}$). These results confirm those in \cite{Comerford2015}, where it was found that dual AGNs have systematically lower X-ray luminosities, at a given [\ion{O}{3}]$\lambda$5007 luminosity, than single AGNs. 

We note that our measurements of \ion{O}{3}$/$H$\beta$ are susceptible to possible amplification by other effects found in mergers, such as star formation and shocks (e.g., \citealt{Rich2011, Kewley2013, Belfiore2016}). In order to best separate AGN from shock-excited gas, follow-up observations, especially with integral field spectroscopy, will be necessary \citep{DAgostino2019}.  Deeper follow-up observations will also allow for a better spectral decomposition of the two X-ray point sources in SDSS J0752+2736 and SDSS J1448+1825; however due to the extreme spatial extent of the [\ion{O}{3}]$\lambda$5007 emission in SDSS J1356+1026 \citep{Comerford2015}, cleanly decomposing the two X-ray point source components in \ion{O}{3} velocity space is most likely not feasible, even with additional observations.

\subsection{The Role of the Merger Environment}
There is reason to believe that galaxy--galaxy interactions can trigger accretion onto central AGNs.  In particular, models show that tidal forces between the galaxies can cause gas to be subject to substantial gravitational torques, resulting in substantial gas flow towards the central SMBHs \citep{Barnes&Hernquist1991, Mihos&Hernquist1996, DiMatteo2008,Hopkins&Hernquist2009, Angles-Alcazar2017}.  In this framework we may expect (i) the fraction of dual AGNs increases as the separation between the two AGN decreases, and (ii) dual AGNs may preferentially reside in gas-rich environments. Regarding (i), such a trend has been found in both simulations and observations. Simulations have been able to probe the smallest separations ($<10$ kpc; \citealt{Blecha2013, Ellison2013, Capelo2017}), while most observations have been constrained to the larger separations ($>10$ kpc, \citealt{Koss2012, Goulding2017}; however, \citealt{Barrows2017} probes separations $<$10 kpc and also find that the fraction of dual AGNs increases as a function of decreasing separation, including SDSS J1126+2944). Regarding (ii), numerical results from \citep{Steinborn2016} have found that dual AGNs are generally in more gas-rich systems; observationally, such a trend was found in \cite{Barrows2018}, where the mean $N_{H}$ value for a sample of dual AGNs was found to be an order of magnitude higher compared to a sample of single AGN.
\par
\begin{figure}
\centering
\includegraphics[width=\linewidth]{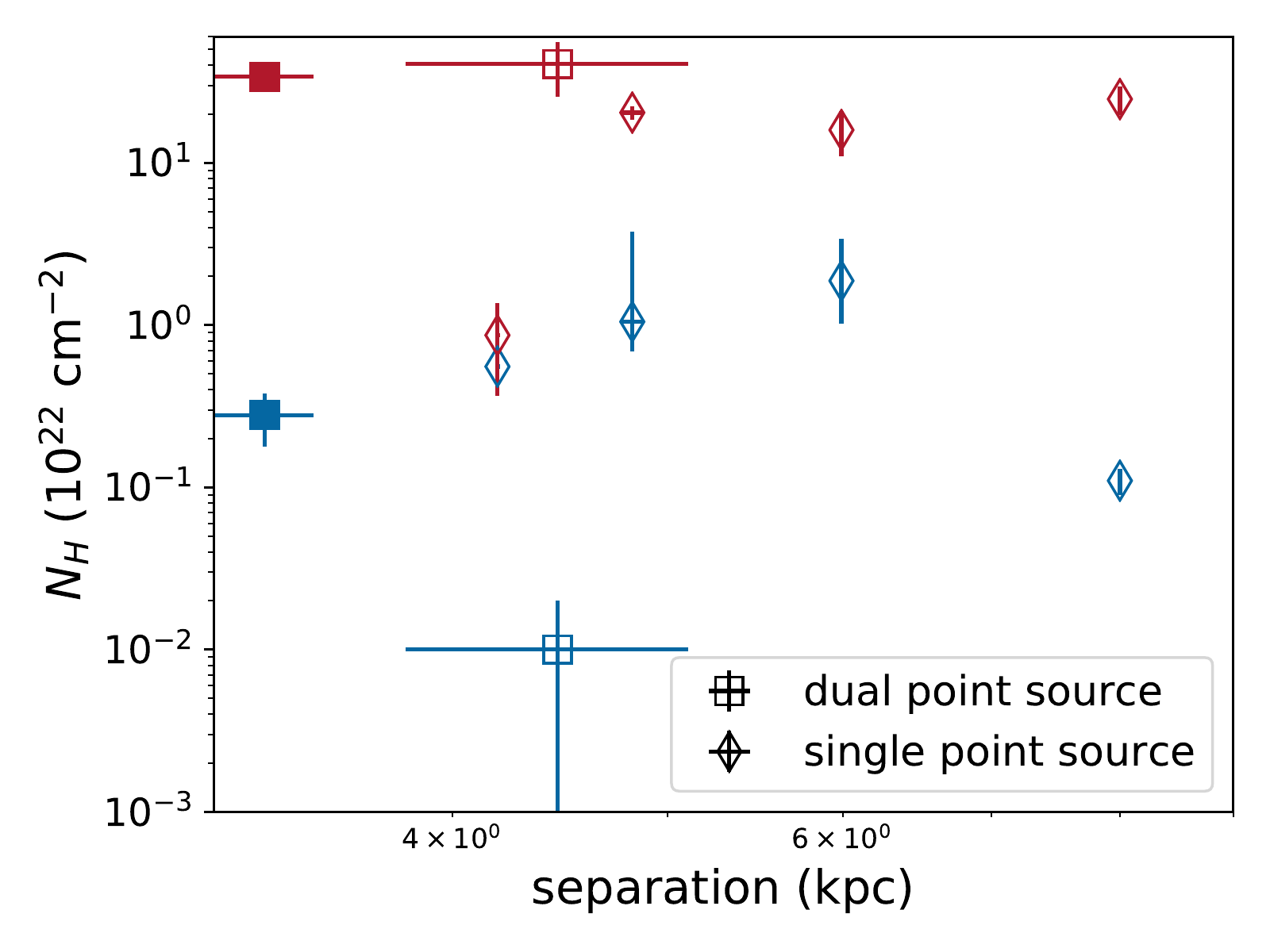}
\caption{Extragalactic column density ($N_{H}$, 10$^{22}$ cm$^{-2}$) vs. separation (kpc) of the six merging systems in our sample using $m_{\mathrm{spec,1}}$ (blue) and $m_{\mathrm{spec,2}}$ (red). We denote the two systems with $BF$ values in favor of the dual point source model with squares (SDSS J1126+2944 and SDSS J1356+1026), where the one confirmed dual AGN in our sample (SDSS J1126+2944) is filled-in. The four other systems (with $BF$ values that favor the single point source model) are denoted with diamonds.} Our data suggest that dual AGN activation may be more common for merging galaxies with smaller separations. Although SDSS J1126+2944 has one of the highest $N_{H}$ values in it's respective gas clump (i.e., $N_{H}$ as determined by $m_{\mathrm{spec,2}}$), we find that the average $N_{H}$ decreases as a function separation, at odds with predictions, and likely a result of selection bias. Given that only \emph{one} of the six merging galaxies are confirmed dual AGNs, future analyses with larger samples will be important to understanding the role of merger environments on SMBH activity.
\label{fig:nHvsR}
\end{figure}

 Taking the six merging galaxies in our sample, as determined visually from the $HST$ observations (SDSS J0841+0101, SDSS J0952+2552, SDSS J1126+2944, SDSS J1239+5314, SDSS J1322+2631, and SDSS J1356+1026; note, this list includes two systems that have $BF$ values strongly in favor of the dual point source model and secondary point sources that meets our AGN luminosity criterion), we plot the separation versus extragalactic column density in Figure~\ref{fig:nHvsR}. For SDSS J1126+2944 and SDSS J1356+1026, we plot separation and error between the X-ray point sources, as estimated by \BAYMAX{}.  For the single X-ray point source systems, we plot the separation and error between the stellar bulges, as measured by \cite{Comerford2015}. We note that the two systems for which we are insensitive to any duality (SDSS J0854+5026 and SDSS J1006+4647) are not merging, and thus are not included.  Here, $N_{H}$ is found by fitting the \emph{Chandra} observations of each system with both $m_{\mathrm{spec,1}}$ and $m_{\mathrm{spec,2}}$.
\par Because the spectroscopically determined extragalactic column density is model-dependent, the $N_{H}$ values for $m_{\mathrm{spec,1}}$ and $m_{\mathrm{spec,2}}$ vary for a given system. By using the simpler spectral model, we are estimating the \emph{average} extragalactic column density surrounding the AGNs. Although the majority of these systems are found to have a statistically better spectral fit using $m_{\mathrm{spec,2}}$, partial covering and/or the Compton scattering fraction in the torus is difficult to estimate.  Thus, the extragalactic column densities estimated with $m_{\mathrm{spec,2}}$ are useful for understanding the magnitude of the column densities in gas clumps, while those estimated with $m_{\mathrm{spec,1}}$ are useful for understanding the average column densities across the system.  The spectral fits are dominated by the emission of the primary X-ray point source, and thus the $N_{H}$ measurements mostly pertain to the environments surrounding the primary X-ray point source.  However, we interpret the $N_{H}$ value as representative of the density of gas being torqued to the center of the gravitational potential well, as a result of the galaxy$–-$galaxy mergers
\par The placement of SDSS J1126+2944 on Fig~\ref{fig:nHvsR} suggests that dual AGNs may prefer systems with both the smallest separations (as previously confirmed by \cite{Comerford2015}) and low average gas-densities (as determined by $m_{\mathrm{spec,1}}$).  Our measurement indicates that dual AGN activation could indeed be more common for merging galaxies with smaller separations, in agreement with both simulations and observations.  However, our measurement of decreasing average $N_{H}$ as a function separation is at odds with predictions, where dual AGNs are expected to reside in environments with higher levels of gas.  These results are most likely a result of selection bias; because these systems were originally selected based on their O[III] $\lambda$5007 emission, our sample of AGNs may generally have lower average extragalactic column densities. 
\par Taking the measured total $L_{\mathrm{[O III]}}$ for each system \citep{Comerford2015}, we find that the systems with the largest $N_{H}$ values tend to have lower [\ion{O}{3}] $\lambda$5007 luminosities.  Indeed, this confirms the findings of \cite{Comerford2015}, where at a given [\ion{O}{3}] luminosity, the hard X-ray luminosity of merging galaxies was found to be lower than non-merging AGNs, likely due to the high $N_{H}$ in dual AGN systems.  In particular, SDSS J1356+1026 has both the lowest average measured $N_{H}$ and the highest measured $L_{\mathrm{[O III]}}$.  All of these trends can be better understood using a larger sample of dual AGN candidates, selected via X-ray diagnostics. In particular, given our sparse data (and that only \emph{one} of the six merging galaxies are confirmed dual AGNs), future analyses with larger samples will be important.

%%%%%%%%%%%%%%%%%%%%%%%%%%%%%%%%%%%%%
%%%%%%%%%%%%% CONCLUSIONS %%%%%%%%%%%%%%%
%%%%%%%%%%%%%%%%%%%%%%%%%%%%%%%%%%%%%
\section{Conclusions}
\label{sec:conclusions}
In this work, we present our analysis using \BAYMAX{}, a tool that uses a Bayesian framework to statistically and quantitatively determine whether a given observation is best described by one or two point sources. We present the results of \BAYMAX{} analyzing a sample of 12 dual AGN candidates, originally targeted due to their double-peaked narrow emission lines.  Each system received follow-up long-slit spectroscopy, targeting the [\ion{O}{3}] $\lambda$5007 emission. Using existing \emph{Chandra} data, we carry out a statistical analysis on the X-ray emission, to determine whether the emission is more likely consistent with a single or a dual point source system. The spatially resolved [\ion{O}{3}] $\lambda$5007 emission components allow for informative priors on the location of the primary and secondary, while complementary $HST$ data allow for further analysis on environments of each system. The main results and implications of this work can be summarized as follows:
\begin{enumerate}
    \item When accounting for contamination from extended diffuse emission, we find that 4 of the 12 systems have $BF$ values strongly favor of a dual point source: SDSS J0752+2736, SDSS J1126+2944, SDSS J1356+1026, and SDSS J1448+1825.  For SDSS J0752+2736 we calculate $\ln{BF}=4.90\pm0.51$; for SDSS J1126+294 we calculate $\ln{BF}=3.54\pm0.43$; for SDSS J1356+1026 we calculate $\ln{BF}=8.70\pm0.70$; and for SDSS J1448+1825 we calculate $\ln{BF}=2.95\pm0.52$. One of these systems, SDSS J0752+2736, has a $BF$ in favor of a dual point source system only when using non-informative priors, while the remaining systems have $BF$ values in favor of a dual point source system when using both informative and non-informative priors on the location of the putative secondary. For the latter case, the $BF$ values are all stronger when using informative priors, defined by the complementary [\ion{O}{3}] $\lambda$5007 observations. 
    \item For the 4 dual AGN candidates, we analyze the strength of each $BF$ value via false-positive tests. For each of the dual AGN candidates, we find there is a $>99.9\%$ chance that the systems are composed of dual point sources.  Based on these runs, we conclude that each system has a ``strong" $BF$.
    \item We estimate the best-fit separation ($r$) and count ratio ($\log{f}$), as well as their uncertainties, for each dual AGN candidate. For SDSS J0752+2736 we find $r=1.5\arcsec\pm0.30\arcsec$ and $\log{f}=-0.47\pm0.36$; for SDSS J1126+2944 we find $r=1.74\arcsec\pm0.33\arcsec$ and $\log{f}=-1.00\pm0.44$; for SDSS J1356+1026 we find $r=2.00\arcsec\pm0.62\arcsec$ and $\log{f}=-0.92\pm0.23$; and for SDSS J1448+1825 we find $r=1.29\arcsec\pm0.52\arcsec$ and $\log{f}=-0.45\pm0.80$.
    \item We investigate the nature of each dual AGN candidate by analyzing each point source's spectrum. Because \BAYMAX{} assigns each count a probability of being associated with different model components, we are capable of fitting the spectrum of each individual X-ray point source in a given system.  We find that the secondary X-ray point sources in SDSS J1126+2944 and SDSS J1356+1026 both meet our AGN luminosity criterion ($L_{2-7~\mathrm{keV, unabs}} > 10^{40}$ erg s$^{-1}$). However, because the softer spectrum of the secondary in SDSS J1356+1026 ($\Gamma \approx 3.0\pm0.64$) is consistent with spectrum of the diffuse emission ($\Gamma \approx 3.2\pm0.25$), we conservatively do not classify this system as a dual AGN. Lastly, although the X-ray emission from SDSS J0752+2736 and SDSS J1448+1825 are better described by dual point sources, the secondaries do not meet our AGN luminosity criterion and are most susceptible to contamination from XRBs.
    \item For the 8 systems that have $BF$ values that favor a single point source, we investigate how the Bayes factor determined by \BAYMAX{} depends on the count ratio of simulated dual AGN systems with comparable counts, separations, and background fractions.  For 2 of these systems, SDSS J0854+5026 and SDSS J1006+4647, we are unable to correctly identify that the emission is consistent with two X-ray point sources, for any count ratio between $0.1$--$1.0$. This is a result of the low number of counts ($\approx$ 13 total counts between $0.5$--$8.0$), as well as small angular separation ($<1\arcsec$) assumed between the primary and secondary. However, for the remaining 6 systems, we are able to correctly identify a dual AGN system for the majority of count ratios analyzed. This corresponds to an upper-limit luminosity threshold $4\times 10^{40} < L_{2-7\mathrm{keV}} <6 \times 10^{41}$ for the secondary AGN. Thus, our dual AGN fraction of 1/12 represents a lower-limit on the true dual AGN fraction of the sample.
    \item We re-plot the \emph{WISE} mid-infrared colors of the four systems with $BF$ values in favor of the dual point source model to test whether our dual AGN candidates lie in an interesting region of IR color-color space. We confirm that only one of the four systems, SDSS J1356+1026, has an AGN-dominated mid-infrared flux. Additionally, SDSS J1356+1026 and SDSS J0752+2736 lie above a less-stringent color-cut that has been found to select both merger-triggered AGNs and dual AGNs.
    \item  We analyze how the [\ion{O}{3}]$/$H$\beta$ and [N~II]/H$\alpha$ ratios compare to the line ratio BPT diagram for the four systems with $BF$ values in favor of the dual point source model. We use available long-slit optical spectroscopic data for SDSS J0752+2736, SDSS J1126+2944, and SDSS J1448+1285 \citep{Comerford2012}; while we use archival Sloan spectral data for SDSS J1356+1026.  
    For SDSS J0752+27364$_{p}$ and SDSS J1356+1026, we find that the line ratios of each system are consistent with AGN photoionization.  For SDSS J1448+1825,  we  find  that  the line  ratios  of  the  primary  and  secondary X-ray point sources are consistent with AGN photoionization; although the X-ray luminosity of this source is below our AGN luminosity criterion, it’s possible that the X-ray emission of the secondary point source is highly obscured. For SDSS J1126+2944, we find that the line ratios of each point source are consistent with AGN photoionization, in agreement with our X-ray analysis.
    \item Lastly, we investigate whether the merger environment plays a role in the triggering of dual AGNs. Taking the six merging galaxies in our sample, we compare the separation  and the extragalactic column density of each system.  Our data suggest that dual AGNs may prefer merger environments with both the smallest separations and $N_{H}$ values. Thus, dual AGN activation may be more common for merging galaxies with smaller separations, in agreement with both simulations and observations. However, given our sparse data (and that only one of the six merging galaxies are dual AGNs), it will be important to study such trends in the future with larger samples.
\end{enumerate}

Using a quantitative and statistical tool, we have confirmed one known dual AGN system (SDSS J1126+2944). Specifically, \BAYMAX{} estimates a Bayes factor strongly in favor of the dual point source model for each system, and our spectral analysis has confirmed that emission from each point source is consistent with that expected from an AGN. In the future, we plan to use \BAYMAX{} on larger samples of \emph{Chandra} observations in order to constrain the rate of dual AGNs across our visible universe. Additionally, using larger samples of dual AGN candidates we can begin to robustly measure the types of environments dual AGNs prefer, allowing for a better understanding of black hole growth and its relation to galaxy--galaxy interactions. \\ 

We thank the referee for thorough and thoughtful comments that strengthened the paper. A.F. and K.G. acknowledge support provided by the National Aeronautics and Space Administration through \emph{Chandra} Award Numbers TM8-19007X, GO7-18087X, and GO8-19078X, issued by the \emph{Chandra} X-ray Observatory Center, which is operated by the Smithsonian Astrophysical Observatory for and on behalf of the National Aeronautics Space Administration under contract NAS8-03060. A.F. also acknowledges support provided by the National Aeronautics and Space Administration through \emph{Chandra} proposal ID 21700319, and the Rackham One-Term Dissertation Fellowship, issued by University of Michigan's Rackham Graduate School. This research has made use of NASA's Astrophysics Data System.

\software{ {\tt CIAO} (v4.8; \citealt{Fruscione2006}), XSPEC (v12.9.0; \citealt{Arnaud1996}), {\tt nestle} (https://github.com/kbarbary/nestle), {\tt PyMC3} \citep{Salvatier2016}, \\ {\tt SAOTrace} (http://cxc.harvard.edu/cal/Hrma/Raytrace/SAOTrace.html), {\tt MARX} (v5.3.3; \citealt{Davis2012})}

\bibliographystyle{aasjournal}
\bibliography{foord.bib}

\end{document}